\documentclass[12pt]{article}
\usepackage{amsmath,amssymb,amsfonts}
\usepackage{graphicx}
\usepackage[round]{natbib}
\usepackage{setspace}
\usepackage{lmodern}
\usepackage[utf8]{inputenc}
\usepackage[T1]{fontenc}
\usepackage{times}
\usepackage{hyperref}
\usepackage{multimedia}
\usepackage{bm}
\usepackage{proba}
\usepackage{enumerate}
\usepackage{natbib}
\usepackage{lscape}
\usepackage{placeins}
\usepackage{longtable}
\usepackage{epstopdf}
\usepackage{graphics}
\usepackage{tikz}
\usepackage{float}
\usepackage{placeins}
\usepackage{longtable}
\usepackage{appendix}
\usepackage{caption}
\usepackage{subcaption}
\usepackage{amstext,bbm}
\usepackage{latexsym}
\usepackage{bigints}
\usepackage{booktabs}
\usepackage{xr}
\usepackage[normalem]{ulem}
\usepackage[amsmath]{ntheorem}
\usepackage{makeidx}
\usepackage{mathtools}
\usepackage[ruled,vlined,linesnumbered,resetcount]{algorithm2e}

\renewcommand{\cite}{\citet}
\usepackage{setspace}
\newcommand{\In}{\mathbbm{1}}

\newtheorem{prop}{Proposition}[section]
\newtheorem{tr}{Theorem}[section]

\definecolor{dgreen}{rgb}{0,0.5,0}
\definecolor{dblue}{rgb}{0,0,0.9}
\definecolor{dred}{rgb}{0.6,0.0,0.1}
\definecolor{dgold}{rgb}{0.5,0.3,0.0}
\definecolor{dvio}{rgb}{0.6,0.3,0.5}
\definecolor{gray}{rgb}{0.5,0.5,0.5}

\newcommand{\bee}{\begin{equation}}
\newcommand{\eee}{\end{equation}}

\newcommand{\wtl}{\widetilde}

\makeatletter \@addtoreset{equation}{section} \makeatother \renewcommand{\theequation}{\thesection.\arabic{equation}}
\bibpunct{(}{)}{,}{a}{,}{,}
\allowdisplaybreaks
\newtheorem{example1}{Example}

\newtheorem{example2}{Example}

\newtheorem{example3}{Example}

\newtheorem{example4}{Example}

\newtheorem{example5}{Example}

\newtheorem{df}{Definition}[section]

\newtheorem{example1case1}{Example}

\input{tcilatex}
\begin{document}

\title{\vspace{-3cm} Revisiting identification concepts in Bayesian analysis
\footnote{First draft: February 2011. The authors gratefully thank the Editor Laurent Linnemer and two anonymous referees for their constructive comments on the previous version of the paper. We acknowledge
helpful comments from Andrew Chesher, Frank Kleibergen and participants to the 2011-Cowles Foundation annual Econometrics Conference, ESEM-2011 in Oslo and Econometric seminar in GREQAM-Marseille. Anna Simoni gratefully acknowledges financial support from Labex ECODEC (ANR-11-LABEX-0047).}}

\author{\begin{tabular}{ccc}
Jean-Pierre Florens\footnote{Toulouse School of Economics, Universit\'{e} de Toulouse Capitole, Toulouse - 21, all\'{e}e de Brienne - 31000 Toulouse (France). Email: jean-pierre.florens@tse-fr.eu} & & Anna Simoni\footnote{CREST, CNRS, ENSAE, \'{E}cole Polytechnique - 5, avenue Henry Le Chatelier, 91120 Palaiseau, France. Email: anna.simoni@ensae.fr (\textit{corresponding author}).}
\end{tabular}}
\maketitle

\vspace{-1cm}
\begin{abstract}
    This paper studies the role played by identification in the Bayesian analysis of statistical and econometric models. First, for unidentified models we demonstrate that there are situations where the introduction of a non-degenerate prior distribution can make a parameter that is nonidentified in frequentist theory identified in Bayesian theory. In other situations, it is preferable to work with the unidentified model and construct a Markov Chain Monte Carlo (MCMC) algorithms for it instead of introducing identifying assumptions. Second, for partially identified models we demonstrate how to construct the prior and posterior distributions for the identified set parameter and how to conduct Bayesian analysis. Finally, for models that contain some parameters that are identified and others that are not we show that marginalizing out the identified parameter from the likelihood with respect to its conditional prior, given the nonidentified parameter, allows the data to be informative about the nonidentified and partially identified parameter.
    The paper provides examples and simulations that illustrate how to implement our techniques.

\vspace{9pt} \noindent {\it Key words:} Minimal Sufficiency, Exact Estimability, Set identification, Dirichlet Process, Capacity functional, Nonparametric models.\par

\end{abstract}\par

\fontsize{10.95}{14pt plus.8pt minus .6pt}\selectfont

\section{Introduction}
This paper investigates Bayesian analysis in models that lack identification. We first revisit theoretical concepts related to identification. We highlight how lack of identification has a different impact for statistical inference depending on whether one develops Bayesian or frequentist inference. Specifically, Bayesian analysis can be carried out without imposing any identifying restrictions. Then, we distinguish between nonidentified and partially identified models. For partially identified models we propose to construct the prior and posterior distributions of a set parameter by using the capacity functional and we introduce the concepts of \emph{prior} (resp. \emph{posterior}) \emph{capacity functional} and \emph{prior} (resp. \emph{posterior}) \emph{coverage function}.\\
\indent As \cite{Lindley1971} remarked, the problem of non identification causes no real difficulty in the Bayesian approach. Indeed, if a proper prior distribution is specified, then the posterior distribution is well-defined. \cite{Kadane1974} observed that identification is a property of the likelihood function which is the same irrespective of whether it is considered from Bayesian or frequentist perspectives. It is however necessary to distinguish between three concepts of identification depending on the level of specification of the model: \emph{sampling} (or \emph{frequentist}) \emph{identification}, \emph{measurable identification} and \emph{Bayesian identification}, see \textit{e.g.} \cite{FlorensMouchartRolin1985,FlorensMouchartRolin1990} and \cite{FlorensMouchart1986}. Sampling identification is defined without the introduction of a $\sigma$-field associated with the parameter space. On the other hand, such a $\sigma$-field is necessary for the other two concepts of identification. In addition, the notion of Bayesian identification requires the introduction of a unique joint probability measure over the sample and the parameters.\\
\indent When a $\sigma$-field associated with the parameter space is introduced, the concept of identification is related to the \emph{minimal sufficient} parameter. Namely, the observed sample brings information only on the \emph{minimal sufficient} parameter and hence, the parameter of the model is identified if it is equal to the minimal sufficient parameter. In other words, the \emph{minimal sufficient} parameter is the smallest $\sigma$-field on the parameter space that makes the sampling probability measurable. So, it is the identified parameter. Conditionally on this parameter, the Bayesian experiment is completely non informative: the prior distribution of a nonidentified parameter is not revised through the information brought by the data so that the conditional posterior and conditional prior distributions (conditioned on the identified parameter) are the same.\\
\indent Equivalently, we say that a model is nonidentified if the parametrization is redundant. It is then natural to wonder why one should introduce a $\sigma$-field larger than the minimal sufficient parameter. In nonexperimental fields, redundant parametrization is usually introduced either as an early stage of model building or as a support for relevant prior information or because the parameter of interest (making \textit{e.g.} the loss function measurable) is larger than the minimal sufficient parameter (see \textbf{Example} \ref{example:1}). In experimental fields, it may be the case that the experimental design will not provide information on all the parameters of a theoretically relevant model, see \citet{FlorensMouchartRolin1990}.\\
\indent This paper makes the following contributions. (I) For nonidentified models, we show that there are situations where the introduction of a non-degenerate prior distribution can make a parameter that is unidentified in frequentist theory identified in Bayesian theory. Specifically, we demonstrate that this is true for nonparametric models with heterogeneity modeled either as a Gaussian process or as a Dirichlet process where the parameter of interest is the (hyper)parameter of the heterogeneity distribution. We show that in these models it is possible to obtain Bayesian identification since the hyperparameter can be expressed as a known function of the identified parameter. We stress that this is not a property of the prior of the nonidentified parameter, but instead it is a property of the conditional prior of the identified parameter, given the unidentified parameter. Such a result is no longer true in parametric models where a degenerate prior for the nonidentified parameter is required to get Bayesian identification, which then is completely artificial. (II) We provide the example of latent variable models to illustrate that it is preferable to conduct Bayesian inference and develop Markov Chain Monte Carlo (MCMC) algorithms for the nonidentified model instead of introducing identifying restrictions. Working with the nonidentified model grants better mixing properties of the MCMC.

(III) We propose a procedure to make Bayesian analysis for partially identified models where the identified parameter is a set, called the \emph{identified set}. For these models we build up a new Bayesian nonparametric approach, based on the Dirichlet process prior, and construct prior and posterior distributions for set parameters. We propose to define these distributions in terms of \emph{prior} and \emph{posterior capacity functionals}. The posterior capacity functional is an appealing tool to build estimators and credible sets for the identified set and, in addition, it can be easily computed either by simulations or in closed-form. (IV) We show that, when the model contains an identified parameter and a parameter of interest that lacks identification, the latter can be identified in the marginal model where the identified parameter has been marginalized out from the likelihood function with respect to its conditional prior given the unidentified parameter. Therefore, the integrated likelihood depends on the unidentified parameter and the prior of the latter is revised by the information brought by the data in the marginal model. To complement our theoretical approach we develop many examples and simulations that show how our method can be implemented.

\indent Our results contribute to show that the Bayesian approach is appealing in models that lack identification for several reasons. First, if the prior distribution on all the parameters is proper, Bayesian analysis of nonidentified and partially identified models is always possible since the posterior distribution always exists. Second, when the parameter in the model is multidimensional, with some components that are identified and others that are nonidentified, data can be marginally informative about the nonidentified parameter. That is, if the parameters are a priori dependent, then after we marginalise out the identified parameter from the likelihood with respect to its conditional prior given the nonidentified parameter, the integrated likelihood will depend on the nonidentified parameter. Third, the issue of non- and/or partial-identification can be reduced (or even eliminated) by introducing an informative prior. Lastly, even when the model is nonidentified or partially identified, Bayesian procedures have computational advantages over frequentist ones. In particular, MCMC algorithms have better mixing properties if they are specified for the nonidentified model instead of imposing identifying restrictions.\\
\indent The paper is organized as follows. In section \ref{s_general_definition} we discuss the three concepts of identification given above and the concept of partial identification. Section \ref{s_identification_by_the_prior} studies models with heterogeneity and models with latent variables. Bayesian analysis of partially identified models is developed in section \ref{s_Bayesian_set_estimation}. Finally, in section \ref{s_marginal_models} we discuss identification by marginalisation for both unidentified and partially identified models. Section \ref{s:conclusion} concludes. Examples and all the proofs are in Appendix E in the Supplement. In the paper we abbreviate ``almost surely'' by ``a.s.'', $P$ will denote the data distribution and $p$ the associated Lebesgue density. For an event $A$, $\In\{A\}$ denotes the indicator function which takes the value $1$ if the event $A$ is satisfied and $0$ otherwise.
\paragraph{Literature Review.}
Our paper is related to two strands of the Bayesian literature, the one focusing on nonidentified models and the Bayesian literature on partially identified models. We provide here a concise review of the previous  contributions that are the most relevant for our paper.\\
\indent Initial discussions of nonidentified models which lay the foundations of nonidentification in a Bayesian experiment in a measure-theoretic framework can be found in \cite{Lindley1971}, \cite{Kadane1974} and \cite{Picci1977}, among others. \cite{FlorensMouchartRolin1985,FlorensMouchartRolin1990} and \cite{FlorensMouchart1986} resume these works and provide further developments. In particular, they provide a rigorous discussion on the difference among \emph{sampling}, \emph{measurable} and \emph{Bayesian identification}. Compared to these contributions, in section \ref{s_general_definition} we provide a unified framework that gather together the concepts and results related to nonidentification that are the most relevant for applied Bayesian analysis in econometrics. We adopt a measure-theoretic framework and present the results of identification in terms of $\sigma$-fields of sets of the parameter space. An aspect that we do not consider in our paper is the prior elicitation process for nonidentified parameters which is investigated for instance in \cite{SanMartinGonzales2010}.\\
\indent As in \cite{Poirier1998}, we emphasise the difference between marginal and conditional uninformativeness of the data for Bayesian analysis of nonidentified models. The main contribution of \cite{Poirier1998} consists in analysing the diverse effect of nonidentification for Bayesian analysis in the two following situations: the case with proper priors, and the case with improper priors. Our paper does not emphasizes this difference between proper and improper priors. Instead, one of our main contributions is to demonstrate the different role played by the prior in nonparametric and parametric models that are nonidentified. The question of nonidentification in nonparametric models has not been explicitly considered in the past Bayesian literature to the best of our knowledge. For these models we prove that Bayesian identification, \textit{i.e.} through the prior, arises in a non-artificial way in some cases.\\
\indent \cite{Gustafson2005} analyses nonidentifiability by taking a nonconventional approach. Instead of contracting the model, which consists in reducing the redundant parametrisation to get identification, he proposes to expand the model to a supermodel that can at best yield identification by the prior. This is an alternative approach to ours.\\
\indent The second strand of literature that relates to our paper is the Bayesian literature on partial identification. It includes relatively few contributions in comparison to the vast frequentist literature on partially identified models. For excellent overviews of frequentist and Bayesian inference for partially identified econometric models see \cite{BontempsMagnac2017}, \cite{MOLINARI2020} and references therein. Most of the previous Bayesian literature is interested in constructing Bayesian procedures that provide asymptotically valid frequentist inferences for partially identified models, see for instance \cite{LiaoJiang2010}, \cite{Gustafson2010,Gustafson2011}, \cite{MS12}, \cite{NoretsTang2013}, \cite{KlineTamer2015}, \cite{chen2017monte}, \cite{LiaoSimoni2019} and \cite{GiacominiKitagawa2018}. They have proposed Bayesian or quasi-Bayesian approaches for constructing (asymptotically) valid frequentist confidence sets of the identified set. Their methods are mostly based either on the limited information likelihood of \cite{CH03} or on the support function. Compared to this literature, we take a different approach that is based on Dirichlet process prior placed on the identified reduced-form parameter. We construct the prior and posterior probabilities for the identified set as prior and posterior capacity functionals, which is new in the literature. In addition, we are not interested in frequentist asymptotic properties of our procedure and conduct a fully Bayesian analysis.\\
\indent Marginal analysis in partially identified models, which we consider in section 5, has been considered in \cite{LiaoJiang2010} but in a way different from ours. They marginalise out the slackness parameters in the partially identified models characterized by moment inequalities while we marginalise out the identified (reduced-form) parameter.

\section{Some General Definition}\label{s_general_definition}

\subsection{Sampling Identification}
\indent In this section we recall basic definitions in the non-Bayesian framework, called in the paper sampling theory approach. Let us consider a statistical model defined by a sampling space $X$ provided with a $\sigma$-field $\mathcal{X}$ and by a collection of probabilities on this space. The collection of sampling probability measures defined on $(X,\mathcal{X})$ is denoted by $(P^{\theta})_{\theta\in\Theta}$ and is in general indexed by a parameter $\theta \in \Theta$, which can be finite dimensional or a functional parameter. Hence, the \emph{sampling statistical model} $\mathcal{E}_{s}$ is defined as $\mathcal{E}_{s}:=\{\Theta,(X,\mathcal{X}),(P^{\theta})_{\theta\in\Theta}\}$.\\
\indent In a small sample approach we observe a finite number of realizations of a random variable taking values in a measurable space  -- for example an \textit{iid} sample $x:=(x_{1},\ldots,x_{n})\in (X,\mathcal{X})$ -- where the sample size is not made explicit and is kept fixed. In an asymptotic approach, $\mathcal{X}$ is provided with a filtration $(\mathcal{X}_{n})_{n\geq 1}$, where $\mathcal{X}_{n}$ represents the information contained in a sample of size $n$. We denote by $\mathcal{X}_{\infty} := \bigvee_{n\geq 1}\mathcal{X}_{n}$ the $\sigma$-field generated by $\bigcup_{n\geq 1}\mathcal{X}_{n}$ and write $\mathcal{X}_{n}\rightarrow \mathcal{X}_{\infty}$ as $n\rightarrow \infty$.\\
\indent Two parameters $\theta_{1}$ and $\theta_{2}$ are said to be \emph{observationally equivalent}, in symbol $\theta_{1}\sim\theta_{2}$, if $P^{\theta_{1}} = P^{\theta_{2}}$. This relation defines an equivalence relation and we denote by $\wtl{\Theta}$ the quotient space $\Theta/\sim$, \textit{i.e.} the elements $\wtl\theta$ of $\wtl{\Theta}$ are the equivalence classes over $\Theta$ by $\sim$. Notice that $\wtl{\theta}$ is a set and the elements of $\wtl{\Theta}$ are sets of parameters. We now define the concept of sampling identification.
    \begin{df}[Sampling identification]
      A real valued function $a(\theta)$ defined on $\Theta$ is \emph{identified} if $\theta_{1}\sim\theta_{2}$ implies $a(\theta_{1}) = a(\theta_{2})$. The sampling model $\mathcal{E}_{s}$ is identified if and only if any real valued function defined on $\Theta$ is identified.
    \end{df}
This definition is equivalent to say that $\theta_{1}\sim\theta_{2}$ implies $\theta_{1} = \theta_{2}$ or to say that any equivalence class is reduced to a singleton. In turn, this is equivalent to say that the sampling model $\mathcal{E}_{s}$ is identified if the mapping $\theta\rightarrow P^{\theta}$ is injective.\\
\indent For any sampling statistical model $\mathcal{E}_{s}$ there exists a canonical identified model $\wtl{\mathcal{E}}_{s}$ defined as the sampling statistical model with parameter space $\wtl{\Theta}:\wtl{\mathcal{E}}_{s}=\{\wtl{\Theta},(X,\mathcal{X}),(P^{\wtl{\theta}})_{\wtl{\theta}\in\wtl{\Theta}}\}$ where $P^{\wtl{\theta}} = P^{\theta}$ for any $\theta\sim\wtl{\theta}$. Equivalently, $\wtl{\mathcal{E}}_{s}$ is the set identified statistical model associated with $\mathcal{E}_{s}$. Thus, one may construct an identified sampling model by selecting a single element in each equivalence class. Let us call \emph{section} a function $\sigma:\wtl{\Theta}\rightarrow\Theta$ such that $\sigma(\wtl{\theta})\in \wtl{\theta}$ and $\sigma(\wtl{\theta})\in\Theta$. By using such a function one may define the identified sampling model
    \begin{displaymath}
      \mathcal{E}_{s,\sigma} := \{\Theta_{\sigma},(X,\mathcal{X}),(P^{\theta})_{\theta\in\Theta_{\sigma}}\},\qquad \Theta_{\sigma} := \sigma(\wtl{\Theta}),
    \end{displaymath}
\noindent where $\Theta_{\sigma}$ is the image of $\wtl{\Theta}$ by $\sigma$ and $P^{\theta} = P^{\sigma(\wtl{\theta})}$. In this context it is natural to look for continuous sections $\sigma(\cdot)$ or to bicontinuous bijections between $\wtl{\Theta}$ and $\Theta_{\sigma}$. In general, it is not possible to devise constructive rules for selecting elements in $\Theta$ for each $\wtl{\theta}\in\wtl{\Theta}$. The existence of $\Theta_{\sigma}$ cannot be proved by using axioms from set theory if we do not have some structure on the parameter space but it must be asserted as an additional axiom called \emph{axiom of choice}, see \textit{e.g.} \cite{KolmogorovFomin1975}.\\
\indent In the sampling theory statistics, a topological structure is needed in the parameter space in order to define statistical decision rules based on convergence, risk or loss function. A canonical topological structure is defined on $\Theta$ and then may be carried on $\wtl{\Theta}$. Let $\rho:\Theta\rightarrow \wtl{\Theta}$ be the canonical application $\theta \rightarrow \rho(\theta) = \wtl{\theta}$. The natural topological structure is the smallest one for which $\rho$ is continuous, \textit{i.e.} $\rho^{-1}(\wtl{O})$ is open in $\Theta$ whenever $\wtl{O} \subset \wtl{\Theta}$ is open. We will not detail the topological aspects of set identification, which is beyond the scope of this paper, and refer to \cite{Husmoller1994} and appendix to chapter 3 in \cite{DellacherieMeyer1975}.

\subsection{Measurable Identification}\label{ss_measurable_Identification}

\indent Let us define a measurable statistical model $\mathcal{E}_{m}$ as $\mathcal{E}_{m} := \{(\Theta,\mathcal{A}), (X,\mathcal{X}), (P^{\theta})_{\theta\in\Theta}\}$, where we use the notation previously introduced and in addition we define $\mathcal{A}$ to be a $\sigma$-field on $\Theta$ such that $P^{\theta}$ is a transition probability. Recall that given two measurable spaces $(\Theta,\mathcal{A})$ and $(X,\mathcal{X})$, the mapping $P^{(\cdot)}(\cdot):\Theta\times\mathcal{X}\rightarrow [0,1]$ is a transition probability if: \textit{(i)} $\forall \theta\in\Theta$, $P^{\theta}(\cdot)$ is a probability measure on $(X,\mathcal{X})$, and \textit{(ii)} $\forall E\in\mathcal{X}$, $P^{(\cdot)}(E)$ is a measurable function on $(\Theta,\mathcal{A})$. The introduction of the $\sigma$-field $\mathcal{A}$ of subsets of $\Theta$ is necessary in order to introduce a joint probability measure on the product space $\Theta\times X$. This probability will be introduced in section \ref{ss_Bayesian_identification}. In this sense, the introduction of a measurable statistical model is a preliminary step for the construction of a Bayesian model.\\
\indent A $\sigma$-field represents an information structure and the parameter of interest is naturally introduced as a $\sigma$-field $\mathcal{A}$ that makes both the transition probability $P^{(\cdot)}(E)$, $\forall E\in\mathcal{X}$, and the loss function of the underlying decision model, $\mathcal{A}$-measurable. We now introduce the concept of sufficient $\sigma$-field which is essential in order to discuss identification.
    \begin{df}[Sufficiency and minimal sufficiency]\label{def_2_2}
      A sub-$\sigma$-field $\mathcal{B}$ of $\mathcal{A}$ is said to be \emph{sufficient} in the model $\mathcal{E}_{m}$ if $P^{(\cdot)}(E)$ is $\mathcal{B}$-measurable for any $E\in\mathcal{X}$. A sub-$\sigma$-field $\mathcal{B}$ of $\mathcal{A}$ is said to be \emph{minimal sufficient} if: \textit{(i)} $\mathcal{B}$ is sufficient and \textit{(ii)} $\mathcal{C}\subset \mathcal{A}$ sufficient implies that $\mathcal{B}\subset\mathcal{C}$.
    \end{df}
The structure of model $\mathcal{E}_{m}$ allows for the possibility of introducing a given distribution on $\Theta$. Here, we discuss identification in model $\mathcal{E}_{m}$ without the specification of a (prior) distribution on $(\Theta,\mathcal{A})$ which will be discussed in the next section. The following proposition introduces the concept of identification in model $\mathcal{E}_{m}$, called \emph{measurable identification}.
    \begin{prop}[Measurable identification]\label{prop:2:1}
      There exists a \emph{minimal sufficient $\sigma$-field} $\mathcal{A}_{*}$ equal to the intersection of all the sufficient $\sigma$-fields. The measurable model $\mathcal{E}_{m}$ is identified if and only if $\mathcal{A}_{*} = \mathcal{A}$.
    \end{prop}
    \begin{df}[Measurable identification of a function]
      A real valued function $a(\cdot)$ defined on $(\Theta,\mathcal{A})$ is \emph{measurably identified} if it is $\mathcal{A}_{*}$-measurable.
    \end{df}
The notion of sampling and measurable identification are identical if $\Theta$ is a measurable subset of $\mathbb{R}^{k}$ provided with the Borelian $\sigma$-field and if $X$ is included in $\mathbb{R}^{n}$ and also provided with the Borelian $\sigma$-field. This equivalence is true more generally and requires that $\Theta$ ``looks like'' a Borelian of $\mathbb{R}$ and that the $\sigma$-field on $X$ is separable or equivalently generated by a countable family of subsets. Such a link between sampling and measurable identification is established in the next theorem where we assume that $(\Theta,\mathcal{A})$ is a Souslin space. We recall that a measurable space $(\Theta,\mathcal{A})$ is a Souslin space if there exists an analytic set $B\subset \mathbb{R}$ and a bimeasurable bijection between $(\Theta,\mathcal{A})$ and $(B,B\cap\mathcal{B})$, where $B\cap\mathcal{B}$ denotes the restriction to $B$ of the Borel $\sigma$-fields $\mathcal{B}$ of $\mathbb{R}$. An analytic set on $\mathbb{R}$ is the projection on $\mathbb{R}$ of a Borelian set in $\mathbb{R}^{2}$. In particular, all Borelian sets are analytic. The property that $(\Theta,\mathcal{A})$ is a Souslin space is clearly true for finite dimensional parameter spaces or more generally for Polish spaces, which include the $L^{2}$ spaces on a real space. The majority of the functional spaces usually considered in statistical and econometric applications are Polish, so the requirement that $(\Theta,\mathcal{A})$ is a Souslin space is almost always satisfied.
    \begin{tr}\label{Theorem:1}
      Let us assume that $(\Theta,\mathcal{A})$ is a Souslin space. Then,
        \begin{itemize}
          \item[(1)] a real valued function $a(\cdot)$ defined on $(\Theta,\mathcal{A})$ is measurable identified if and only if $a(\cdot)$ is constant on the equivalence class $\wtl{\Theta}$;
          \item[(2)] if $\mathcal{X}$ is separable, a model is sampling identified if and only if it is measurable identified.
        \end{itemize}
    \end{tr}
\noindent The first point of the theorem provides an additional caracterisation of measurable identification. The second part establishes that sampling and measurable identification are equivalent when $\mathcal{X}$ is separable.

\subsection{Bayesian Identification}\label{ss_Bayesian_identification}

\indent A Bayesian model consists of a measurable statistical model and a measure on $(\Theta,\mathcal{A})$ which can be either proper (if it is a probability measure) or improper, called \emph{prior distribution} and denoted by $\mu$. For simplicity we will consider a probability measure. Then, $\mu$ and $P^{\theta}$ generate a unique measure on $(\Theta\times
X,\mathcal{A}\otimes\mathcal{X})$ denoted by $\Pi$:
    \begin{displaymath}
      \Pi(A\times E) = \int_{A}P^{\theta}(E)\mu(d\theta) = \int_{E}\mu^{x}(A)P(dx),\quad A\in\mathcal{A},\: E\in\mathcal{X}
    \end{displaymath}
\noindent or, equivalently $\Pi = \mu\otimes P^{\theta} = P\otimes \mu^{x}$, where $P$ is the marginal probability measure on $(X,\mathcal{X})$ called the predictive distribution and $\mu^{x}$ is the posterior distribution. We assume that there exists a regular version of the conditional probability on $\Theta$ given $X$, that is, $\mu^{x}(\cdot)$ is a transition probability (see definition in section \ref{ss_measurable_Identification}). The Bayesian model is then defined by the following probability space
    \begin{displaymath}
      \mathcal{E}_{b} := \{\Theta\times X,\mathcal{A}\otimes\mathcal{X},\Pi\}.
    \end{displaymath}
\noindent In the following, we may use the notation $\mu(\cdot|x)$ instead of $\mu^{x}(\cdot)$ when it is more appropriate. Moreover, by abuse of notation we use $\mu(\theta|x)$ (resp. $\mu(\theta)$) to denote both the posterior (resp. the prior) distribution and its Lebesgue density function. The sampling (resp. predictive) density function with respect to the Lebesgue measure is denoted by $p(x|\theta)$ (resp. $p(x)$). Before defining the concept of Bayesian identification, we recall the definition of sufficiency and minimal sufficiency in the Bayesian model.
    \begin{df}[Bayesian sufficiency]
      A sub $\sigma$-field $\mathcal{B}$ of $\mathcal{A}$ is \emph{sufficient} in the Bayesian model if and only if $\mathcal{A} \perp\mathcal{X}|\mathcal{B}$.
    \end{df}
\indent The conditional independence of the previous definition has two equivalent characterizations: \textit{(i)} for every positive function $t : \mathcal{X}\rightarrow\mathbb{R}_{+}$, $\mathbf{E}(t(x)|\mathcal{A}) = \mathbf{E}(t(x)|\mathcal{B})$, $\Pi-a.s.$
\noindent provided that the conditional expectations exist; \textit{(ii)} for every positive function $a : \mathcal{A} \rightarrow \mathbb{R}_{+}$, $\mathbf{E}(a(\theta)|\mathcal{X}\vee\mathcal{B})=\mathbf{E}(a(\theta)|\mathcal{B})$, $\Pi-a.s.$
\indent The first characterization \textit{(i)} weakens the concept of sufficient $\sigma$-field because the property is only required almost surely with respect to the prior probability. If $\mathcal{A}$ is generated by a function $a(\cdot)$ and $\mathcal{B}$ by a function $b(\cdot)$ then \textit{(i)} means that the likelihood functions $p(x|a)$ and $p(x|b)$ are \textit{a.s.} equal. The second characterization \textit{(ii)} says that the conditional prior and posterior distributions are a.s. equal given a sufficient $\sigma$-field. Equivalently, \textit{(ii)} means that the posterior distribution $\mu(a|x,b)$ of $a$ is \textit{a.s.} equal to the prior $\mu(a|b)$.\\
\indent It may be proven that there exists a \emph{minimal Bayesian sufficient $\sigma$-field} in $\mathcal{A}$, denoted by $\mathcal{A}_{*}^{\mu}$ and defined as follows.
    \begin{df}[Minimal Bayesian sufficiency]
      A \emph{minimal Bayesian sufficient $\sigma$-field} $\mathcal{A}_{*}^{\mu}$ (also called \emph{$\mu$-a.s. minimal sufficient $\sigma$-field}) is the $\sigma$-field generated by all the versions of $\mathbf{E}(t(x)|\mathcal{A})$ for any positive $\mathcal{X}$-measurable function $t : \mathcal{X}\rightarrow\mathbb{R}_{+}$, provided the expectation exists. This $\sigma$-field $\mathcal{A}_{*}^{\mu}$ is called the projection of $\mathcal{X}$ on $\mathcal{A}$ and is also denoted by $\mathcal{AX}$.
    \end{df}
\noindent The minimal Bayesian sufficient $\sigma$-field $\mathcal{A}_{*}^{\mu}$ is a $\sigma$-field on the parameter space and depends on the prior $\mu$. An equivalent definition of $\mathcal{A}_{*}^{\mu}$ is that $\mathcal{A}_{*}^{\mu}$ is the smallest $\sigma$-field which makes the sampling probabilities $P^{(\cdot)}(E)$ measurable, $\forall E\in\mathcal{X}$, completed by the null sets of $\mathcal{A}$ with respect to $\mu$. It may be easily verified that $\overline{\mathcal{A}_{*}}\cap\mathcal{A} = \mathcal{AX} \equiv \mathcal{A}_{*}^{\mu}$ where $\overline{\mathcal{A}_{*}}$ is the $\sigma$-field generated by $\mathcal{A}_{*}$ and all the null sets of $\mathcal{A}\vee\mathcal{X}$, and $\overline{\mathcal{A}_{*}}\cap\mathcal{A}$ is the $\sigma$-field generated by $\mathcal{A}_{*}$ and all the null sets of $\mathcal{A}$ with respect to $\mu$.\\
\indent We are now ready to give the definition of Bayesian identification.
    \begin{df}[Bayesian identification]\label{def_2_6}
      The Bayesian model $\mathcal{E}_{b}$ is \emph{Bayesian identified} if and only if $\mathcal{AX} = \mathcal{A}$.
    \end{df}
\indent More generally, the concept of identification may be introduced for a sub-$\sigma$-field (\textit{i.e.} a parameter) $\mathcal{B}\subset\mathcal{A}$:
\begin{df}[Bayesian identified parameter]\label{def_2_7}
  $\mathcal{B}$ is an $\mathcal{E}_b$-identified parameter if $\mathcal{BS} = \mathcal{B}$.
\end{df}
We recall that $\mathcal{BS}$ denotes the projection of $\mathcal{S}$ on $\mathcal{B}$, that is,
$$\mathcal{BS}:=\sigma(\{\mathbf{E}[s|\mathcal{B}]; s \textrm{ belongs to the set of positive random variables defined on }(X,\mathcal{X})\}),$$
where $\sigma(\{G\})$ denotes the $\sigma$-field generated by the set $G$.\\
\indent Definition \ref{def_2_6} means that in a Bayesian identified model, $\mathcal{A}_{*}^{\mu}$ is almost surely equal to $\mathcal{A}$, that is, $\overline{\mathcal{A}_{*}}\cap\mathcal{A} = \mathcal{A}$ or: $\forall A\in\mathcal{A}$, $\exists B\in\mathcal{A}_{*}$ such that $\mu(A\bigtriangleup B) = 0$, where $\bigtriangleup$ denotes the symmetric difference. This property shows that Bayesian identification is an almost sure measurable identification with respect to the prior. We also have the following characterization of Bayesian identification, see \cite[Theorem 4.6.21]{FlorensMouchartRolin1990}.
    \begin{tr}\label{tr_Bayes_ident}
      If $(\Theta,\mathcal{A})$ is a Souslin space (see definition before Theorem \ref{Theorem:1}) and if $\mathcal{X}$ is separable, then the Bayesian model is identified if and only if $\exists \: \Theta_{0} \subset \mathcal{A}$ such that \textit{(i)} $\mu(\Theta_{0}) = 0$ and \textit{(ii)} the sampling model restricted to $\Theta - \Theta_{0}$ is identified, \textit{i.e.} $\theta\rightarrow P^{\theta}$ is injective on $\Theta - \Theta_{0}$.
    \end{tr}
\indent It is clear from the theorem that sampling identification (and equivalently measurable identification) implies Bayesian identification but the reverse is not true. For instance, a degenerate prior that puts mass one on specific values of the parameter space makes Bayesian identified a model that is not sampling identified as the following trivial example shows. Suppose that we observe a random variable $x$ from the sampling model $x|\theta_1,\theta_2 \sim \mathcal{N}(\theta_1 + \theta_2,1)$, $\theta = (\theta_1,\theta_2)' \in \Theta = \mathbb{R}^2$, and that we endow the parameter $\theta$ with the degenerate prior $\theta\sim\mathcal{N}((\theta_{10},\theta_{20})',\iota \iota')$, where $\iota = (1,1)'$. This prior gives probability zero to all the value in $\mathbb{R}^2$ but the values on the line $\theta_2 = \theta_1$. Therefore, $\Theta_0 = \mathbb{R}^2\setminus \{\theta\in\mathbb{R}^2; \theta_2 = \theta_1\}$ and the sampling distribution is injective on $\Theta - \Theta_0 = \{\theta\in\mathbb{R}^2; \theta_2 = \theta_1\}$. This model is not sampling identified nor measurable identified. However, this specification of the prior makes the model Bayesian identified.\\

\indent Before concluding this section we point out the following relationships existing between the three concepts of identification that we have seen. \textit{(i)} Measurable identification implies Bayesian identification for any $\mu$. \textit{(ii)} Measurable identification implies sampling identification if and only if $\mathcal{A}$ is separating, that is, all the atoms of $\mathcal{A}$ are singletons.

\subsection{Identification and Bayesian consistency}

In this section we present the important connection that exists between the concepts of identification and of \emph{exact estimability} in the Bayesian models. We start by defining exact estimability.
    \begin{df}[Exact estimability]
      In a Bayesian model $\mathcal{E}_{b}$, let $\mathcal{B}$ be a sub-$\sigma$-field of $\mathcal{A}$. Then, $\mathcal{B}$ is \emph{exactly estimable} if and only if $\mathcal{B}\subset \overline{\mathcal{X}}$ where $\overline{\mathcal{X}}$ is the $\sigma$-field generated by $\mathcal{X}$ and all the null sets of the product space $\Theta \times X$ with respect to $\Pi$. Equivalently, $\mathcal{B}$ is \emph{exactly estimable} if $\mathcal{B}\perp\mathcal{A}|\overline{\mathcal{X}}$.
    \end{df}
\noindent The $\sigma$-field $\overline{\mathcal{X}}$ is a $\sigma$-field on the product space $\Theta \times X$ and not a $\sigma$-field on the sampling space $X$. So, it is possible that $\mathcal{B}\subset \overline{\mathcal{X}}$ but that $\mathcal{B} \not\subset \mathcal{X}$. For instance, consider $x_i|\theta \sim^{iid}\mathcal{N}(\theta,1)$ and $\mathcal{B} = \sigma(\{\theta\})$. Consider a prior for $\theta$ that puts all its mass on the sample mean $\bar{x}$, then we clearly have that $\mathcal{B}\subset \overline{\mathcal{X}}$ but $\mathcal{B} \not\subset \mathcal{X}$. This situation is artificial in small sample, but it describes well what it happens asymptotically if $\bar{x}$ is a consistent estimator since $\bar{x}\rightarrow \theta$, $\Pi$-a.s. without requiring a degenerate prior, see also Definition \ref{def:2:9} and the discussion below it.\\
\indent The inclusion $\mathcal{B}\subset \overline{\mathcal{X}}$ means that for any positive random variable $a$ defined on $(\Theta,\mathcal{B})$, the posterior expectation $\mathbf{E}(a|\mathcal{X}) = a$ $\Pi$-a.s., provided that the conditional expectation exists. This means that a posteriori -- that is, after observing the sample -- we know $a$ $\Pi$-a.s. Thus, any sub-$\sigma$-field of $\mathcal{A}\cap\overline{\mathcal{X}}$ is exactly estimable. The next proposition states a link between exact estimability and Bayesian identification.
    \begin{prop}\label{prop:2:2}
      Any exactly estimable parameter $\mathcal{B}\subset \mathcal{A}\cap\overline{\mathcal{X}}$ is Bayesian identified.
    \end{prop}

\indent The result in the proposition, together with the following result, allows to show that in an \textit{i.i.d.} experiment, the minimal sufficient $\sigma$-field is Bayesian identified.
    \begin{tr}\label{tr:1}
      In an \textit{i.i.d.} model, the minimal sufficient $\sigma$-field of $\mathcal{A}$, $\mathcal{AX}$, is exactly estimable.
    \end{tr}
Moreover, if $\mathcal{AX}$ is exactly estimable, then the following theorem shows that the reverse implication of Proposition \ref{prop:2:2} holds.
    \begin{tr}\label{tr_exact_estimation_identification}
      Let $\{(\Theta,\mathcal{A}),(X,\mathcal{X}),\Pi\}$ be a Bayesian model such that the Bayesian minimal sufficient $\sigma$-field $\mathcal{AX}$ is exactly estimable, \textit{i.e.} $\mathcal{AX}\subset \overline{\mathcal{X}}$. Then, $\mathcal{B}\subset \mathcal{A}$ is exactly estimable if and only if $\mathcal{B}$ is a.s. identified, \textit{i.e.} $\mathcal{B}\subset\mathcal{AX}$.
    \end{tr}

\indent The concept of exact estimability has a particular interest in asymptotic models. Let $x^{(n)}:=(x_1,\ldots,x_n)$ denote a sequence of $n$ observations and $x^{(\infty)}$ denote the sample of infinite size. The $\sigma$-field generated by $x^{(n)}$ (resp. $x^{(\infty)}$) is denoted by $\mathcal{X}_{n}$ (resp. $\mathcal{X}_{\infty}$). For any function $a(\cdot)$ of the parameter $\theta$, the martingale convergence theorem implies that $\mathbf{E}(a|\mathcal{X}_{n})\rightarrow \mathbf{E}(a|\mathcal{X}_{\infty})$ a.s. This convergence is a.s. with respect to the joint distribution $\Pi$ and also with respect to the predictive probability $P$. This is one concept of \emph{Bayesian consistency} for which the convergence must be taken with respect to the joint probability distribution $\Pi$. Another concept of Bayesian consistency is convergence with respect to the sampling measure $P^{\theta}$, that is, $\mathbf{E}(a|\mathcal{X}_{n})\rightarrow a(\theta)$, $P^{\theta}$- a.s. This convergence does not follow from the previous argument and there are cases in which it is not be verified, see \textit{e.g.} \cite{DiaconisFreedman1986} and \cite{FlorensSimoni2010}.\\
\indent From the above concept of Bayesian consistency, if $\mathcal{B}\subset \overline{\mathcal{X}}_{\infty}$, we have $\mathbf{E}(a|\mathcal{X}_{n})\rightarrow a$ $\Pi$-a.s. for any $\mathcal{B}$-measurable positive function $a$ of $\theta$, that is, the posterior mean of $a$ converges a.s. to the true model. This is a consequence of the asymptotically exact estimability which we now define.
    \begin{df}[Asymptotically exact estimability]\label{def:2:9}
      Let $\mathcal{E}_{\infty}$ be the sequential Bayesian experiment defined as $\mathcal{E}_{\infty} = (\Theta \times X,\mathcal{A}\otimes\mathcal{X},\Pi, \mathcal{X}_{n}\rightarrow\mathcal{X}_{\infty})$ and $\mathcal{B}$ be a sub-$\sigma$-field of $\mathcal{A}$ such that $\mathcal{B}\subset \overline{\mathcal{X}}_{\infty}$. Then, $\mathcal{B}$ is said to be \emph{asymptotically exactly estimable} in $\mathcal{E}_{\infty}$.
    \end{df}
This definition and the martingale convergence theorem imply that if $\mathcal{B}\subset \mathcal{A}$ is asymptotically exactly estimable then, for every integrable real random variable $a$ defined on $\mathcal{B}$, $\mathbf{E}(a|\mathcal{X}_{n})\rightarrow a$ $\Pi$-\textit{a.s.}\\
\indent In a Souslin space, \emph{asymptotic exact estimability} means that the posterior distribution is asymptotically a Dirac measure on a function of the sample, see \cite[Theorem 4.8.3]{FlorensMouchartRolin1990}. The intuition is that, if $\mathbf{E}(a|\mathcal{X}_{n})\rightarrow a$ $\Pi$-a.s., then also $\mathbf{E}(a^{2}|\mathcal{X}_{n})\rightarrow a^{2}$ $\Pi$-a.s. which implies that the posterior variance of $a$ converges to $0$. This clearly explains that the posterior distribution concentrates around the $\Pi$-a.s. limit of its mean. More generally, an integrable real random variable $a(\cdot)$ defined on $\mathcal{B}$ is asymptotically exactly estimable if there exists a strongly consistent sequence of estimators of $a(\theta)$, \textit{i.e.} if there exists a set of random variables $(t_{n})_{n\in\mathbb{N}}$, each defined on $\mathcal{X}$, such that $t_{n}\rightarrow a(\theta)$, $\Pi$-a.s. In sampling theory, a necessary condition for the existence of such a sequence is the identifiability of the parameter.

\subsection{Nonidentified and partially identified models}\label{sss_nn_ident_partially_ident}
Between the two extreme situations of identification and non-identification there are models that are \textit{partially identified}. Partial identification arises when the combination of available data and assumptions that are plausible for the model only allows to place the population parameter of interest $\gamma$ within a proper subset $\Gamma_{I}$ of the parameter space $\Gamma$ called identified set. If one has very precise prior information, identification can be restored by specifying a prior distribution degenerated at some point in $\Gamma_I$ as we explain in section \ref{ss_models_hyperparameters} - Model 5. However, this strategy can be unsuitable if one is concerned with robustness of the Bayesian procedure. Therefore, a more appealing approach consists in working directly with the parameter $\Gamma_I$ which is a set and: first, define the prior and posterior distribution of $\Gamma_I$ (see section \ref{s_Bayesian_set_estimation}); second, define a prior for $\gamma$ inside this set (see section \ref{ss:marginal:set:estimation}).\\
\indent In a partially identified model, usually, there is an identified parameter, say $\theta$, and a partially identified parameter which is the parameter of interest, denoted by $\gamma$. The latter enters the model as a supplementary parameter and is linked to $\theta$ through a relation of the form $A(\theta,\gamma)\in A_{0} \subset \Phi$ where $\Phi$ is a suitably defined space and $A$ is a given function of $(\theta, \gamma)$. Such relation characterizes the identified set as $\Gamma_{I}:=\{\gamma;\:A(\theta,\gamma)\in A_{0} \in \Phi\}$ which may not be a singleton. The function $(\theta, \gamma)\mapsto A(\theta, \gamma)$ may be a parametric likelihood function, flat over a region of $\gamma$s, and $A_0$ is its maximum value. Alternatively, $\Gamma_{I}$ may be a set of moment restrictions as in \textbf{Example} \ref{example:1} below. See \cite{chen2017monte} for more details on Bayesian analysis of these examples.
\begin{example}[Moment Restrictions] \label{example:1}
  A natural construction of sampling econometric model which may lead to identification issues is the following. The econometrician first considers a sampling model perfectly identified $\{\Theta,(X,\mathcal{X}),(P^{\theta})_{\theta\in \Theta}\}$ but she completes her specification by considering another parameter $\gamma\in \Gamma$. This parameter $\gamma$ is the natural parameter arising in economic models and is related to $\theta$. The relation linking $\theta$ and $\gamma$ may be of the form $A(\theta,\gamma) = 0$ or $A(\theta,\gamma) \geq 0$ or more generally $A(\theta,\gamma) \in A_{0}$, $A_{0}\in\Phi$ where $\Phi$ is a suitably defined space and $A$ is a given function of $(\theta, \gamma)$. The parameter $\gamma$ may be identified or not depending on the relation $A$. For example, the relation $A$ may be a set of moment restrictions specified with respect to the distribution $F$ which generates the data. In that case $\theta=F$ the parameter $\gamma$ could be characterized through the moment condition
    \begin{equation}\label{eq_GMM_moment_condition}
      A(\theta,\gamma) := \mathbf{E}_{F}(h(x,\gamma)) = 0,
    \end{equation}
\noindent where $h$ is a known moment function and $\mathbf{E}_{F}(\cdot)$ denotes the expectation taken with respect to $F$. The parameter $\gamma$ is identified if a unique solution to equation \eqref{eq_GMM_moment_condition} exists. Condition (\ref{eq_GMM_moment_condition}) may be extended into $\mathbf{E}_{F}(h(x,\gamma)) \geq 0$ which in general defines a set of $\gamma$ solutions to these inequalities for a given $F$.
\end{example}
\indent To develop Bayesian analysis it is important to understand that the specification of the prior distribution differs in nonidentified models and in partially identified models. In partially identified models, the prior distribution is naturally decomposed in the marginal prior on $\theta$ and the conditional prior on $\gamma$ given $\theta$ which incorporates the link between $\theta$ and $\gamma$: $x|\theta,\gamma \sim P^{\theta}$, $\theta\sim \mu_{\theta}$ and $\gamma|\theta\sim \mu_{\gamma}(\cdot|\theta)$. \\
\indent On the other hand, models with nonidentified parameters arise when $\theta$ is partitioned in $\theta = (\beta,\gamma)$ where $\beta$ is identified and $\gamma$ is not. For instance, $\gamma$ can be the parameter of the distribution of the heterogeneity parameter $\beta$. Therefore, the prior is naturally decomposed in the prior on $\beta$ conditional on $\gamma$, $\mu_{\beta}(\cdot|\gamma)$, and the prior $\mu_{\gamma}$ on $\gamma$. This decomposition of the prior distribution is the reverse of the decomposition made for partially identified models.
%
\section{Bayesian analysis of nonidentified models}\label{s_identification_by_the_prior}
This section is made of two subsections that describe two different options one has to make Bayesian inference for nonidentified models. We first discuss and construct models that may satisfy Bayesian identification even if they are not sampling identified nor measurable identified. This happens when the function $\theta \mapsto P^{\theta}$ is not injective but the prior assigns a mass zero on the nonidentified component of the model, as it has been shown in Theorem \ref{tr_Bayes_ident}. A second option, discussed in section \ref{ss:latent:variable:models}, consists in working with the nonidentified model without introducing any identifying priors or identifying assumptions. The beauty of the Bayesian approach is that, as long as the prior is proper, one can make Bayesian inference because the posterior distribution is well defined and then MCMC algorithms can be designed to simulate from the posterior even in a nonidentified model.
%
\subsection{Models with heterogeneity and hyperparameters}\label{ss_models_hyperparameters}
\indent This section considers models where the parameter of interest $\gamma$ is a parameter of the distribution of a heterogeneity parameter $\beta$. More precisely, the parameter space $(\Theta,\mathcal{A})$ contains two subparameter spaces $B$ and $\Gamma$ with the associated sub-$\sigma$-fields $\mathcal{B}$ and $\mathcal{G}$, where $\mathcal{B}$ is the $\mu$-a.s. minimal sufficient $\sigma$-field in the parameter space. The $\sigma$-field $\mathcal{G}$ is not identified in the sense that $\mathcal{G}\nsubseteq\mathcal{B}$. The sampling distribution only depends on $\mathcal{B}$ and the Lebesgue data density, if it exists, satisfies $p(x|\beta,\gamma) = p(x|\beta)$ a.s., where $\beta\in(B,\mathcal{B})$ and $\gamma\in (\Gamma,\mathcal{G})$. The posterior satisfies $\mu(\gamma|x,\beta) = \mu(\gamma|\beta)$ a.s. An interesting situation, known as local identification, arises when the $\sigma$-field $\mathcal{G}$ is not identified only for particular values taken on by the parameter $\mathcal{B}$, that is, $p(x|\beta=\bar{\beta},\gamma) = p(x|\beta=\bar{\beta})$ for some value $\bar{\beta}$, see \textit{e.g.} \cite{DREZE1983}, \cite{KleibergenVanDijk1998}, \cite{HoogerheideEtAL2007} and \cite{KleibergenMavroeidis2011}. We do not analyze this situation in this paper.\\
\indent The prior is naturally specified as the product of the conditional distribution of $\mathcal{B}$ given $\mathcal{G}$ and the marginal on $\mathcal{G}$. In this case, $\mathcal{G}$ is interpreted as a parameter of the prior on $\mathcal{B}$, usually called an \emph{hyperparameter}, see \textit{e.g.} \cite{Berger1985}, or \emph{latent variable}. Even if this model is not sampling (nor measurable) identified, it may satisfy Bayesian identification if the prior is conveniently specified as it has been shown in Theorem \ref{tr_Bayes_ident}. This identification by the prior is fully artificial in the finite-dimensional parameter case where a degenerate prior is required to get identification and we do not consider this case. Instead, we consider nonparametric and asymptotic settings where identification can be obtained without a degenerate prior. The main argument that we use to get Bayesian identification in this way is the following.
    \begin{prop}
      Let us consider the Bayesian model $\mathcal{E}_{b} = \{\Theta\times X,\mathcal{A}\otimes\mathcal{X},\Pi\}$ and a sub-$\sigma$-field $\mathcal{B}\subset \mathcal{A}$ such that $\mathcal{B} = \mathcal{AX}$. Then, any sub-$\sigma$-field $\mathcal{G}\subset \mathcal{A}$ is identified if and only if $\mathcal{G}\subset \mathcal{B}$, $\mu$-a.s. or, equivalently, if any $\mathcal{G}$-measurable function is $\mu$-a.s. equal to a $\mathcal{B}$-measurable function.
    \end{prop}
\noindent Notice that the \emph{a.s.} in the proposition are related to the prior and that $\mathcal{B}$ is the minimal sufficient $\sigma$-field. The proposition states that any sub-$\sigma$-field $\mathcal{G}\subset \mathcal{A}$ is identified if and only if it is almost surely included in $\mathcal{B}$. Intuitively, the result of the proposition holds if the conditional prior distribution on $\mathcal{G}$ given $\mathcal{B}$ is degenerate into a Dirac measure on a $\mathcal{B}$-measurable function. No requirement are made on the marginal prior on $\mathcal{G}$. We now discuss four models where Bayesian identification can be obtained without a degenerate marginal prior on $\gamma$ and a fifth example where the model is partially identified, and contrarily to the previous cases, Bayesian identification can be only obtained artificially with a degenerate marginal prior.
\paragraph{Model 1: unobserved heterogeneity (or incidental parameter).}
The incidental parameter problem typically arises with panel data models when a regression model includes an agent specific intercept $\beta_{i}$ which is a latent variable. The general conditional model can be simplified as
    \begin{displaymath}
      x_{it}|\beta_{i},\gamma \sim \: ind \:\mathcal{N}(\beta_{i},\sigma^{2}),\qquad i = 1,\ldots,\infty,\quad t=1,\ldots,T
    \end{displaymath}
\noindent where $\sigma^{2}$ is known and $\gamma$ is a hyperparameter characterizing the distribution of $\beta_{i}$. If $\beta_{i}$ is \textit{i.i.d.} across individuals then $\gamma$ is the common parameter in the population. Let $\beta:=(\beta_{1},\beta_{2},\ldots)$. The $\mu$-\textit{a.s.} \emph{minimal sufficient $\sigma$-field} $\mathcal{B}$ is generated by $\beta$, that is, $\beta$ is the identified parameter. Instead of estimating the distribution of the heterogeneity parameter $\beta_{i}$ one could be satisfied with the estimation of the common parameter $\gamma$. Unfortunately, $\gamma$ is not measurably identified even if it may be Bayesian identified. To illustrate this, let us specify a prior probability for $\beta$ and $\gamma$ as follows:
    \begin{eqnarray*}
      \beta_{i}|\gamma & \sim & i.i.d. \:\mathcal{N}(\gamma,\sigma_{0}^{2}),\qquad i = 1,\ldots,\infty\\
      \gamma & \sim & \mu_{\gamma},
    \end{eqnarray*}
\noindent where $\sigma_{0}^{2}$ is known and $\mu_{\gamma}$ is a probability measure. By the strong law of large numbers, $\gamma = \lim_{n\rightarrow\infty}\frac{1}{n}\sum_{i=1}^{n}\beta_{i}$ \textit{a.s.} with respect to both the conditional prior distribution of $\beta$ given $\gamma$ and the joint prior distribution of $(\beta,\gamma)$. Therefore, the conditional distribution of $\gamma|\{\beta_i\}_{i=1}^{\infty}$ puts all its mass on $\lim_{n\rightarrow\infty}\frac{1}{n}\sum_{i=1}^{n}\beta_{i}$ and the
model is fully Bayesian identified. We stress that this property does not depend on the marginal prior on $\gamma$, which does not have to be degenerate on some value, but crucially depends on the prior of $\beta$ given $\gamma$.
%
\paragraph{Model 2: Gaussian process and hyperparameter in the mean.}
  Let $L^{2}[0,1]$ denote the space of square integrable functions defined on $[0,1]$ with respect to the uniform distribution, endowed with the $L^{2}$-inner product $\langle\cdot,\cdot\rangle$ and the induced norm $\|\cdot\|$. We consider a sample space $X=L^{2}[0,1]$. The sample is made of a single observation which is a trajectory $x$ from a Gaussian process in $X$: $x|\beta,\gamma \sim \mathcal{GP}(\beta,\Sigma)$ where $\beta\in X$ and $\Sigma : X\rightarrow X$ is a covariance operator which is bounded, linear, positive definite, self-adjoint and trace-class. It follows that $\mathbf{E}||x||^{2}<\infty$. The parameter $\beta$ is measurably identified but the parameter $\gamma$ is not and it may be interpreted as a low dimensional parameter that controls the distribution of $\beta$.\\
\indent Suppose that $\beta$ and $\gamma$ are two random variables with values in $(B,\mathcal{B})$ and $(\mathbb{R}_{+},\mathcal{R})$, respectively, with $B = L^{2}[0,1]$ and $\mathcal{B}$ and $\mathcal{R}$ are the associated Borel $\sigma$-field. The joint prior distribution on $(B\times\mathbb{R}_{+},\mathcal{B}\otimes\mathcal{R})$ is specified as:
    \begin{eqnarray*}
      \beta|\gamma & \sim & \mathcal{GP}(\gamma\beta_{0}, \sigma_{0}^{2}\Omega), \;\;\qquad \beta_{0}\in B=L^{2}[0,1],\;\; \qquad\Omega:B\rightarrow B\\
      \gamma & \sim & \mathcal{N}(\gamma_{0},\sigma_{0}^{2}), \qquad \gamma_{0},\: \sigma_{0}^{2}\in\mathbb{R}_{+},
    \end{eqnarray*}
\noindent where $\beta_{0}$, $\gamma_{0}$, $\sigma_{0}^{2}$ and $\Omega$ are known hyperparameters. The covariance operator $\Omega$ is bounded, linear, positive definite, self-adjoint and trace-class and its eigensystem is denoted by $(\lambda_{j},\varphi_{j})_{j\geq 1}$. By definition of Gaussian processes in Hilbert space, $\beta|\gamma \sim \mathcal{GP}(\gamma\beta_{0}, \sigma_{0}^{2}\Omega)$ if and only if $\langle \beta,\varphi_{j}\rangle |\gamma \sim \mathcal{N}(\langle \beta_{0},\varphi_{j}\rangle \gamma, \sigma_{0}^{2}\lambda_{j})$ for all $j\geq 1$. Hence, it follows that
$$\left.\frac{\langle \beta,\varphi_{j}\rangle}{\sigma_{0}\sqrt{\lambda_{j}}}\right|\gamma \sim \mathcal{N}\Big(\langle \beta_{0},\varphi_{j}\rangle/(\sigma_{0}\sqrt{\lambda_{j}})\gamma,1\Big).$$
Let us write $\beta = \beta_0\gamma + U$, where $U\sim\mathcal{GP}(0,\sigma_0^2 \Omega)$. Let $\gamma^*(\beta)$ denote the minimizer of the least squares criterion: $\gamma^*(\beta):=\arg\min_{\gamma}\sum_{j=1}^{\infty} (\langle\beta,\varphi_{j}\rangle - \langle\beta_0,\varphi_{j}\rangle\gamma)^2/(\sigma_{0}^{2}\lambda_{j})$ which takes the form
    \begin{equation}\label{eq:1}
      \gamma^*(\beta) = \frac{\sum_{j=1}^{\infty}\frac{\langle\beta,\varphi_{j}\rangle \langle\beta_{0},\varphi_{j}\rangle}{\sigma_{0}^{2}\lambda_{j}}}{\sum_{j=1}^{\infty}
      \frac{\langle\beta_{0},\varphi_{j}\rangle^{2}}{\sigma_{0}^{2}\lambda_{j}}} = \gamma + \frac{\sum_{j=1}^{\infty}\xi_{j}\frac{\langle\beta_{0},\varphi_{j}\rangle}{\sqrt{\lambda_{j}}}}{\sum_{j=1}^{\infty}
      \frac{\langle\beta_{0},\varphi_{j}\rangle^{2}}{\sigma_{0}\lambda_{j}}},
    \end{equation}
\noindent where $\xi_{j} := \langle U,\varphi_{j}\rangle / (\sigma_0\sqrt{\lambda_j}) \sim i.i.d. \; \mathcal{N}(0,1)$, $j\geq 1$. If $\langle\Omega^{-1/2}U,\Omega^{-1/2}\beta_{0}\rangle =0$ then the second term on the right hand side of \eqref{eq:1} is zero and so $\gamma^*(\beta) = \gamma$. It follows that we can write $\gamma$ as a function of $\beta$ and we have exact estimability and Bayesian identification of $\gamma$, that is, $\gamma\in\overline{\mathcal{B}}$. As in \textbf{Model 1}, this property does not depend on the marginal prior on $\gamma$ but crucially depends on the prior of $\beta$ given $\gamma.$ The condition $\langle\Omega^{-1/2}U,\Omega^{-1/2}\beta_{0}\rangle =0$ means that the scaled error term and the prior mean functions have to be orthogonal in $L^2[0,1]$.

\paragraph{Model 3: Gaussian process and hyperparameter in the variance.}
Let us consider the same setting as in Model 2 but suppose that now we are interested in the common variance parameter $\sigma^2$. That is, the prior distribution for $(\beta,\sigma^2)\in(B\times\mathbb{R}_{+},\mathcal{B}\otimes\mathcal{R})$ is specified as:
    \begin{eqnarray*}
      \beta|\sigma^{2} & \sim & \mathcal{GP}(0,\sigma^{2}\Omega), \qquad \textrm{and}\qquad \sigma^{2} \sim \mathcal{I}\Gamma (\nu_{0},\sigma_{0}^{2}),
    \end{eqnarray*}
\noindent where $\mathcal{I}\Gamma$ denotes an inverse-gamma distribution and $\Omega:X\rightarrow X$ is a known covariance operator which is bounded, linear, positive definite, self-adjoint and trace-class. The hyperparameters $\nu_{0}$ and $\sigma_{0}^{2}$ are known. Knowledge of $\Omega$ implies knowledge of its eigensystem $(\lambda_{j},\varphi_{j})_{j\geq 1}$.\\
\indent By definition, $\beta|\sigma^2\sim \mathcal{GP}(0,\sigma^{2}\Omega)$ if and only if $\left\{\langle\beta,\varphi_{j}\rangle/\sqrt{\lambda_{j}}\right\}_{j\geq 1}|\sigma^2\sim^{i.i.d.}\mathcal{N}(0,\sigma^{2})$. By the strong law of large numbers we have
    \begin{displaymath}
      \lim_{J\rightarrow\infty}\frac{1}{J}\sum_{j=1}^{J}\frac{\langle\beta,\varphi_{j}\rangle^{2}}{\lambda_{j}} = \sigma^{2},\quad \mu-a.s.
    \end{displaymath}
\noindent This convergence is valid with respect to both the conditional prior $\mu(\beta|\sigma^{2})$ of $\beta$ given $\sigma^{2}$ and the joint prior $\mu(\beta,\sigma^{2})$ of $(\beta,\sigma^{2})$ and shows that $\sigma^{2}$ is a $\mu-a.s.$ function of $\beta$. Therefore, $\sigma^{2}$ is Bayesian identified.
%
\paragraph{Model 4: Bayesian nonparametric and Dirichlet process.}
  We observe an $n$-sample from the following sampling model $x_{i}|F,G\sim\:i.i.d.\: F$, $i=1,\ldots,n$, where $F$ and $G$ are two probability distributions -- for instance on $\mathbb{R}$. We can interpret $F$ as a nuisance parameter and $G$ is the only parameter of interest which characterizes the distribution of $F$. The parameter $F$ generates a $\sigma$-field $\mathcal{B}$ that is measurable identified while $G$ generates a $\sigma$-field $\mathcal{G}$ that is not measurable identified. We specify a nonparametric Dirichlet process prior for $F$ with parameters $n_{0}$ and $G$: $F|G \sim \mathcal{D}ir(n_{0},G)$ and $G \sim \mu$, where $\mu$ denotes a probability measure on the space of distributions that generate almost surely a diffuse distribution, \textit{i.e.} $G(x) = 0$, $\forall x$. The $F$ generated from the above Dirichlet process can be equivalently generated by using the stick-breaking representation as $F=\sum_{j}\alpha_{j}\delta_{\xi_{j}}$, where $\{\xi_{j}\}_{j\geq 1}$ are independent draws from $G$, $\alpha_{j} = v_{j}\prod_{k=1}^{j}(1 - v_{k})$ with $\{v_{j}\}_{j\geq 1}$ independent draws from a Beta distribution $\mathcal{B}e(1,n_{0})$ and $\{v_{j}\}_{j\geq 1}\perp\{\xi_{j}\}_{j\geq 1}$, see Appendix D in the Supplement. Then, $\lim_{J\rightarrow \infty}\frac{1}{J}\sum_{j=1}^{J}\delta_{\xi_{j}} = G$,\quad $\mu-a.s.$ which proves that $G$ is a $\mu$-a.s. function of $F$ and so, it is Bayesian identified.
%
\paragraph{Model 5: Moment conditions and partially identified models.}\label{ss_partially:identified:models}
  Consider the setting of \textbf{Example} \ref{example:1} where the parameter of interest $\gamma\in(\Gamma,\mathcal{G})$ is characterized through a moment condition of the type $\mathbf{E}_{F}(h(x,\gamma))\geq 0$ for a known function $h$ and a data generating process $F$. We observe $n$ realisations $(x_{1},\ldots,x_{n})$ of $n$ \textit{i.i.d.} random variables from $F$ and specify for $F$ a Dirichlet process prior with parameters $n_{0}$ and $F_{0}$:
    \begin{eqnarray*}
      x_{i}|F & \sim & \: i.i.d.\: F, \qquad i=1,\ldots,n,\\
      F & \sim & \: \mathcal{D}ir(n_{0},F_{0}).
    \end{eqnarray*}
  \noindent The parameter $F$ generates the $\sigma$-field $\mathcal{B}$ and is measurable identified. Suppose that $\gamma$ is partially identified. Despite of this, $\gamma$ can be Bayesian identified by specifying a conditional prior $\mu(\gamma|F)$ for $\gamma$, given $F$, degenerate on a given functional of $F$. For example, if the restriction $\mathbf{E}_{F}(h(x,\gamma)) \geq 0$ writes $\gamma\in[\phi_{1}(F),\phi_{2}(F)]$ for two functionals $\phi_{1}$, $\phi_{2}$ of $F$, then $\mu(\gamma|F)$ could be specified as a Dirac on $(\phi_{1}(F) + \phi_{2}(F))/2$. We stress that Bayesian identification in this model is obtained artificially and in a different way than in Models 1-4 above because of the construction of a conditional degenerate prior for the partially identified parameter $\gamma$, given the identified one. Instead, in Models 1-4 we have specified a marginal prior for the unidentified parameter that was not degenerate and Bayesian identification arose because of the specification of the conditional prior for the identified parameter.\\
  \indent This artificial way to obtain Bayesian identification through a degenerate prior can be reprehensible as it can be seen against the logic of partial identification. In section \ref{s_Bayesian_set_estimation} we will proceed without imposing Bayesian identification.

\subsection{Latent variable models}\label{ss:latent:variable:models}
In this section we take the example of latent variable models to illustrate that it is possible to make Bayesian inference directly for the nonidentified model without introducing an identifying prior or identifying assumptions. Because these models are well known in the literature we discuss them briefly.\\
\indent Consider discrete observable outcomes $y_i$ arising from the model $y_i = g(z_i)$, where $g(\cdot)$ is a given function and $z_i$ is a latent variable satisfying: $z_i = x_i'\delta + u_i$, where $x_i$ is an observable real-valued random vector, $\delta$ is the parameter vector of interest, and $u_i$ is an unobservable component with unrestricted variance. Identification problems appear if the $g(\cdot)$ function is invariant to location or scale transformations of $z$. Differently from frequentist analysis, imposing identifying restrictions is not necessary if one wants to conduct Bayesian analysis. With proper priors, one obtains the posterior for the nonidentified model, constructs an MCMC algorithm to simulate from this posterior and then marginalises to obtain the posterior of the identified parameters. It is often the case that it is easier to construct MCMC algorithms for unrestricted parameters and they have better mixing properties than MCMC algorithms for the corresponding identified model, see \textit{e.g.} \cite{vanDykMeng2001}.\\
\indent The simpler example of a latent variable model is the binary logit / probit, where the observable outcome $y_i$ is binary and modeled as: $y_i = \In\{z_i>0\}$ with $z_i$ a real-valued random variable. If $u_i$ follows a Gaussian (resp. Logistic) distribution we have the probit (resp. logit) model. The identification problem arises because the indicator function is invariant to scale transformation of $z_i$: multiplication of $z_i$ by a positive constant does not change the likelihood. Identification can be restored by imposing for instance the restriction $Var(u_i) = 1$. However, Bayesian analysis does not require any identifying restrictions. Other example of latent variable models are provided in Appendix A.

\section{Bayesian analysis of partially identified models}\label{s_Bayesian_set_estimation}
This section considers partially identified models as described in section \ref{sss_nn_ident_partially_ident}. Let us consider a model with an identified parameter $\theta$ and another parameter $\gamma$ characterized by the condition
    \begin{equation}\label{eq_moment_equation}
      A(\theta,\gamma)\in A_{0},
    \end{equation}
\noindent where $\gamma\in\Gamma\subset\mathbb{R}^{d_{\gamma}}$, $A: (\Theta,\mathcal{A},\mu) \times\Gamma\rightarrow \Phi$, for some finite or infinite dimensional space $\Phi$ and $A_{0}$ is a subset of $\Phi$. The function $A(\theta,\gamma)$ is usually either a likelihood function or a moment function and condition \eqref{eq_moment_equation} can contain both inequalities and equalities, see \cite{chen2017monte} for interesting examples. The parameter $\gamma$ is the parameter of interest and, depending on the relation \eqref{eq_moment_equation}, it may be only partially identified which means that for a given $\theta$ there may exist more than one value of $\gamma$ in $\Gamma$ satisfying \eqref{eq_moment_equation} but not all the values in $\Gamma$ satisfy the condition.\\
\indent Let $\Gamma_{I} := \Gamma_{I}(\theta) = \{\gamma\in\Gamma; A(\theta,\gamma) \in A_{0}\}$ be the \emph{identified set} which is a proper subset of $\Gamma$ and depends on $\theta$. The model is point identified if $\Gamma_{I}$ is a singleton and is partially identified otherwise. Inference of partially identified models can focus either on $\gamma$, or on the set $\Gamma_I$ or on both. In this section, we focus on $\Gamma_I$ and so do not endow $\gamma$ with a prior. A prior on $\gamma$ will be introduced in section \ref{s_marginal_models}.\\
\indent In the Bayesian analysis, the quantity $A(\theta,\gamma)$ in \eqref{eq_moment_equation} is a random element, where the randomness comes from $\theta$: for each $\gamma\in\Gamma$, $A(\cdot,\gamma)$ is a function on $(\Theta,\mathcal{A})$ which takes values in the state space $(\Phi,\mathcal{B}(\Phi))$, where $\mathcal{B}(\Phi)$ denotes the Borel $\sigma$-field associated with $\Phi$. Therefore, condition \eqref{eq_moment_equation} has to hold \emph{a.s.} with respect to the prior distribution of $\theta$. For a realisation of $\theta$ in $\Theta$, the set $\Gamma_{I}$ is constructed on the basis of the trajectory $\gamma \mapsto A(\theta,\gamma)$, $\gamma\in\Gamma$. Therefore, $\Gamma_{I}(\theta)$ has to be interpreted as a random set where the randomness comes from $\theta$. This differs from the frequentist analysis of partially identified models where $\Gamma_I$ is non random.\\
\indent The next definition describes a random closed set. For this, let $\mathcal{C}$ (resp. $\mathcal{O}$) denote a family of closed (resp. open) subsets of $\Gamma$.
    \begin{df}[Random closed set]
      Let $(\Theta, \mathcal{A},\mu)$ be a complete probability space and $\Gamma$ be a locally compact space. The multivalued function $\Gamma_{I}(\cdot):\Theta \rightarrow \mathcal{C}$ is called a \emph{random closed set} if, for every compact set $K$ in $\Gamma$, $\{\theta; \: \Gamma_{I}(\theta)\cap K \neq \varnothing\}\in\mathcal{A}.$
    \end{df}
\noindent The multivalued function $\Gamma_{I}(\cdot):\Theta \rightarrow \mathcal{O}$ is called a \emph{random open set} if its complement is a random closed set. The following proposition gives a condition that guarantees that $\Gamma_{I}$ is a random closed set. A regular closed set $C$ is such that $C$ coincides with the closure of its interior, see \cite{Molchanov2005}.
    \begin{prop}\label{prop:1}
      Assume that the multivalued function $\Gamma_{I}(\cdot):\Theta \rightarrow \mathcal{C}$ has realizations \emph{a.s.} equal to regular closed subsets of $\Gamma$. If the set $A_0$ in \eqref{eq_moment_equation} is such that $A_{0}\in\mathcal{B}(\Phi)$ then $\Gamma_{I}(\cdot)$ is a closed random set.
    \end{prop}

\indent We assume in the following that $\Gamma_{I}(\theta)$ is a closed random set. Our analysis can be extended to the case where $\Gamma_{I}(\theta)$ is an open random set. It is convenient to introduce the stochastic process $\{g_{\theta}(\gamma)\}_{\gamma\in \Gamma}$ associated with condition \eqref{eq_moment_equation}, where for every $\gamma\in\Gamma$
    \begin{equation}\label{eq:2}
      g_{\theta}(\gamma) := \mathbbm{1}\{A(\theta,\gamma)\in A_{0} \subset \Phi\}.
    \end{equation}
\noindent So, $g_{\theta}(\gamma)$ is the indicator of the random set $\Gamma_{I}(\theta)\subset\Gamma$ and $\gamma$ is the index of the stochastic process $g_{(\cdot)}(\cdot):(\Theta,\mathcal{A},\mu)\times\Gamma \rightarrow \{0,1\}$. When $\Gamma_{I}(\theta)$ is a separable random set, see \cite[Definition 4.8]{Molchanov2005}, then it can be completely described through its indicator function.

\subsection{Construction of the prior and posterior capacity functionals}\label{ss_construction_prior_post_capacity_functionals}

\indent A Bayesian analysis requires the specification of a prior distribution for the random set $\Gamma_{I}$. Since $\Gamma_{I}$ depends on $\theta$, we propose to construct such a prior by first specifying a prior distribution for $\theta$ and then recovering from it the prior for $\Gamma_{I}$. Suppose $X \subseteq \mathbb{R}$ and let $F$ denote the data distribution. We suppose that $\theta$ may be written as a measurable functional of the distribution $F$, that is, there exists a measurable functional $\phi(\cdot)$ such that $\theta = \phi(F)$. For instance, $\theta=\mathbf{E}_{F}(x) = \int x dF(x)$.\\
\indent In our analysis $F$ is not restricted to belong to some parametric class. Thus, we specify a Dirichlet process prior for $F$ and write $F\sim\mathcal{D}ir(n_{0},F_{0})$ where $n_{0}\in\mathbb{R}_{+}$ and $F_{0}$ is a diffuse probability measure on $\mathbb{R}$, \textit{i.e.} $F_{0}(y) = 0$, $\forall y\in\mathbb{R}$, see \textit{e.g.} \cite{Ferguson1973} and \cite{Florens2002}. The prior distribution for $\Gamma_{I}(\theta)$ is obtained from this prior for $F$ through the \emph{prior capacity functional}. Define $\mathcal{K}$ as the family of compact subsets of $\Gamma$. The \emph{prior capacity functional} $T_{\Gamma_{I}}:\mathcal{K} \mapsto[0,1]$ is given by
    \begin{displaymath}
      T_{\Gamma_{I}}(K) := P \{K \: \cap \: \Gamma_{I}(\theta) \neq \varnothing\}, \quad K\in\mathcal{K},
    \end{displaymath}
\noindent where the probability $P$ is determined by the prior distribution of $\theta$ which in turns is determined by the Dirichlet process prior on $F$: $F\sim \mathcal{D}ir(n_{0},F_{0})$ since $\theta=\phi(F)$.\\
\indent When the prior capacity functional is defined on singletons instead of on $\mathcal{K}$, \textit{i.e.} $K = \{\gamma\}$ for some $\gamma\in\Gamma$, then it is called \emph{prior coverage function} of $\Gamma_{I}$ and denoted by $p_{\scriptscriptstyle{\Gamma_{I}}}(\gamma)$. In particular, $\forall \gamma \in\Gamma$
    \begin{equation}\label{coverage_function}
      p_{\scriptscriptstyle{\Gamma_{I}}}(\gamma) := \mathbf{E}(g_{\theta}(\gamma)) = Prob(g_{\theta}(\gamma) = 1) = P(\{\gamma\}\cap\Gamma_{I}(\theta) \neq \varnothing) =P(A(\theta,\gamma)\in A_{0})\in[0,1],
    \end{equation}
\noindent where $g_{\theta}(\cdot)$ is the stochastic process defined in \eqref{eq:2} and $P$ and $\mathbf{E}$ are the probability and expectation, respectively, taken with respect to the prior of $\theta$. The appealing fact of the prior coverage function with respect to the prior capacity functional is that $p_{\scriptscriptstyle{\Gamma_{I}}}(\gamma)$ can be represented graphically in an easier way than $T_{\Gamma_{I}}(K)$ (at least if $\Gamma\subset\mathbb{R}$), see for instance Figures \ref{fig_Example_1_Dirichlet_random_set}, \ref{fig_Example_3_Dirichlet_random_set} and 9 where we represent $p_{\scriptscriptstyle{\Gamma_{I}}}(\cdot)$ and the posterior coverage function $p_{\scriptscriptstyle{\Gamma_{I}}}(\cdot|x)$. We remark that the prior coverage function $p_{\scriptscriptstyle{\Gamma_{I}}}(\gamma)$ does not characterize the distribution of the stochastic process $g_{\theta}(\gamma)$ which is instead characterized by its finite-dimensional distributions.\\
\indent We detail now the construction of the prior and posterior capacity functionals with the help of a generic example. Suppose $\theta := (\theta_{1},\theta_{2})' \in\mathbb{R}^2$, $A(\theta, \gamma) = (\theta_{1} - \gamma, \gamma - \theta_{2})'$ for some $\gamma\in\mathbb{R}$ and $A_{0} = (-\infty, 0]\times (-\infty, 0]$. Hence, the condition $A(\theta,\gamma)\in A_{0}$ writes as $\gamma \in [\theta_{1},\theta_{2}]$ and $\Gamma_{I} = [\theta_{1},\theta_{2}]$. We recall that the aim is to make inference on the identified set $\Gamma_I$ and not on the partially identified parameter. To start with, suppose that we specify a parametric prior for $\theta$. For instance, $\theta_{1}\sim \mathcal{U}[0,1]$ and $\theta_{2}\sim\mathcal{U}[1,2]$. For any $K\in\mathcal{K}$, write $K = [K_{1},K_{2}]$, $\bar{K} = [\bar{K}_{1},\bar{K}_{2}] := K \cap[0,1]$ and $\bar{\bar{K}} = [\bar{\bar{K}}_{1},\bar{\bar{K}}_{2}] := K \cap[1,2] $. The prior capacity functional then is:
    \begin{displaymath}
      T_{\Gamma_{I}}(K) = \left\{\begin{array}{ccc}
        0 & \textrm{if} & K\cap[0,2] = \varnothing\\
        \bar{K}_{2} - \bar{K}_{1} & \textrm{if} & K\cap [0,1]\neq\varnothing \textrm{ and } K\cap [1,2]=\varnothing\\
        \bar{K}_{2} - \bar{K}_{1} + \bar{\bar{K}}_{2} - \bar{\bar{K}}_{1} & \textrm{if} & K\cap [0,1]\neq\varnothing \textrm{ and } K\cap [1,2]\neq\varnothing\\
        \bar{\bar{K}}_{2} - \bar{\bar{K}}_{1} & \textrm{if} & K\cap [0,1]=\varnothing \textrm{ and } K\cap [1,2]\neq\varnothing\\
      \end{array}\right.,
    \end{displaymath}
\noindent while the prior coverage function is given by $p_{\scriptscriptstyle{\Gamma_{I}}}(\gamma) = P(\gamma \in [\theta_{1},\theta_{2}]) = \gamma \mathbbm{1}\{\gamma\in[0,1]\} + (2 - \gamma) \mathbbm{1}\{\gamma\in[1,2]\}$, where $P$ is the distribution with respect to the prior of $\theta$.\\
\indent With this intuition in mind, let us move to the nonparametric Bayesian approach. This approach is based on a Dirichlet process prior and requires to write $\theta =(\theta_{1},\theta_{2})'$ as $\theta_{i} = \phi_{i}(F)$, for a measurable functional $\phi_{i}$, $i=1,2$. If we observe realizations of a random vector $Y$ from a distribution $F$ then the Bayesian model writes $Y|F \sim F$, $F \sim \mathcal{D}ir(n_{0},F_{0})$.
\noindent The probability measure $F_{0}$ should be chosen such that $\phi_{2}(F) > \phi_{1}(F)$, $\mu$-\textit{a.s.} By using the stick-breaking representation of the Dirichlet process, see Appendix D in the Supplement, the prior capacity functional of $\Gamma_{I} = [\phi_{1}(F),\phi_{2}(F)]$ is given by
    \begin{equation}\label{eq_prior_capacity_functional_interval}
      T_{\Gamma_{I}}(K) = P\left\{K\cap \left[\phi_{1}\left(\sum_{j=1}^{\infty}\alpha_{j}\delta_{\xi_{j}}\right),\phi_{2}\left(\sum_{j=1}^{\infty}\alpha_{j}\delta_{\xi_{j}}\right)\right] \neq \varnothing \right\}, \quad K\in\mathcal{K},
    \end{equation}
\noindent where $\{\xi_{j}\}_{j\geq 1}$ are independent draws from $F_{0}$, $\delta_{\xi_{j}}$ denotes the Dirac mass in $\xi_{j}$, $\alpha_{j} = v_{j}\prod_{l=1}^{j-1}(1 - v_{l})$ with $\{v_{l}\}_{l\geq 1}$ independent draws from a Beta distribution $\mathcal{B}e(1,n_{0})$ and $\{v_{j}\}_{j \geq 1}$ are independent of $\{\xi_{j}\}_{j\geq 1}$. In Appendix D we recall how to simulate $\phi_{i}(F)$, $i=1,2$, from the prior and posterior distribution by using this representation. The prior coverage function of $\Gamma_{I}$ is: for every $\gamma\in\Gamma$,
    \begin{eqnarray}
      p_{\scriptscriptstyle{\Gamma_{I}}}(\gamma) = P\Big(\gamma\in[\phi_{1}(F),\phi_{2}(F)]\Big) & = & P\Big(\phi_{1}\Big(\sum_{j=1}^{\infty}\alpha_{j}\delta_{\xi_{j}}\Big) \leq \gamma \leq \phi_{2}\Big(\sum_{j=1}^{\infty}\alpha_{j}\delta_{\xi_{j}}\Big)\Big)\label{prior_Dirichlet_random_set}
    \end{eqnarray}
\noindent where we have taken $K = \{\gamma\}$ and $P$ is the prior distribution of $\theta$, $\{\alpha_j\}$ and $\{\xi_j\}$. In general we do not have an analytic form for $T_{\Gamma_{I}}$ and $p_{\scriptscriptstyle{\Gamma_{I}}}$ but we have a perfect knowledge of them since we can easily simulate from $T_{\Gamma_{I}}$ and $p_{\scriptscriptstyle{\Gamma_{I}}}$ by using the stick-breaking representation of the Dirichlet process.\\
\indent After observing an $n$-sample of $Y$, $(y_{1}, \ldots,y_{n})$, one computes the posterior distribution of $F$ as
    \begin{eqnarray*}
      F|y_{1},\ldots,y_{n} & \sim & \mathcal{D}\Big(n_{0} + n, \frac{n_{0}}{n_{0} + n}F_{0} + \frac{n}{n_{0} + n}F_{n}\Big),
    \end{eqnarray*}
\noindent where $F_{n}(\cdot) := \frac{1}{n}\sum_{j}\delta_{y_{j}}(\cdot)$ denotes the empirical cumulative distribution. If the true data distribution $F$ is such that $\phi_{2}(F) > \phi_{1}(F)$ then the same is true for the distribution $F$ generated by the posterior. The posterior capacity functional is denoted by $T_{\Gamma_{I}}(K|\{y_{i}\}_{i=1}^{n})$ and given by: $\forall K\in\mathcal{K}$,
    \begin{align*}
      T_{\Gamma_{I}}(&K|\{y_{i}\}_{i=1}^{n}) = \\
      &P\left\{K\cap \left[\phi_{1}\Big(\rho\sum_{j=1}^{n}\beta_{j}\delta_{y_{j}} + (1 - \rho)\sum_{j=1}^{\infty}\alpha_{j}\delta_{\xi_{j}}\Big),\phi_{2}\Big(\rho\sum_{j=1}^{n}\beta_{j}\delta_{y_{j}} + (1 - \rho)\sum_{j=1}^{\infty}\alpha_{j}\delta_{\xi_{j}}\Big)\right] \neq \varnothing\right\},
    \end{align*}
\noindent where $\{\alpha_{j}\}$, $\{\delta_{\xi_{j}}\}$, $\{\delta_{y_{j}}\}$ and $\{\xi_{j}\}$, are as above, $\rho$ is drawn form a Beta distribution $\mathcal{B}e(n,n_{0})$ independently of the other quantities and $(\beta_{1}, \ldots, \beta_{n})$ are drawn from a Dirichlet distribution with parameters $(1,\ldots, 1)$ on the simplex $S_{n-1}$ of dimension $(n-1)$. For every $\gamma\in\Gamma$, the posterior coverage function $p_{\scriptscriptstyle{\Gamma_{I}}}(\gamma|\{y_{i}\}_{i=1}^{n})=P\Big(\gamma\in[\phi_{1}(F),\phi_{2}(F)]\Big|\{y_{i}\}_{i=1}^{n}\Big)$ is: $\forall \gamma\in\Gamma$,
    \begin{equation}\label{posterior_Dirichlet_random_set}
      P\left(\phi_{1}\Big(\rho\sum_{j=1}^{n}\beta_{j}\delta_{y_{j}} + (1 - \rho)\sum_{j=1}^{\infty}\alpha_{j}\delta_{\xi_{j}}\Big)\leq \gamma \leq \phi_{2}\Big(\rho\sum_{j=1}^{n}\beta_{j}\delta_{y_{j}} + (1 - \rho)\sum_{j=1}^{\infty}\alpha_{j}\delta_{\xi_{j}}\Big)\right).
    \end{equation}
\indent For simplicity we have presented only the case where the condition $A(\theta,\gamma)\in A_{0}$ writes as $\gamma\in[\phi_{1}(F),\phi_{2}(F)]$ but our nonparametric method can be generalized to the case where $\Gamma_{I}$ is not an interval. In that case, if a Dirichlet process prior is specified for $F$, the prior capacity functional of $\Gamma_{I}$ is given by: $\forall K\in\mathcal{K}$,
    \begin{eqnarray}
      T_{\Gamma_{I}}(K) = P(K\cap\Gamma_{I}\neq \varnothing) = P\left(K\cap\left\{\gamma\in\Gamma;A(\phi\Big(\sum_{j=1}^{\infty}\alpha_{j}\delta_{\xi_{j}}\Big),\gamma)\in A_{0}\right\}\neq \varnothing\right)\label{prior_Dirichlet_random_set}
    \end{eqnarray}
\noindent and the posterior capacity functional is: $\forall K\in\mathcal{K}$,
    \begin{align}
      T_{\Gamma_{I}}(&K|\{y_{i}\}_{i=1}^{n}) = P(K\:\cap\:\Gamma_{I}\neq \varnothing|\{y_{i}\}_{i=1}^{n}) = \nonumber\\
      & P\left(\left.K\cap\left\{\gamma\in\Gamma;A(\phi\Big(\rho\sum_{j=1}^{n}\beta_{j}\delta_{y_{j}} + (1 - \rho)\sum_{j=1}^{\infty}\alpha_{j}\delta_{\xi_{j}}\Big),\gamma)\in A_{0}\right\}\neq \varnothing\right|\{y_{i}\}_{i=1}^{n}\right),
    \end{align}
\noindent where $\{\alpha_{j}\}$, $\{\delta_{\xi_{j}}\}$, $\{\delta_{y_{j}}\}$, $\{\xi_{j}\}$, $\rho$ and $(\beta_{1},\ldots,\beta_{n})$ are as above.\\
\indent Once the posterior capacity functional is available, an estimator for $\Gamma_{I}(\theta)$ can be easily constructed. One possibility is to fix $\theta$ equal to its posterior mean or median, denote it by $\widehat\theta$, and take the corresponding $\Gamma_{I}(\widehat\theta)$ as an estimator for $\Gamma_{I}$. In alternative, one could construct a closed set $C_{\alpha}$ that satisfies the following condition
    \begin{equation}\label{eq_posterior_credible_region}
      P(\Gamma_{I}(\theta)\subset C_{\alpha}|\{y_{i}\}_{i=1}^{n}) \geq \alpha
    \end{equation}
\noindent for some $\alpha\in[0,1]$. This is the usual posterior credible region. A similar estimator is proposed \textit{e.g.} by \cite{LiaoSimoni2019} and \cite{chen2017monte}. We remark that the probability in (\ref{eq_posterior_credible_region}) is determined by the posterior Dirichlet process and is the \emph{posterior containment functional} evaluated at $C_{\alpha}$.\\

\subsection{Examples}\label{ss_examples_set_estimation}
\indent In this section we provide two examples where we use our proposed Bayesian nonparametric method described in section \ref{ss_construction_prior_post_capacity_functionals}. Other two examples are developed in Appendix B in the Supplement. For simplicity, we only focus on the prior and posterior coverage functions, which are easy to represent graphically.

\begin{example}[\textbf{Interval Censored Data}.]\label{Example:2}
  This example is motivated by interval responses in survey data. Let $Y$ be the real random variable of interest that is unobserved but is known to lie in the interval $[Y_{1},Y_{2}]$ a.s. with respect to the sampling distribution, where $Y_{1}$ and $Y_{2}$ are two observable real random variables. The probability distributions of $Y_1$ and $Y_2$ are unknown and denoted by $F_1$ and $F_2$, respectively. We denote with $\mathbf{E}_{F_{1}}(Y_{1})$ and $\mathbf{E}_{F_{2}}(Y_{2})$ the expectation taken with respect to $F_1$ and $F_2$, respectively. Let $\gamma:=\mathbf{E}(Y)\in\Gamma = \mathbb{R}$ be the parameter of interest. Since $Y\in[Y_{1},Y_{2}]$ a.s., the condition $A(\theta,\gamma)\in A_{0}$ takes the form
    \begin{equation}
      \mathbf{E}_{F_{1}}(Y_{1}) \leq \gamma \leq\mathbf{E}_{F_{2}}(Y_{2})
    \end{equation}
\noindent \textit{a.s.} with respect to the prior distribution $\mu$ of $\theta$, where $\theta := \phi(F) = (\mathbf{E}_{F_{1}}(Y_{1}),\mathbf{E}_{F_{2}}(Y_{2}))'$ and $F := (F_{1},F_{2})'$ is the joint distribution of $(Y_1, Y_2)$. More precisely, for every $\gamma\in\Gamma$, $A(\theta,\gamma) = [\mathbf{E}_{F_{1}}(Y_{1}) - \gamma, \gamma - \mathbf{E}_{F_{2}}(Y_{2})]$ and $A_{0} = (-\infty, 0]\times(-\infty,0]$ which is an element of the Borel $\sigma$-field of subsets of $\Phi=\mathbb{R}^{2}$. Hence, $\Gamma_{I} = [\mathbf{E}_{F_{1}}(Y_{1}), \mathbf{E}_{F_{2}}(Y_{2})]$ is a random closed set, the parameter $\gamma$ is partially identified and our object of interest becomes the identified set $\Gamma_I$.\\
\indent We compute the prior and posterior coverage function of the identified set $\Gamma_{I}$ by specifying a Dirichlet process prior for $\theta$. Let us assume that $Y_{1} \perp Y_{2}|F$. For $F_0^1$ and $F_0^2$ two probability measures, the Bayesian hierarchical model is
    \begin{eqnarray}
      (y_{11},\ldots,y_{1n})|F & \sim \:iid & F_{1} \nonumber\\
      (y_{21},\ldots,y_{2n})|F & \sim \:iid & F_{2} \nonumber\\
      F_{1} & \sim & \mathcal{D}ir(n_{0}^{1},F_{0}^{1}),\qquad n_0^1\in\mathbb{R}_+ \nonumber\\
      F_{2} & \sim & \mathcal{D}ir(n_{0}^{2},F_{0}^{2}),\qquad n_0^2\in\mathbb{R}_+,\label{eq:3}
    \end{eqnarray}
\noindent where $(y_{11},\ldots,y_{1n})$ and $(y_{21},\ldots,y_{2n})$ denote two $n$-samples of realisations of $Y_{1}$ and $Y_{2}$, respectively. If the probability measures $F_{0}^{1}$ and $F_{0}^{2}$ have disjoint supports, that is, $\max Supp(F_{0}^{1}) < \min Supp(F_{0}^{2})$ then, $\mathbf{E}_{F_{2}}(Y_{2})>\mathbf{E}_{F_{1}}(Y_{1})$ $\mu$-\emph{a.s.} For every $\gamma\in\Gamma$, let $g_{\theta}(\gamma):=\mathbbm{1}\{\gamma\in[\mathbf{E}_{F_{1}}(Y_{1}),\mathbf{E}_{F_{2}}(Y_{2})]\}$. For every $\gamma\in\Gamma$ the prior coverage function is $\mathbf{E}(g_{\theta}(\gamma))$, where the expectation is taken with respect to the prior $\mu$ of $F$, and can be represented as, $\forall \gamma\in\Gamma$,
    \begin{eqnarray*}
      p_{\scriptscriptstyle{\Gamma_{I}}}(\gamma) = P\Big(\gamma\in[\mathbf{E}_{F_{1}}(Y_{1}),\mathbf{E}_{F_{2}}(Y_{2})]\Big) & = & P\Big(\sum_{j}\alpha_{j}^{1}\xi_{j}^{1}\leq \gamma\leq \sum_{j}\alpha_{j}^{2}\xi_{j}^{2}\Big)
    \end{eqnarray*}
\noindent where, for $i=1,2$, $\{\xi_{j}^{i}\}_{j\geq 1}$ are independent draws from $F_{0}^{i}$, $\alpha_{j}^{i} = v_{j}^{i}\prod_{l=1}^{j-1}(1 - v_{l}^{i})$ with $\{v_{l}^{i}\}_{l\geq 1}$ independent draws from a Beta distribution $\mathcal{B}e(1,n_{0}^{i})$ and $\{v_{j}^{i}\}_{j\geq1}$ are independent of $\{\xi_{j}^{i}\}_{j\geq 1}$. For $i=1,2$, let $\mathcal{F}_{i}(\cdot)$ denote the cumulative distribution function of $\mathbf{E}_{F_{i}}(Y_{i})$. Hence, $p_{\scriptscriptstyle{\Gamma_{I}}}(\gamma)$ takes the following values: \textit{(i)} $\forall \gamma\notin [\min Supp(F_{0}^{1}), \max Supp(F_{0}^{2})]$, $p_{\scriptscriptstyle{\Gamma_{I}}}(\gamma) = 0$; \textit{(ii)} $\forall \gamma\in [\max Supp(F_{0}^{1}), \min Supp(F_{0}^{2})]$, $p_{\scriptscriptstyle{\Gamma_{I}}}(\gamma) = 1$; \textit{(iii)} $\forall \gamma\in Supp(F_{0}^{1})$, $p_{\scriptscriptstyle{\Gamma_{I}}}(\gamma) = P\Big(\sum_{j}\alpha_{j}^{1}\xi_{j}^{1}\leq \gamma \Big) \equiv \mathcal{F}_{1}(\gamma)$; \textit{(iv)} $\forall \gamma\in Supp(F_{0}^{2})$, $p_{\scriptscriptstyle{\Gamma_{I}}}(\gamma) = P\Big(\gamma \leq \sum_{j}\alpha_{j}^{2}\xi_{j}^{2}\Big) \equiv 1 - \mathcal{F}_{2}(\gamma)$.\\
\indent The posterior distributions of $F_{1}$ and $F_{2}$ are $F_{i}|(y_{i1},\ldots,y_{in}) \sim \mathcal{D}ir(n_{*}^{i},F_{*}^{i})$, $i=1,2$,
\noindent where for $i=1,2$, $n_{*}^{i} = n_{0}^{i} + n$, $F_{*}^{i} = \frac{n_{0}^{i}}{n_{0}^{i} + n}F_{0}^{i} + \frac{n}{n_{0}^{i} + n}F_{n}^{i}$ and $F_{n}^{i}(\cdot) := \frac{1}{n}\sum_{j}\delta_{y_{ij}}(\cdot)$ is the empirical distribution of the sample $(y_{i1},\ldots,y_{in})$. The posterior coverage function of $\Gamma_{I}$ is, $\forall \gamma\in\Gamma$
    \begin{displaymath}
      p_{\scriptscriptstyle{\Gamma_{I}}}(\gamma|\{y_{1i}\}, \{y_{2i}\}) = P\Big(\gamma\in[\mathbf{E}_{F_{1}}(Y_{1}),\mathbf{E}_{F_{2}}(Y_{2})]\Big|\{y_{1i}\}, \{y_{2i}\}\Big)
    \end{displaymath}
\noindent and can be represented as:
    \begin{displaymath}
      P\Big(\rho_{1}\sum_{j=1}^{n}\beta_{1j}y_{j1} + (1 - \rho_{1})\sum_{j=1}^{\infty}\alpha_{j}^{1}\xi_{j}^{1}\leq \gamma \leq \rho_{2}\sum_{j=1}^{n}\beta_{2j}y_{2j} + (1 - \rho_{2})\sum_{j=1}^{\infty}\alpha_{j}^{2}\xi_{j}^{2}\Big),
    \end{displaymath}
\noindent where, for $i=1,2$, $\alpha_{j}^{i}$ and $\xi_{j}^{i}$ are as above, $\rho_{i}$ is drawn from a Beta distribution $\mathcal{B}e(n,n_{0}^{i})$ independently of the other quantities and $(\beta_{i1},\ldots,\beta_{in})$ are drawn from a Dirichlet distribution with parameters $(1, \ldots, 1)$ on the simplex $S_{n-1}$ of dimension $(n-1)$.\\
\indent A simulation exercise allows to visualize the prior and posterior coverage functions of $\Gamma_{I}$. We generate an $n$-sample of realizations of $Y_{1}$ and $Y_{2}$ from the following distributions\footnote{This data generating process is the same used in \cite{LiaoJiang2010} in their Example 5.1.}:
    \begin{displaymath}
      Y_{1}\sim\mathcal{N}(0,0.1),\qquad Y_{2}\sim\mathcal{N}(5,0.1).
    \end{displaymath}
\noindent The parameters are fixed as follows: $n = 1000$, $n_{0}^{1}=10$, $n_{0}^{2}=20$, $F_{0}^{1}= \mathcal{N}(0,1)$ and $F_{0}^{2}= \mathcal{N}(10,1)$. The supports of $F_{0}^{1}$ and $F_{0}^{2}$ are not disjoint. However, since the tails of a normal density function are very thin, the prior probability that $\mathbf{E}_{F_{2}}(Y_{2}) < \mathbf{E}_{F_{1}}(Y_{1})$ is very small. The true identified set in our simulation is $[\mathbf{E}_{F_{1}}(Y_{1}), \mathbf{E}_{F_{2}}(Y_{2})] = [0,5]$.\\
\indent In Figure \ref{fig_Example_1_Dirichlet_random_set} we represent the subset $[-3,12]$ of $\Gamma$ on the horizontal axis and we evaluate $p_{\scriptscriptstyle{\Gamma_{I}}}(\cdot)$ and $p_{\scriptscriptstyle{\Gamma_{I}}}(\cdot|\{y_{1i}\}, \{y_{2i}\})$ over a grid of values on this intervals. Then, we draw $1000$ times from the prior and posterior distribution of $(F_{1},F_{2})$ and for every value $\gamma$ in the grid of $[-3,12]$ we count the number of simulated $\Gamma_{I}$s that contain this value $\gamma$. Figure \ref{fig_Example_1_Dirichlet_random_set} shows the prior and posterior coverage functions for each value of $\gamma\in[-3,12]$. Figure \ref{fig_Example_1_Dirichlet_Figure_2} displays the intervals $\Gamma_{I}$ drawn from the prior and posterior distributions (on the vertical axis) against the true interval $[0,5]$ (on the horizontal axis). It also plots the true interval $[0,5]$ against itself in red. Both figures show that the posterior concentrates on the true $\Gamma_I = [0,5]$.

\begin{figure}[!h]
  \centering
  \subfloat[{Prior coverage function.}]{
      \includegraphics[width=0.4\linewidth]{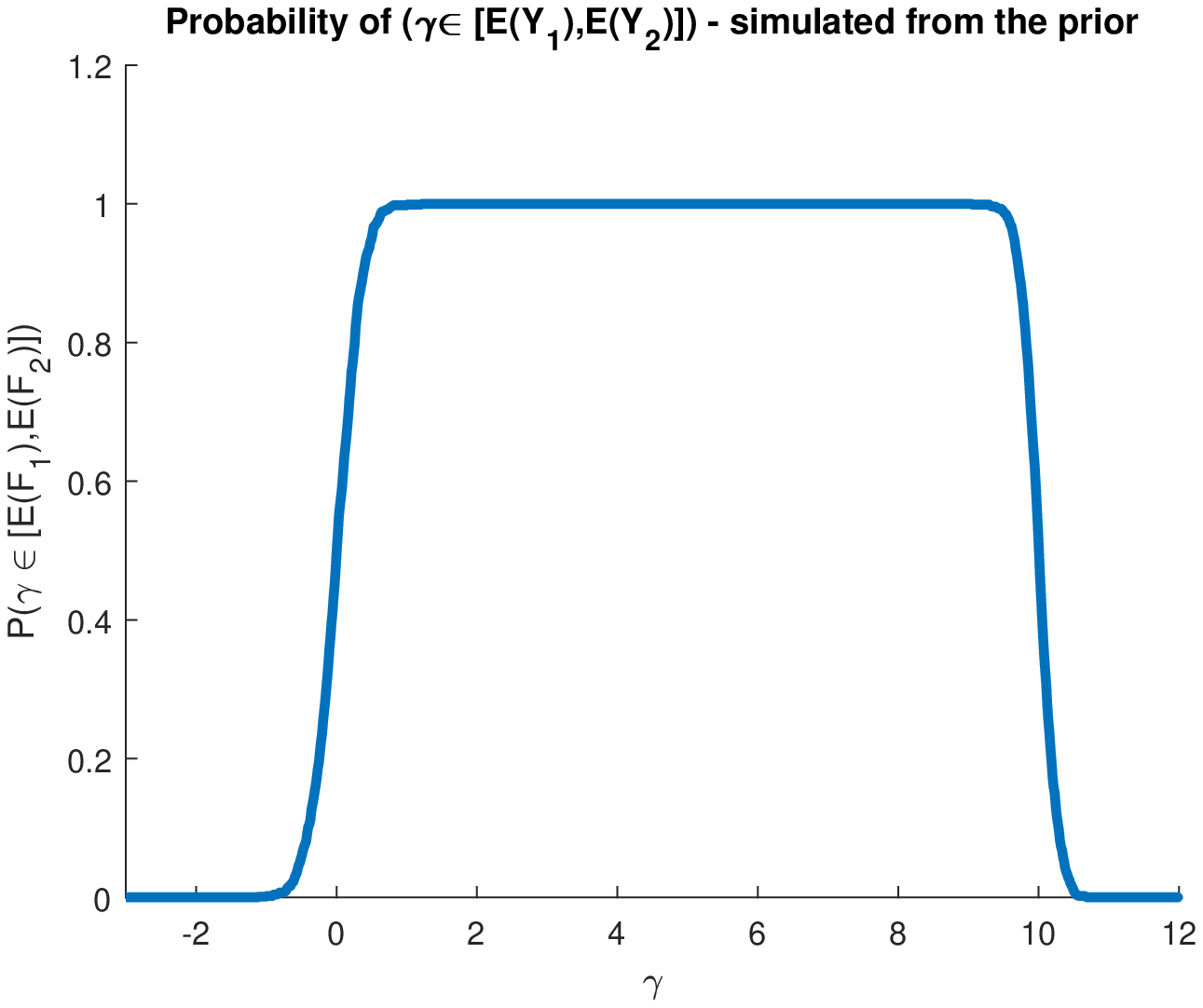}}
  \hspace{0.1\linewidth}
  \subfloat[{Posterior coverage function.}]{
      \includegraphics[width=0.4\linewidth]{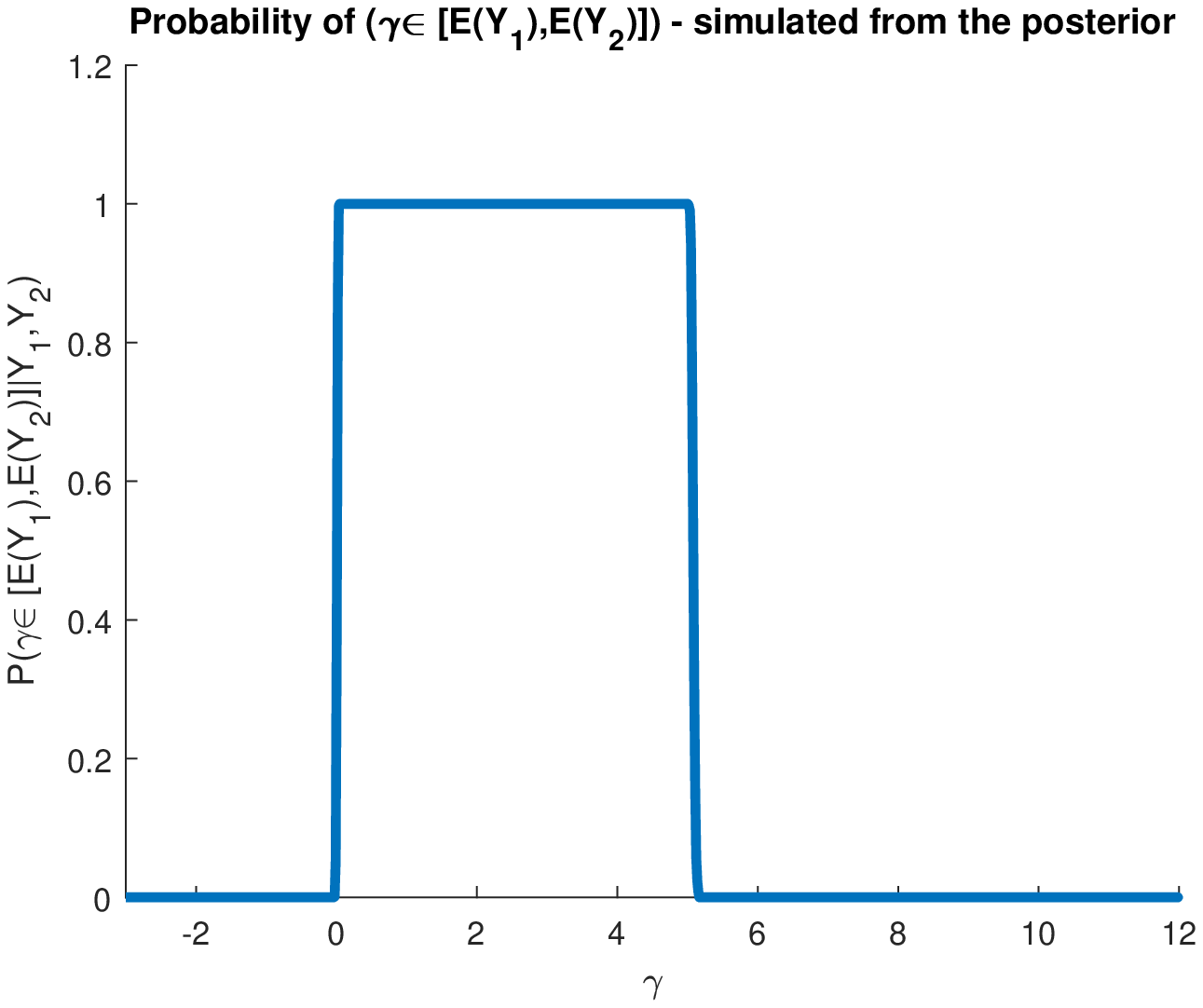}}
  \caption{{\small Interval Censored Data. Prior and posterior coverage functions $p_{\scriptscriptstyle{\Gamma_{I}}}(\cdot)$ and $p_{\scriptscriptstyle{\Gamma_{I}}}(\cdot|\{y_{1i}\}, \{y_{2i}\})$ of $\Gamma_{I}$. The true $\Gamma_I$ is $[0,5]$.}}
  \label{fig_Example_1_Dirichlet_random_set}
\end{figure}

\begin{figure}[!h]
  \centering
  \subfloat[{Prior}]{\label{Example_1_fig_2_a}
      \includegraphics[width=0.4\linewidth]{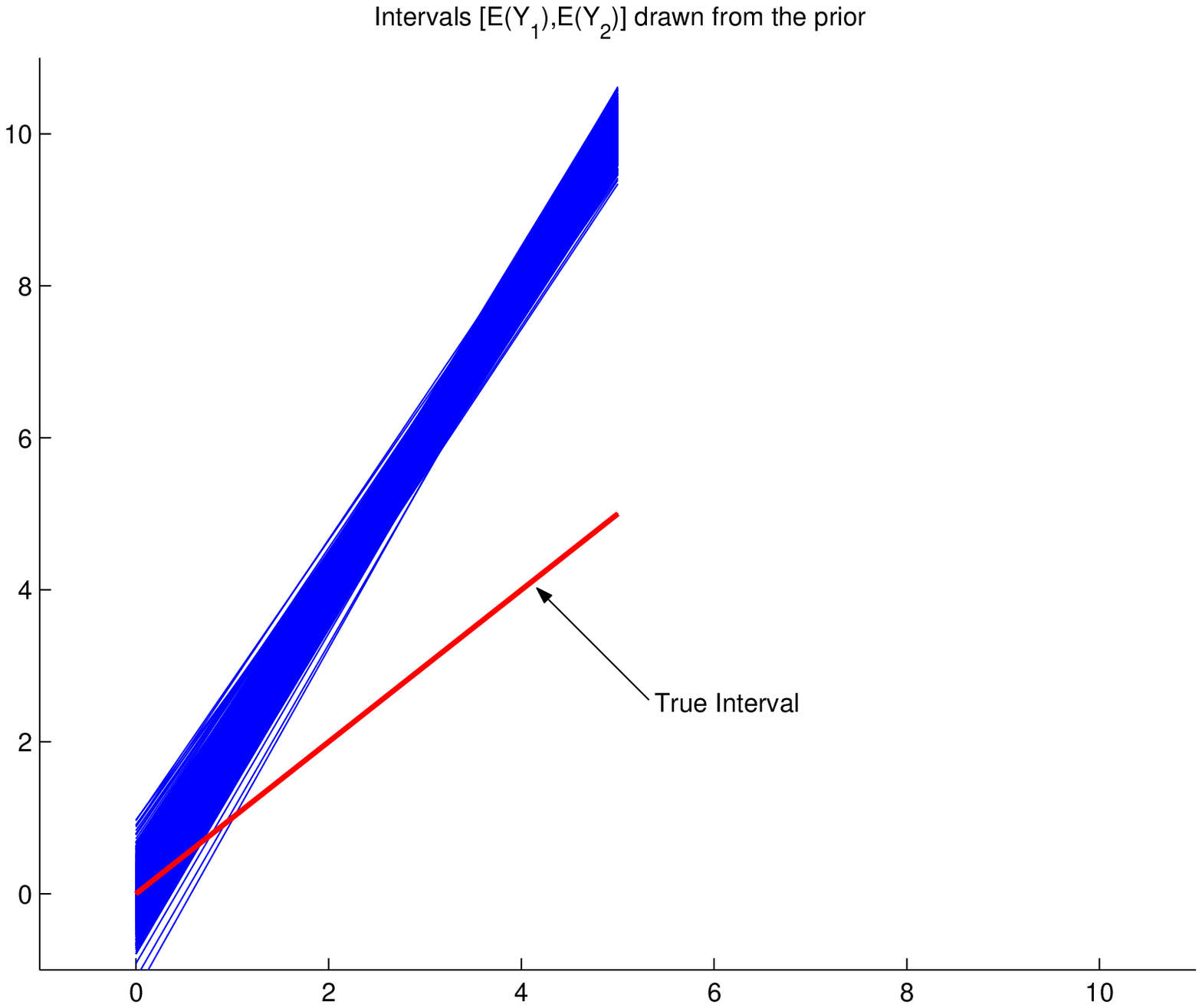}}
  \hspace{0.1\linewidth}
  \subfloat[{Posterior}]{\label{Example_1_fig_2_b}
      \includegraphics[width=0.4\linewidth]{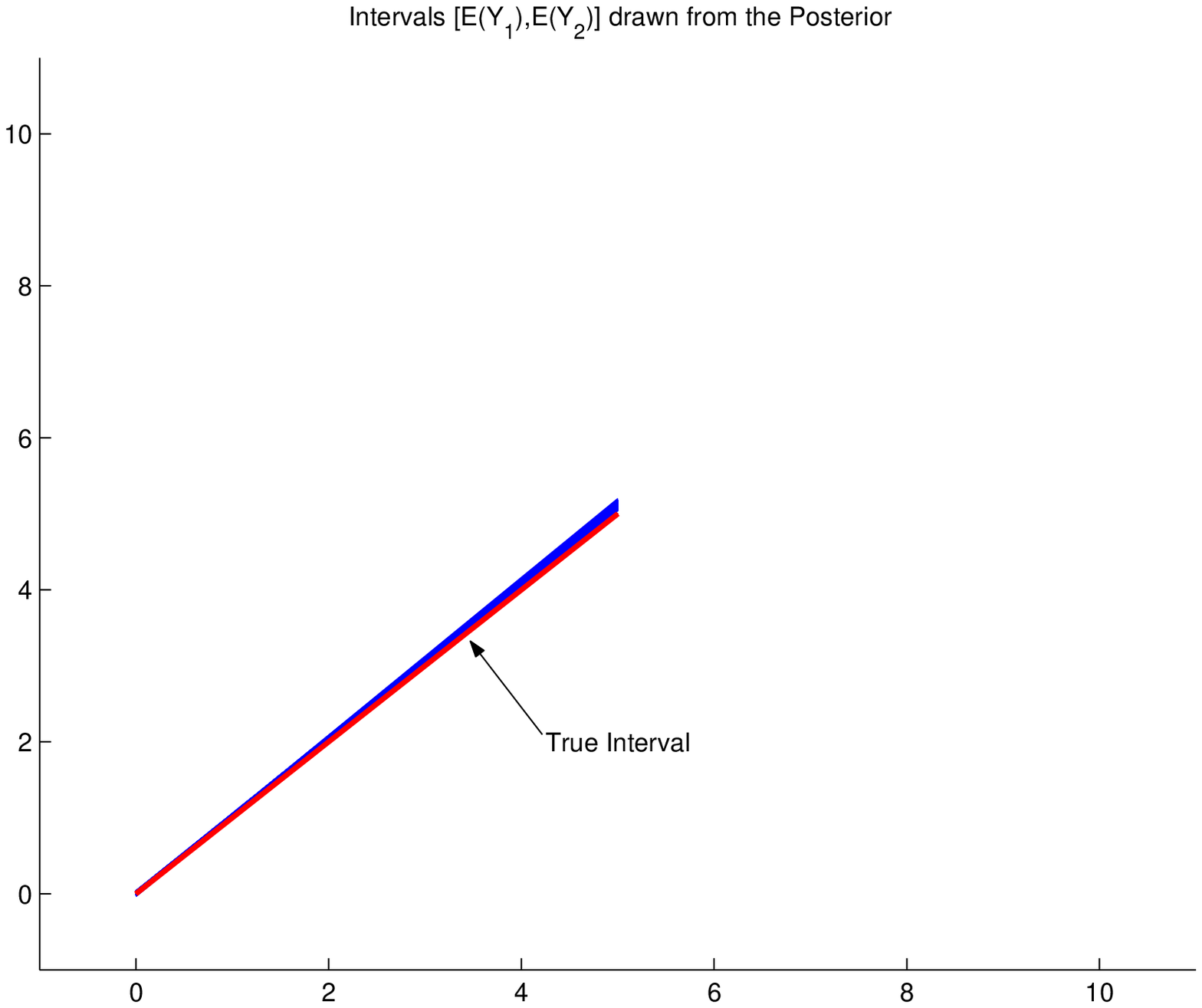}}
  \caption{{\small Interval Censored Data. Representation of the intervals $\Gamma_I(\theta) = [\mathbf{E}_{F_{1}}(Y_{1}), \mathbf{E}_{F_{2}}(Y_{2})]$ drawn from the prior (panel \ref{Example_1_fig_2_a}) and posterior (panel \ref{Example_1_fig_2_b}) against the true interval $\Gamma_I = [0,5]$ (in red).}}
  \label{fig_Example_1_Dirichlet_Figure_2}
\end{figure}
\end{example}
%
\begin{example}[\textbf{Linear Regression with errors in Regressors}.]\label{Example:3}
  This is the well-known linear errors-in-variables structural model considered in \cite{Frisch1934} and \cite{KlepperLeamer1984}. For simplicity, we focus here on the univariate linear regression model. Let $Y$ be an observable random variable satisfying the model $Y = \gamma\xi + u$ where $\xi$ is an unobservable random variable such that $\mathbf{E}(\xi) = 0$, $Var(\xi)= \tau^{2}$ and for which only realizations affected by an error are available: $Z = \xi + v$. The error terms $(u,v)$ are zero-mean jointly distributed random variables, independent of $\xi$, and with a variance-covariance matrix $\Sigma$ which may be diagonal.\\
\indent Suppose $\Sigma = diag (\sigma_{u}^{2},\sigma_{v}^{2})$ for simplicity and that $\mathbf{E}(\xi)$ is known, so that the structural parameters are $\gamma$, $\sigma_{u}^{2}$, $\sigma_{v}^{2}$ and $\tau^{2}$ while $\xi$ is the incidental parameter. Denote $\sigma_{yz} := Cov(Y,Z)$, $\sigma_{zz} := Var(Z)$ and $\sigma_{yy} := Var(Y)$. Due to endogeneity of $Z$, the parameter $\gamma$ lies in the identified set $\Gamma_{I} = [\theta_{1},\theta_{2}]$ given by
    \begin{multline}\label{ex_regressors_with_errors}
      [\theta_{1},\theta_{2}] = \Big[\min\Big(\frac{\sigma_{yz}}{\sigma_{zz}}, \frac{\sigma_{yy}}{\sigma_{yz}}\Big), \max\Big(\frac{\sigma_{yz}}{\sigma_{zz}},\frac{\sigma_{yy}}{\sigma_{yz}}\Big)\Big]\\
      = \Big[\min \Big(\frac{\gamma\tau^{2}}{\tau^{2} + \sigma_{v}^{2}}, \gamma + \frac{\sigma_{u}^{2}}{\gamma \tau^{2}}\Big), \max \Big(\frac{\gamma\tau^{2}}{\tau^{2} + \sigma_{v}^{2}}, \gamma + \frac{\sigma_{u}^{2}}{\gamma \tau^{2}}\Big)\Big],
    \end{multline}

\noindent where $\frac{\sigma_{yz}}{\sigma_{zz}}$ is the coefficient of the regression line of $Y$ on $Z$ and $\frac{\sigma_{yz}}{\sigma_{yy}}$ is the coefficient of the reverse regression line of $Z$ on $Y$, both without intercept. This can be seen by running the two regressions $y = \gamma(Z - v) + u$ and $Z = \frac{1}{\gamma}(Y - u) + v$. The first regression gives: $\gamma = \sigma_{yz}/(\sigma_{zz} - \sigma_{v}^2) \geq \sigma_{yz}/\sigma_{zz}$ while the second regression gives: $\gamma = (\sigma_{yy} - \sigma_u^2)/\sigma_{yz} \leq \sigma_{yy}/\sigma_{yz}$.\\
\indent We specify a Dirichlet process prior on the joint probability distribution $F_{yz}$ of $(Y,Z)$. In the simulation exercise we generate the data as: for $i = 1, \ldots,n$,
    \begin{eqnarray}
      \xi_{i} & \sim & i.i.d.\;\mathcal{N}(0,1), \qquad (u_{i},v_{i})' \; \sim \; i.i.d.\;\mathcal{N}_{2}(0,I_{2}),\label{simulation:regression:errors:regressors:1}\\
      Z_{i} & = & \xi_{i} + v_{i}, \nonumber\\
      Y_{i} & = & \gamma \xi_{i} + u_{i},\quad \gamma = 1, \nonumber
    \end{eqnarray}

\noindent where $I_{2}$ is the $2$-dimensional identity matrix. Since $\gamma > 0$ the identified set is $[\frac{\sigma_{yz}}{\sigma_{zz}},\frac{\sigma_{yy}}{\sigma_{yz}}] = [1/2,2]$. We specify the prior on $F_{yz}$ as $F_{yz} \sim \mathcal{D}ir(n_{0}, F_{0})$ with $n_{0} =20$ and
    \begin{equation}
      F_{0} = \mathcal{N}_{2}\Big(\left(\begin{array}
        {c} 0\\0
      \end{array}\right), \left(\begin{array}
        {cc}2 & 0.9\\0.9 & 2
      \end{array}\right)\Big).\label{simulation:regression:errors:regressors:2}
    \end{equation}

The sample size is $n=1000$ and we draw $1000$ intervals $\Gamma_{I}$ from the prior and posterior distribution of $F$. The results are shown in Figures \ref{fig_Example_3_Dirichlet_random_set} and \ref{fig_Example_3_Dirichlet_Figure_2}.

\begin{figure}[!h]
  \centering
  \subfloat[{Prior coverage function}]{
      \includegraphics[width=0.4\linewidth]{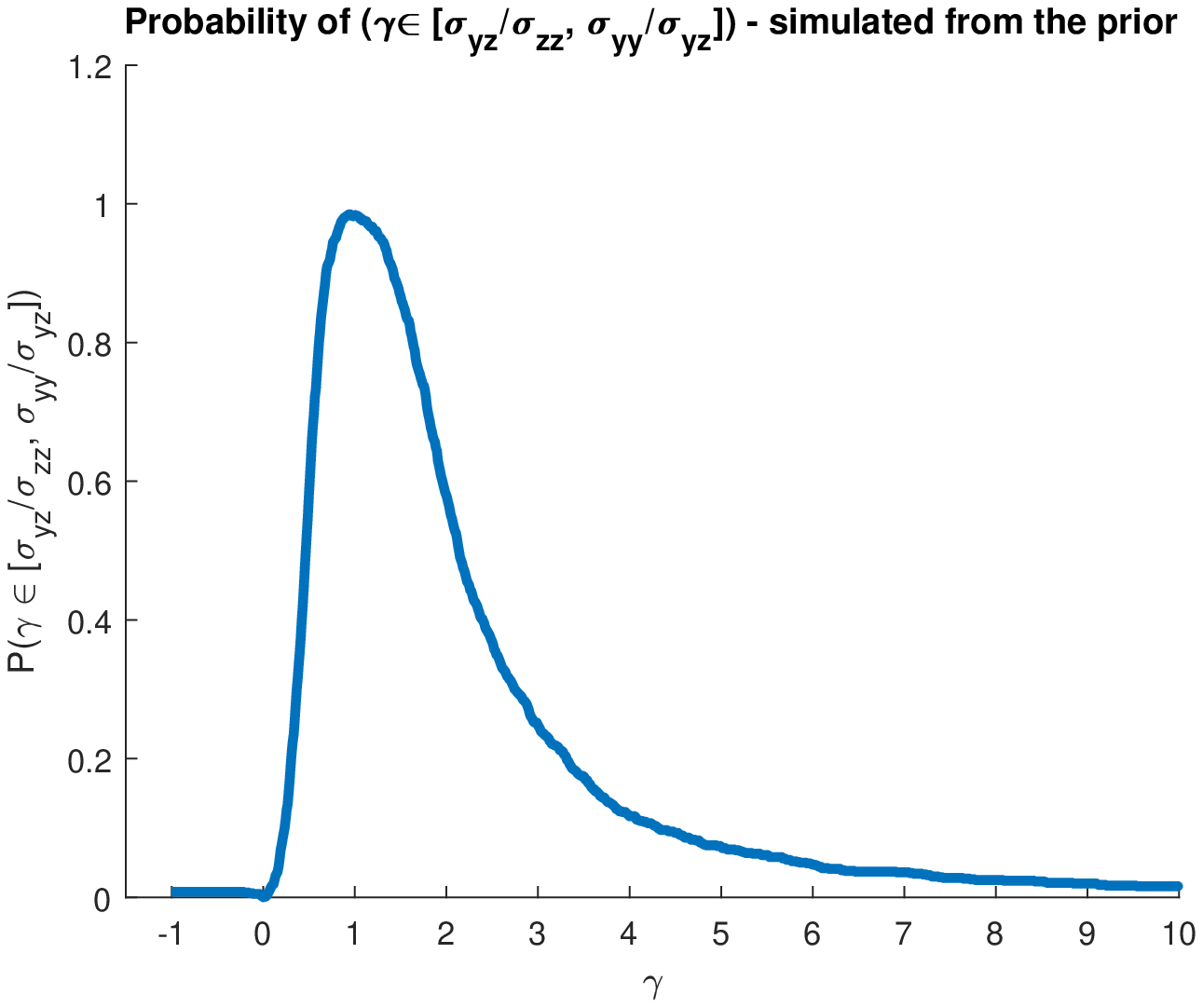}}
  \hspace{0.1\linewidth}
  \subfloat[{Posterior coverage function}]{
      \includegraphics[width=0.4\linewidth]{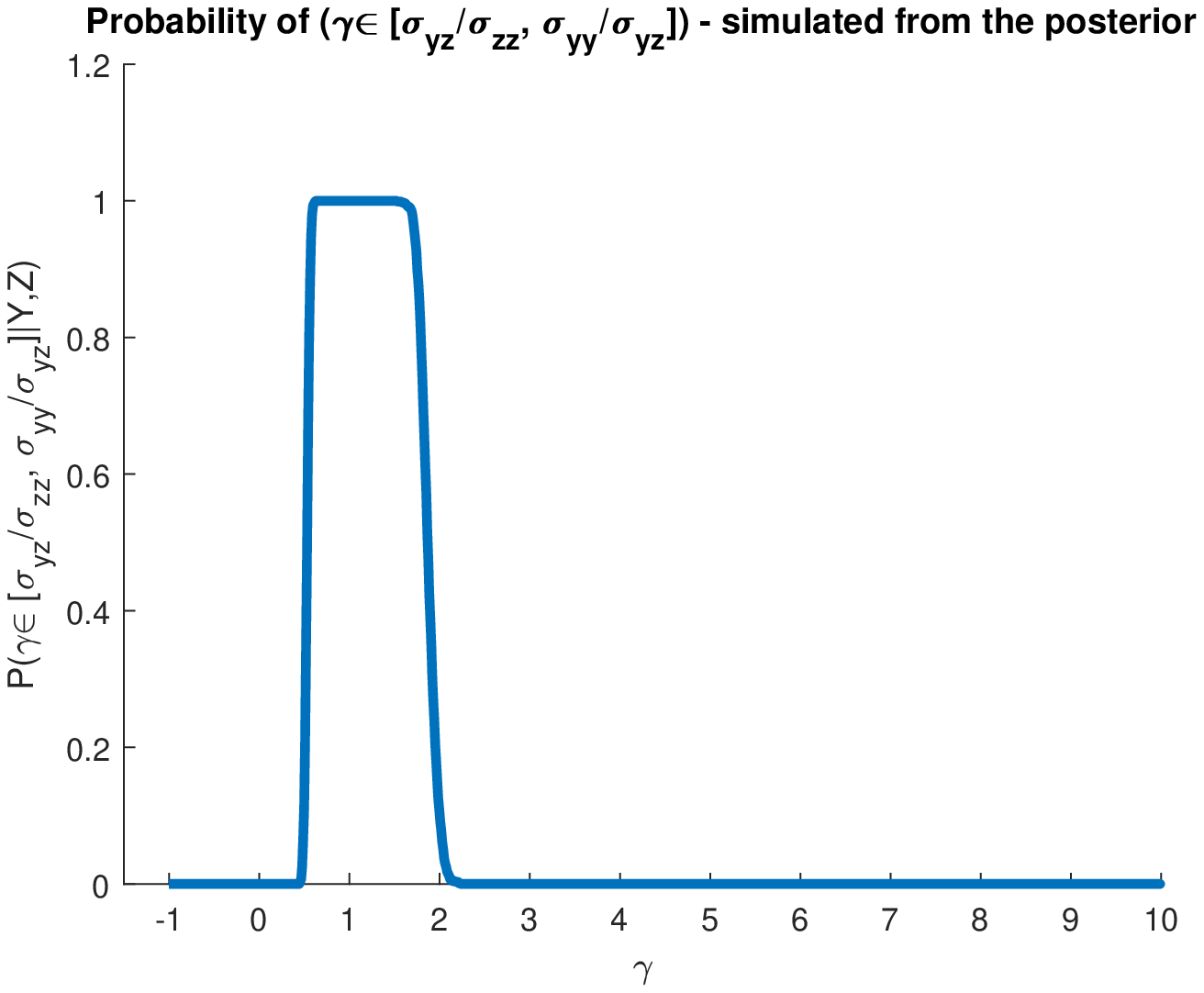}}
  \caption{{\small Linear Regression with errors in Regressors. Prior and posterior coverage functions $p_{\scriptscriptstyle{\Gamma_{I}}}(\cdot)$ and $p_{\scriptscriptstyle{\Gamma_{I}}}(\cdot|\{y_{i}\}, \{z_{i}\})$ of $\Gamma_{I}$. The true $\Gamma_I$ is $[1/2,2]$.}}
  \label{fig_Example_3_Dirichlet_random_set}
\end{figure}

\begin{figure}[!h]
  \centering
  \subfloat[{Prior}]{\label{Example_3_fig_2_a}
      \includegraphics[width=0.4\linewidth]{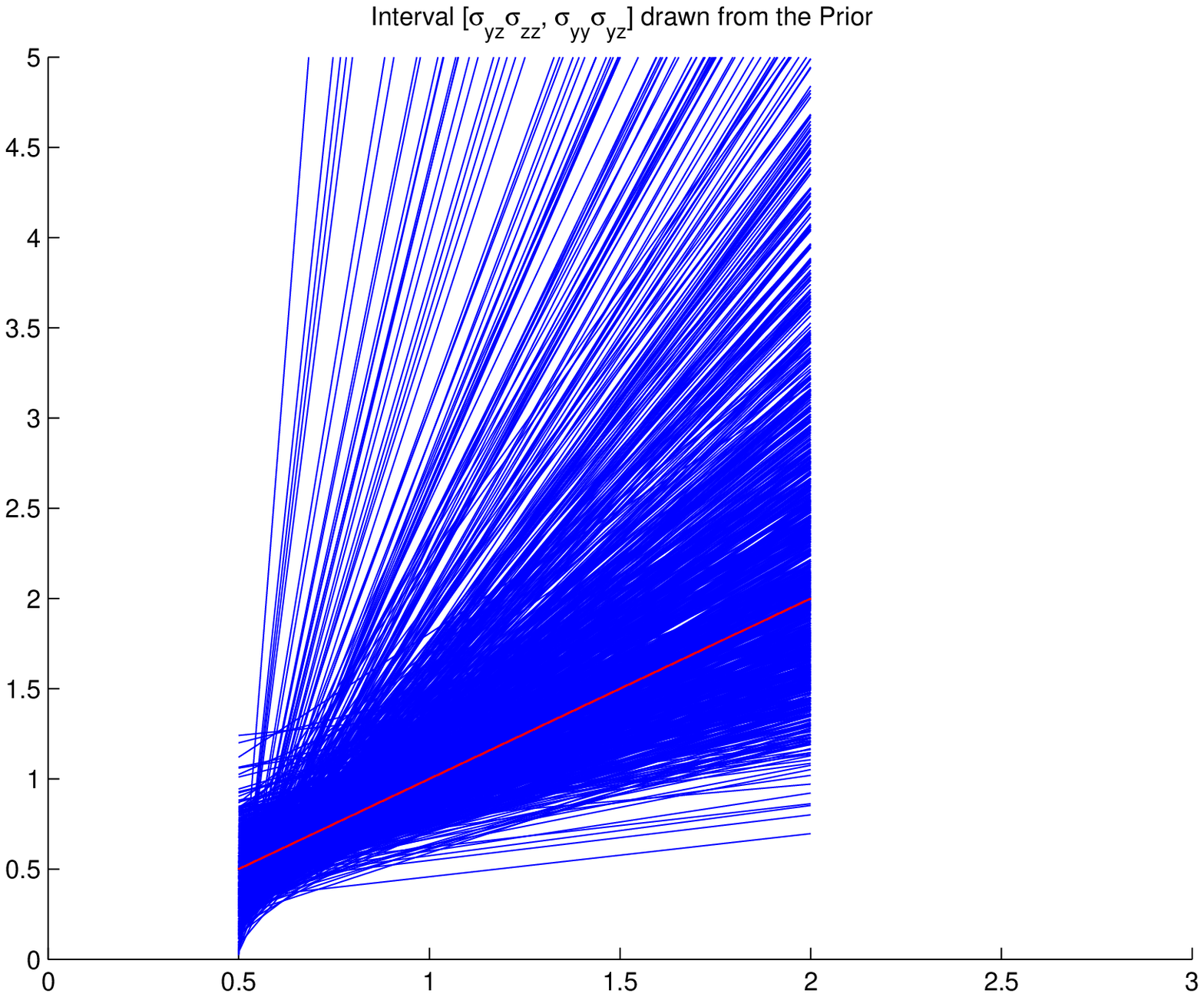}}
  \hspace{0.1\linewidth}
  \subfloat[{Posterior}]{\label{Example_3_fig_2_b}
      \includegraphics[width=0.4\linewidth]{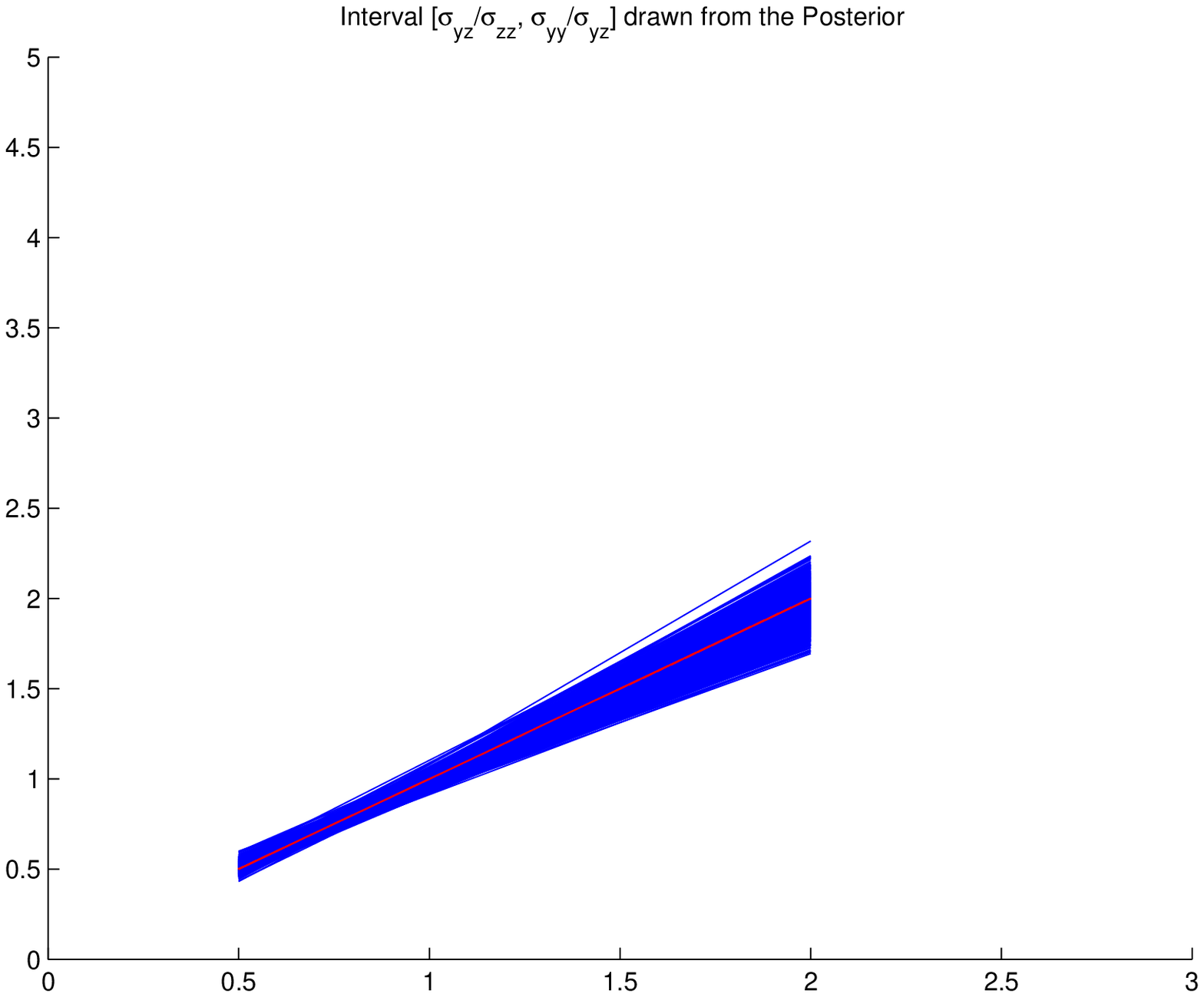}}
  \caption{{\small Linear Regression with errors in Regressors. Representation of the intervals $\Gamma_{I}(\theta) = \left[\frac{\sigma_{yz}}{\sigma_{zz}},\frac{\sigma_{yy}}{\sigma_{yz}}\right]$ drawn from the prior (panel \ref{Example_3_fig_2_a}) and posterior (panel \ref{Example_3_fig_2_b}) against the true interval $[1/2,2]$ (in red).}}
  \label{fig_Example_3_Dirichlet_Figure_2}
\end{figure}
\end{example}
%
\section{Marginal Identification}\label{s_marginal_models}
In sections \ref{s_identification_by_the_prior} and \ref{s_Bayesian_set_estimation} we have considered statistical models where no parameter is marginalized out. We refer to these models as full models. When the parameter of interest is a sub-parameter $\gamma$ of the whole model parameter one might want to perform the analysis by getting rid of the parameters of the model that are not of interest. Therefore, it is natural to examine the marginal model in this sub-parameter. 

\subsection{Marginal Identification of nonidentified sub-parameters}
\indent Let $\theta := (\beta,\gamma)$ denote the whole parameter of the model where $\beta$ is identified and $\gamma$ is the parameter of interest that is nonidentified. For instance, $\gamma$ is the parameter related to a latent variable as in section \ref{ss_models_hyperparameters}. One can specify the prior for $\theta$ as $\mu(\theta) = \mu(\beta|\gamma)\mu(\gamma)$. Hence, the marginal model is obtained by integrating out the parameter $\beta$ in the original model with respect to the prior $\mu(\beta|\gamma)$:
    \begin{displaymath}
      P^{\gamma}(E) = \int P^{\beta}(E)\mu(d\beta|\gamma),\quad \forall E\in\mathcal{X},
    \end{displaymath}
\noindent where $P^{\gamma}(\cdot)$ denotes the integrated sampling distribution which depends on $\gamma$. The corresponding Lebesgue density function (or marginal likelihood) writes $p(x|\gamma) = \int p(x|\beta,\gamma)\mu(\beta|\gamma) d\beta$. The predictive density $p(x)$, obtained by integrating out $\gamma$ from $p(x|\gamma)$ with respect to the prior, is the same as in the full model and the marginal posterior $\mu(\gamma|x)$ of $\gamma$ is obtained by marginalizing the joint posterior $\mu(\beta,\gamma|x)$ with respect to $\beta$.

\begin{example}[\textbf{Classical model of hyperparameter}]\label{Example:5}
  Let us consider the following Bayesian model $x_{i}|\beta,\gamma  \sim iid\:\mathcal{N}(\beta,\sigma^{2})$, $i=1,\ldots,n$, $\beta|\gamma \sim \mathcal{N}(\gamma,\sigma_{0}^{2})$, $\gamma \sim \mathcal{N}(\gamma_{0},\tau_{0}^{2})$, where $\gamma$ is the hyperparameter of the prior distribution and $\sigma^{2},\sigma_{0}^{2},\gamma_{0}$, and $\tau_{0}^{2}$ are known parameters. The parameter $\gamma$ is unidentified in the sampling model. The minimal sufficient $\sigma$-field $\mathcal{AX}$ is almost surely equal to the $\sigma$-field generated by $\beta$. The marginal model is $(x_1, \ldots, x_n)' |\gamma \sim \mathcal{N}\Big(\gamma \iota,\sigma^{2}I_{n} + \sigma_{0}^{2}\iota\iota'\Big)$, where $\iota = (1,\ldots,1)'$ is $n\times 1$ and $I_{n}$ is the $n$-dimensional identity matrix. Therefore, $\gamma$ is identified in the marginal model.
\end{example}

\indent Note that, in the previous example, even if $\gamma$ is identified in the marginal model it is not exactly estimable. In fact, Theorem \ref{tr:1} does not apply because the marginal model is not \emph{i.i.d.} Exact estimability would hold only if the conditional distribution of $\gamma$ given $\beta$ was a degenerated Dirac measure on a deterministic function of $\beta$.\\
\indent In the setting of Example \ref{Example:5}, $\beta$ can be interpreted as an heterogeneity parameter whose distribution depends on an unidentified parameter $\gamma$, see \textit{e.g.} \cite{HeckamnSigner1984}. In the frequentist setting, the conditional distribution of $\beta|\gamma$ is part of the data generating process while in the Bayesian setting the conditional distribution of $\beta|\gamma$ is the prior.\\
\indent The next theorem considers the asymptotic behavior of a sub-parameter which might be nonidentified in the full model. It states that the posterior mean of a sub-parameter $c$ converges $\Pi$-\textit{a.s.} to the conditional prior mean given the identified parameter. Remark that the \textit{a.s.} in the theorem is with respect to the joint distribution.
    \begin{tr}\label{tr_consistency_marginal_model}
      Let us consider a Bayesian model $\{\Theta \times X, \mathcal{A}\otimes \mathcal{X},\Pi\}$ with a filtration $\mathcal{X}_{n}\rightarrow
      \mathcal{X}_{\infty}$ and where the identified parameter $\mathcal{AX}_{\infty}$ is asymptotically exactly estimable, that is, $\mathcal{AX}_{\infty}\subset\overline{\mathcal{X}}_{\infty}$. Let $c$ be an integrable function defined on $\mathcal{A}$, then: $\mathbf{E}(c|\mathcal{X}_{n}) \rightarrow \mathbf{E}(c|\mathcal{AX}_{\infty})$, $\Pi-a.s.$
    \end{tr}

\subsection{Marginal identification in partially identified models}\label{ss:marginal:set:estimation}
\indent Consider now the partially identified model of section \ref{s_Bayesian_set_estimation}. Let $Y$ be an observable random variable with distribution $F$. The parameter $F$ is identified and suppose that there is  another parameter $\theta$ of the model that is identified and that can be written as $\theta = \phi(F)$ for some functional $\phi$. The parameter of interest is denoted by $\gamma$ and is related to $\theta$ by relation (\ref{eq_moment_equation}): $A(\theta,\gamma) \in A_{0}\subset\Phi$. Let $\Gamma$ be provided with a $\sigma$-field $\mathcal{G}$. Hence, the parameter space is $(\Theta \times \Gamma, \mathcal{A}\otimes\mathcal{G})$. By using the structural relation \eqref{eq_moment_equation} we now specify a restricted prior $\mu(\gamma|\theta)$ for $\gamma$ conditional on $\theta$. In particular, $\mu(\gamma|\theta)$ has support equal to the set of the $\gamma$s that satisfy the constraint $A(\theta,\gamma)\in A_{0}$ in \eqref{eq_moment_equation} for a given $\theta\in \Theta$.\footnote{In alternative, we may relax this constraint on the support of $\mu(\gamma|\theta)$ into a constraint on the hyperparameter of the distribution of $\gamma$, as illustrated by the examples below.} The marginal posterior of $\gamma$ is: $\forall \Gamma_{1}\in\mathcal{G},$
    \begin{eqnarray*}
      \mu(\Gamma_{1}|y) & = & \int_{\Gamma_{1}}\int_{\Theta} \mu(\theta|y)\mu(\gamma|\theta)d\theta d\gamma = \int_{\Gamma_{1}}d\gamma\int_{\{F;\phi(F)\in\Theta\}} \mu(\gamma|\phi(F))\mu(dF|y) ,
    \end{eqnarray*}
\noindent where $\mu(\theta|y)$ denotes the posterior distribution of $\theta$ and $\mu(dF|y)$ denotes the posterior distribution of $F$. In the second equality we have written the integral in terms of $\mu(dF|y)$ to stress the fact that the prior distribution of $\theta$ is recovered from the Dirichlet process prior for $F$ as described in section \ref{s_Bayesian_set_estimation}. It is clear that $\gamma$ is identified in the marginal model since its marginal prior distribution is updated by the data. In addition, Theorem \ref{tr_consistency_marginal_model} above applies also to the case of partially identified models.\\
\indent While the marginal posterior density $\mu(\gamma|y)$ of $\gamma$ is not usually known in closed-form -- in particular if $\theta$ is infinite dimensional -- one can easily simulate from it. For this, one first simulates $\theta$ given $y$ from $\mu(\theta|y)$ and then, for each draw of $\theta$, one simulates $\gamma$ given $\theta$ from $\mu(\gamma|\theta)$. This simulation scheme produces draws from $\mu(\gamma|y)$ and this is due to the lack of identification of $\gamma$ which implies that $\gamma \perp y|\theta$, see section \ref{ss_Bayesian_identification}. Having a marginal posterior distribution $\mu(\gamma|y)$ of the parameter of interest $\gamma$ is important for instance in a decision problem setting. In fact, knowledge of $\mu(\gamma|y)$ allows to select the most likely value of $\gamma$ or the region inside $\Gamma_{I}$ with the highest posterior probability. Such a selection is clearly affected by the choice of the prior on $\gamma$.
\subsection{Examples}\label{ss:exmple:marginal:identification}
\indent In this section we develop further Example \ref{Example:3} of section \ref{ss_examples_set_estimation} by endowing the parameter $\gamma$ with a conditional prior distribution given $\theta=\phi(F)$ that we denote by $\mu_{\gamma}^{F}$: $\gamma|\theta\sim \mu_{\gamma}^{F} := \mu(\gamma|\phi(F))$. Additional examples are developed in Appendix C in the Supplement. In our simulations we consider four different specifications for $\mu_{\gamma}^{F}$, where the hyperparameters $a_0$ and $b_0$ are specified in each specific example:
    \begin{enumerate}
      \item[(I).] $\gamma|F \sim  \mathcal{N}(\gamma_{0},\tau_{0}^{2})$, $\gamma_{0} = \frac{\wtl{\gamma}_{0}^{1} + \wtl{\gamma}_{0}^{2}}{2 c_0}$, $\tau_{0}\in\mathbb{R}$ and we discard the draws of $\gamma$ that do not belong to the interval $[a_0,b_0]$;
      \item[(II).] $\gamma|F \sim \mathcal{N}(0,\sigma_{0}^{2})$ truncated to the interval $[a_0,b_0]$;
      \item[(III).] $\gamma|F \sim \mathcal{U}[a_0,b_0]$ (flat prior);
      \item[(IV).] $\gamma|F \sim \mathcal{B}e\Big(a_0,b_0,p,q\Big)$, that is, a Beta prior distribution with support $[a_0,b_0]$ and shape parameters $p$ and $q$. The corresponding probability density function is:
            \begin{displaymath}
              \mu(\gamma|F) = \frac{\Big(\gamma - a_0\Big)^{p-1}\Big(b_0 - \gamma\Big)^{q - 1}}{B(p,q)\Big(b_0 - a_0\Big)^{p+q-1}}
            \end{displaymath}
          \noindent where $B(p,q)$ is the beta function.
    \end{enumerate}
\setcounter{example}{2}

\begin{example}[\textbf{Linear Regression with errors in Regressors} (\textit{continued}).]
  Suppose that we are not only interested in the identified region $\Gamma_{I} := [\theta_{1},\theta_{2}]$, where $\theta_1,\theta_2$ are defined in \ref{ex_regressors_with_errors}, but also in the parameter $\gamma$ itself. The marginal posterior distribution of $\gamma$ is informative about the areas of the identified region $\Gamma_{I}$ where the parameter is more likely. The prior distribution of $\theta := (\theta_{1},\theta_{2})$ is obtained from a Dirichlet process prior for the joint probability distribution $F_{yz}$ of $(Y,Z)$ denoted by $\mu_{F}$, see section \ref{s_Bayesian_set_estimation}. The Bayesian hierarchical model is as in \eqref{simulation:regression:errors:regressors:1}-\eqref{simulation:regression:errors:regressors:2} completed with the specification of a prior for $\gamma$ conditional on $F_{yz}$: $\gamma|F_{yz} \sim \mu_{\gamma}^{F_{yz}}$. In our simulation exercise we consider the four specifications (I)-(IV) for $\mu_{\gamma}^{F_{yz}}$ given above with: $a_0 = \min\Big(\frac{\sigma_{yz}}{\sigma_{zz}}, \frac{\sigma_{yy}}{\sigma_{yz}}\Big)$, $b_0 = \max\Big(\frac{\sigma_{yz}}{\sigma_{zz}},\frac{\sigma_{yy}}{\sigma_{yz}}\Big)$, $c_0 = 1$, and $\wtl{\gamma}_{0}^{1} = a_0$, $\wtl{\gamma}_{0}^{2} = b_0$. Since $\gamma$ is nonidentified in the full model, but identified in the marginal model, its posterior distribution depends on the data only through $F$. This means that the moments $\sigma_{yz}$, $\sigma_{zz}$ and $\sigma_{yy}$ in the prior for $\gamma$, computed from the Dirichlet process prior for $F_{yz}$ are replaced with the posterior means of $\sigma_{yz}$, $\sigma_{zz}$ and $\sigma_{yy}$ in the posterior distribution, computed from the posterior of the Dirichlet process for $F_{yz}$.\\
\indent We generate an $n$-sample of observations of $(Y_{1},Y_{2})$ as in \eqref{simulation:regression:errors:regressors:1}. The parameters are fixed as follows: $n = 1000$, $F_0$ is specified as in \eqref{simulation:regression:errors:regressors:2}, $\tau_{0}^{2} = 1$, $\sigma_{0}^{2}=2$, $p=1$ and $q=0.5$. The true identified set is $\Gamma_{I} = [1/2,2]$.\\
\indent We draw $1000$ times from the marginal prior and posterior distributions of $\gamma$. The simulation scheme is the following: for each $1\leq j\leq 1000$, draw $F_{yz}^{(j)}$ from the prior $\mu(F_{yz})$ (resp. the posterior $\mu(F_{yz}|\{y_{i},z_i\}_{i=1}^{n})$), compute $\theta_{i}^{(j)}$, $i=1,2$ and draw $\gamma^{(j)}$ from $\mu_{\gamma}^{F_{yz}^{(j)}}$ (resp. $\mu_{\gamma}^{F_{yz}^{(j)}}(\gamma|F_{yz}^{(j)},\{y_{i},z_i\}_{i=1}^{n})$). Figure \ref{fig_Example_Regression_errors_Regressors_Dirichlet_2nd} shows the histograms of the marginal prior (in blue) and posterior (in red) distribution of $\gamma$. Each panel corresponds to one of the four specifications for $\mu_{\gamma}^{F_{yz}}$. We see that the marginal posterior distribution is much more concentrated on the true identified set than the corresponding prior.
\begin{figure}[!h]
  \centering
  \subfloat[{$\gamma|F_{yz} \sim \mathcal{N}(\gamma_{0}, 1)$ by discarding the draws that are not in $[\theta_1,\theta_2]$}.]{
      \includegraphics[width=0.4\linewidth]{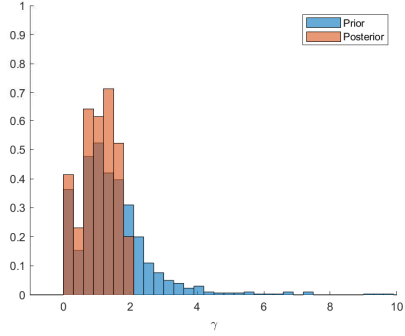}}
  \hspace{0.1\linewidth}
  \subfloat[{$\gamma|F_{yz} \sim \mathcal{N}(0, 2)$ truncated to $[\theta_1,\theta_2]$}.]{
      \label{density_constraint}
      \includegraphics[width=0.4\linewidth]{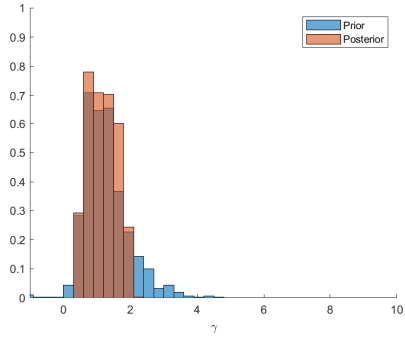}}
  \subfloat[{$\gamma|F_{yz} \sim \mathcal{U}[\mathbf{E}_{F_{1}}(Y_{1}),\mathbf{E}_{F_{2}}(Y_{2})]$}.]{
      \includegraphics[width=0.4\linewidth]{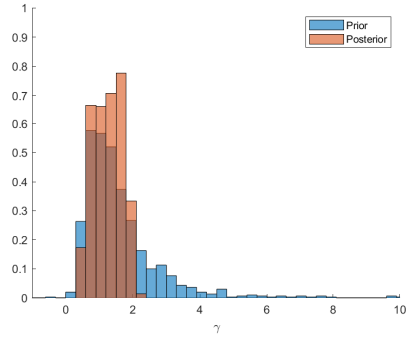}}
  \hspace{0.1\linewidth}
  \subfloat[{$\gamma|F_{yz} \sim \mathcal{B}eta(\mathbf{E}_{F_{1}}(Y_{1}),\mathbf{E}_{F_{2}}(Y_{2}),2,2)$}.]{
      \label{density_constraint}
      \includegraphics[width=0.4\linewidth]{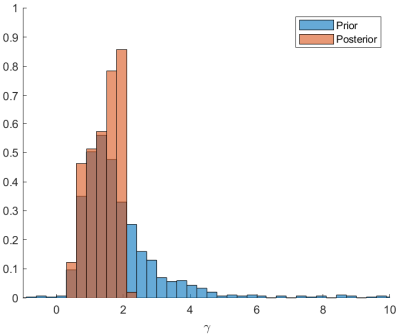}}
  \caption{{\small Linear Regression with errors in Regressors. Histograms of the prior (in blue) and posterior (in red) probability distributions. The true identified set is $\Gamma_{I} = [1/2,2]$.}}
  \label{fig_Example_Regression_errors_Regressors_Dirichlet_2nd}
\end{figure}
\end{example}

\setcounter{section}{5}
\section{Conclusions}\label{s:conclusion}
This paper studies theoretical properties and implementation of the Bayesian approach in various models that lack identification. As examples of unidentified models, we analyse nonparametric models with heterogeneity modeled either as a Gaussian process or as a Dirichlet process where the parameter of interest is the (hyper)parameter of the heterogeneity distribution which is unidentified. We also analyse unidentified latent variable and partially identified models.\\
\indent In partially identified models we propose to construct the prior and posterior of the identified set through the prior and posterior capacity functionals. The prior capacity functional is obtained as the transformation of a Dirichlet process, so that our approach is completely nonparametric. The proposed procedure is appealing since, even if the posterior capacity functional has a complicated expression, simulating from it is simple.\\
\indent Finally, we discuss models that have some parameters that are identified and others that are not. For these models, we show that the parameter that is unidentified or partially identified in the full model is identified in the marginal model.


\begin{spacing}{1}
\setlength{\bibsep}{0.2cm}

\bibliography{AnnaBib}
\end{spacing}

\end{document}


\title{Supplement to:\\Revisiting identification concepts in Bayesian analysis}

\author{\begin{tabular}{ccc}
Jean-Pierre Florens\footnote{Toulouse School of Economics, Universit\'{e} de Toulouse Capitole, Toulouse - 21, all\'{e}e de Brienne - 31000 Toulouse (France). Email: jean-pierre.florens@tse-fr.eu} & & Anna Simoni\footnote{CREST, CNRS, ENSAE, \'{E}cole Polytechnique - 5, avenue Henry Le Chatelier, 91120 Palaiseau, France. Email: anna.simoni@ensae.fr (\textit{corresponding author}).}
\end{tabular}}
\maketitle
\appendix
\setcounter{figure}{8}
\section*{Appendix}
\section{Other latent variable models for section 3.2}\label{App:s:Ex:latent:models}
\paragraph{Multinomial probit model.}
The observable outcome $y_i$ can take on one of the $p$ values $\{1,\ldots,p\}$ and $z_i$ is a continuous $p$-dimensional random vector. The function $g(\cdot)$ is defined as: $$g(z_i) = \sum_{j=1}^p j \In\{\max_k z_{ik} = z_{ij}\}.$$ The identification problem arises because the $g(\cdot)$ function is invariant to both location and scale transformations of $z_i$: if we add the same constant to all the $p$ components of $z_i$ or if we scale each component by the same positive constant, the likelihood does not change. An MCMC algorithm to draw from the posterior of the nonidentified models without having to impose identifying restrictions is given in \cite{McCullochRossi1994}.
\paragraph{Multivariate probit model.}
The observable outcome $y_i$ is a $p$-dimensional vector where each component is a binary random variable: $y_{ij} \in\{0,1\}$, for $j=1,\ldots,p$. The latent vector is also $p$-dimensional and $g(z_i) = (\In\{z_{i1}>0\}, \ldots, \In\{z_{ip}>0\})'$. This model is nonidentified because one can multiply each component of $z_i$ by a different positive constant without changing the likelihood function. To conduct Bayesian analysis for this model one has to choose whether to use the nonidentified or the identified model. Even if the nonidentified model has larger dimension than the identified model (because it does not include identifying restrictions) it will produce an MCMC algorithm with better mixing properties than MCMC algorithm for the corresponding identified model. In particular, one can use the MCMC algorithm for the nonidentified model in \cite{EdwardsAllenby2003}.

\section{Further examples for section 4.2}\label{App:s:Ex:1}
\setcounter{example}{4}
\begin{example}[\textbf{Interval Regression Model}]
  This example is a generalization of Example 2 in the manuscript. Let $X$ be a vector of regressors and suppose the conditional expectation of $Y$ is linear: $\mathbf{E}(Y|X)=X'\gamma$, where $\gamma$ is the parameter vector of interest. The random variable $Y$ is unobservable but it is known that with probability one
    \begin{displaymath}
      \mathbf{E}(Y_{1}|X)\leq X'\gamma \leq \mathbf{E}(Y_{2}|X)
    \end{displaymath}
\noindent where $Y_{1}$ and $Y_{2}$ are two observable real random variables with unknown probability distribution. These conditional moment inequalities are then transformed in unconditional moment inequalities by using a vector $Z$ of either positive transformations of $X$ or positive instrumental variables:
    \begin{equation}\label{example_4_2_moment_inequalities}
      \mathbf{E}_{F}(Y_{1}Z)\leq \mathbf{E}_{F}[ZX]'\gamma \leq \mathbf{E}_{F}(Y_{2}Z), \quad \mu- \textrm{a.s.},
    \end{equation}
\noindent where $\mathbf{E}_{F}(\cdot) := \mathbf{E}(\cdot|F)$ denotes the expectation taken with respect to the joint distribution $F$ of $(Y_{1},Y_{2},X,Z)$. For a probability measure $F_0$, the nonparametric Bayesian model is
    \begin{eqnarray}
      (y_{1i},y_{2i},x_{i},z_{i})|F & \sim \:iid & F, \quad i = 1, \ldots, n, \nonumber\\
      F & \sim & \mathcal{D}ir(n_{0},F_{0}), \qquad n_0\in\mathbb{R}_+ \label{eq:4}
    \end{eqnarray}
\noindent where $w_i := (y_{1i},y_{2i},x_{i},z_{i})_i$ are realizations of $W := (Y_{1},Y_{2},X,Z)$. The identified set is $\Gamma_{I}(F) = \{\gamma\in\Gamma; \:\mathbf{E}_{F}(Y_{1}Z)\leq \mathbf{E}_{F}[ZX]'\gamma \leq \mathbf{E}_{F}(Y_{2}Z)\}$. For every $\gamma \in\Gamma$, its prior coverage function $p_{\scriptscriptstyle{\Gamma_{I}}}(\gamma)$ is given by
    \begin{eqnarray*}
      P\Big(\mathbf{E}_{F}(Y_{1}Z)\leq \mathbf{E}_{F}[ZX]'\gamma \leq \mathbf{E}_{F}(Y_{2}Z)\Big) & = & P\Big(\sum_{j}\alpha_{j}\xi_{j}^{y_{1}}\xi_{j}^{z}\leq \sum_{j}\alpha_{j}\xi_{j}^{z}\xi_{j}^{x'}\gamma\leq \sum_{j}\alpha_{j}^{2}\xi_{j}^{y_{2}}\xi_{j}^{z}\Big)
    \end{eqnarray*}
\noindent where $\xi_{j} := (\xi_{j}^{y_{1}}, \xi_{j}^{y_{2}}, \xi_{j}^{x'}, \xi_{j}^{z'})'\sim \:iid\: F_{0}$ and $\{\alpha_{k}\}_{k\geq 1}$ are computed as in (4.4). Remark that $\xi_{j}^{x}$ and $\xi_{j}^{z}$ have the same dimension as $X$ and $Z$, respectively. The posterior of $F$ is $F|(w_{1},\ldots,w_{n}) \sim \mathcal{D}ir(n_{*},F_{*})$,
%
%
\noindent where for $i=1,2$, $n_{*} = n_{0} + n$, $F_{*} = \frac{n_{0}}{n_{0} + n}F_{0} + \frac{n}{n_{0} + n}F_{n}$, $F_{n}(\cdot) := \frac{1}{n}\sum_{j}\delta_{w_{j}}(\cdot)$ is the empirical distribution of the sample $(w_{1},\ldots,w_{n})$. The posterior coverage function of $\Gamma_{I}$ is given by
    \begin{multline*}
      P\Big(\mathbf{E}_{F}(Y_{1}Z)\leq \mathbf{E}_{F}[ZX]'\gamma \leq \mathbf{E}_{F}(Y_{2}Z)\Big|\{w_{i}\}_{i=1}^{n}\Big) = \\
      P \Big(\rho\sum_{j}\beta_{j}y_{1j}z_{j} + (1 - \rho)\sum_{j}\alpha_{j}\xi_{j}^{y_{1}}\xi_{j}^{z}\leq \left[\rho\sum_{j}\beta_{j}z_{j}x_{j}' + (1 - \rho)\sum_{j}\alpha_{j}\xi_{j}^{z}\xi_{j}^{x'}\right]\gamma \leq \\
      \rho\sum_{j}\beta_{j}y_{2j}z_{j} + (1 - \rho)\sum_{j}\alpha_{j}^{2}\xi_{j}^{y_{2}}\xi_{j}^{z}\Big|\{w_{i}\}_{i=1}^{n}\Big),
    \end{multline*}
\noindent where $\rho$ and $(\beta_{1},\ldots, \beta_{n})$ are defined as in (4.6).\\
\indent A simulation exercise allows to visualize the prior and posterior coverage function of $\Gamma_{I}$. We generate an $n$-sample of realizations of $(Y_{1},Y_{2},X,Z)$ as: for $i=1,\ldots,n$,
    \begin{eqnarray}
      x_{i} & = & z_{i} + u_{i},\qquad z_{i} \;\sim i.i.d. \;\mathcal{U}[0,1],\qquad u_{i}\sim \; i.i.d. \; \mathcal{N}(0,1), \nonumber\\
      y_{1i} & = & \theta_{1}x_{i} + \varepsilon_{1i}, \qquad \theta_{1} = 2, \qquad \varepsilon_{1i}\sim i.i.d. \;\mathcal{N}(0,0.1), \nonumber\\
      y_{2i} & = & \theta_{2}x_{i} + \varepsilon_{2i}, \qquad \theta_{2} = 6, \qquad \varepsilon_{2i}\sim i.i.d. \;\mathcal{N}(0,0.1). \label{eq:5}
    \end{eqnarray}
\noindent Since $\mathbf{E}_{F}(ZX) = \mathbf{E}_{F}(Z^2) = 1/12>0$ we can divide by $\mathbf{E}_{F}(ZX)$ to get the true identified set given by $\Gamma_{I} = \Big[\frac{\mathbf{E}_{F}(Y_{1}Z)}{\mathbf{E}_{F}(ZX)}, \frac{\mathbf{E}_{F}(Y_{2}Z)}{\mathbf{E}_{F}(ZX)}\Big] = [2,6]$. The parameters of the Dirichlet process are fixed as $n_{0} = 20$ and
    \begin{equation}\label{ex:3:Dir:hyperparameter}
      F_{0} = \mathcal{N}_{4}\Big(\left(\begin{array}
        {c}0\\4\\0\\0.5
      \end{array}\right),\left(\begin{array}
        {cccc}0.1 & 0 & 0.2 & \frac{3}{2}\\0 & 0.1 & 0.2 & 3\\ 0.2 & 0.2 & 0.1 & 0.5\\ \frac{3}{2} & 3 & 0.5 & 0.1
      \end{array}\right)\Big).
    \end{equation}
\noindent Moreover, we set: $n = 1000$ and draw $1000$ intervals $\left[\frac{\mathbf{E}_{F}(Y_{1}Z)}{\mathbf{E}_{F}(ZX)}, \frac{\mathbf{E}_{F}(Y_{2}Z)}{\mathbf{E}_{F}(ZX)}\right]$ from the prior and posterior distributions of $F$. In Figure \ref{fig_Example_2_Dirichlet_random_set} we represent the prior and posterior coverage functions of the random interval $\Gamma_I = \left[\frac{\mathbf{E}_{F}(Y_{1}Z)}{\mathbf{E}_{F}(ZX)}, \frac{\mathbf{E}_{F}(Y_{2}Z)}{\mathbf{E}_{F}(ZX)}\right]$ for each value of $\gamma$ in a grid of $[-1,20]$. Figure \ref{fig_Example_2_Dirichlet_Figure_2} displays the intervals $\Gamma_I$ drawn from the prior and posterior distribution of $\Gamma_I$ (on the vertical axis) against the true interval $[2,6]$ (on the horizontal axis).\\

\begin{figure}[!h]
  \centering
  \subfloat[{Prior coverage function}]{
      \includegraphics[width=0.4\linewidth]{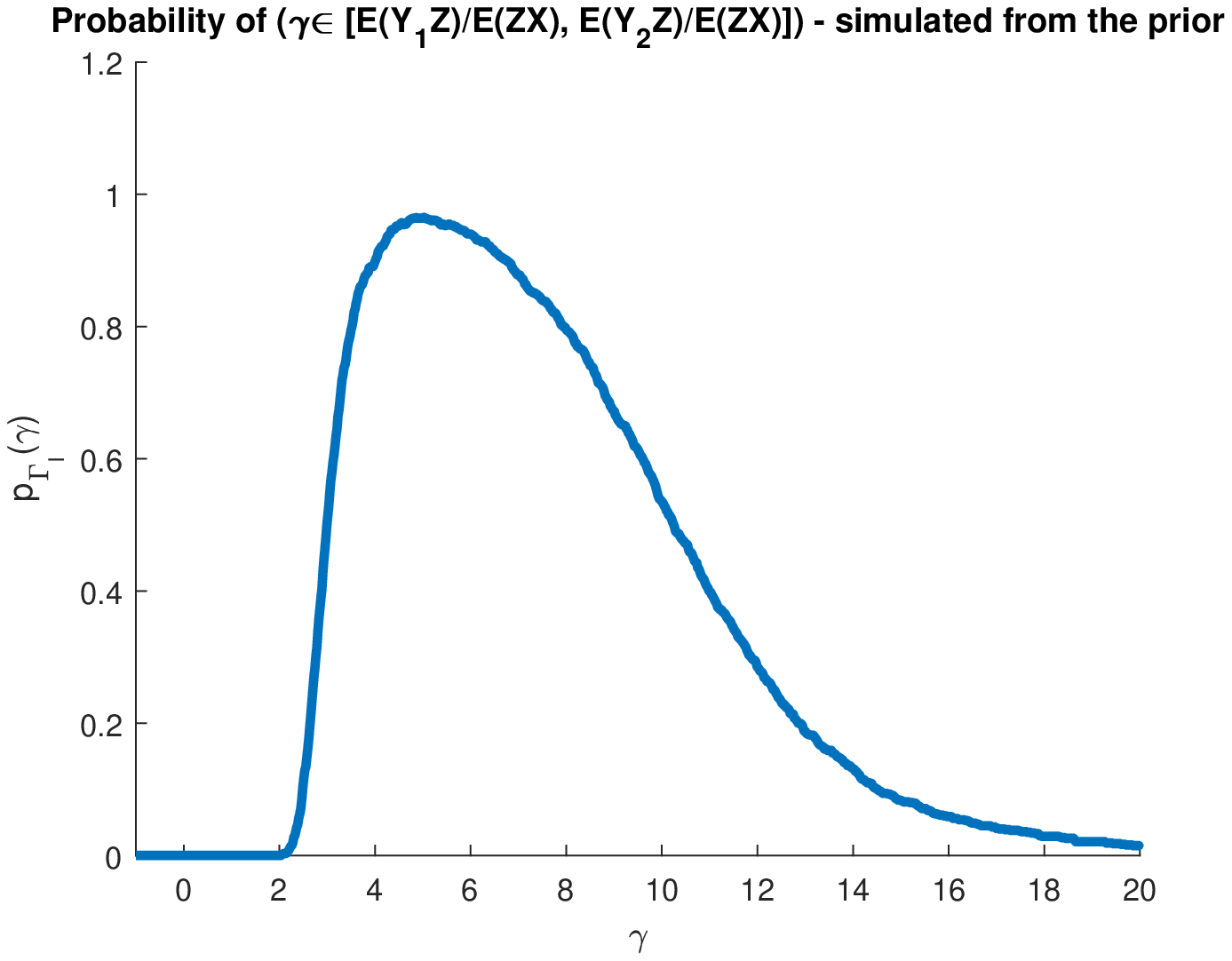}}
  \hspace{0.1\linewidth}
  \subfloat[{Posterior coverage function}]{
      \includegraphics[width=0.4\linewidth]{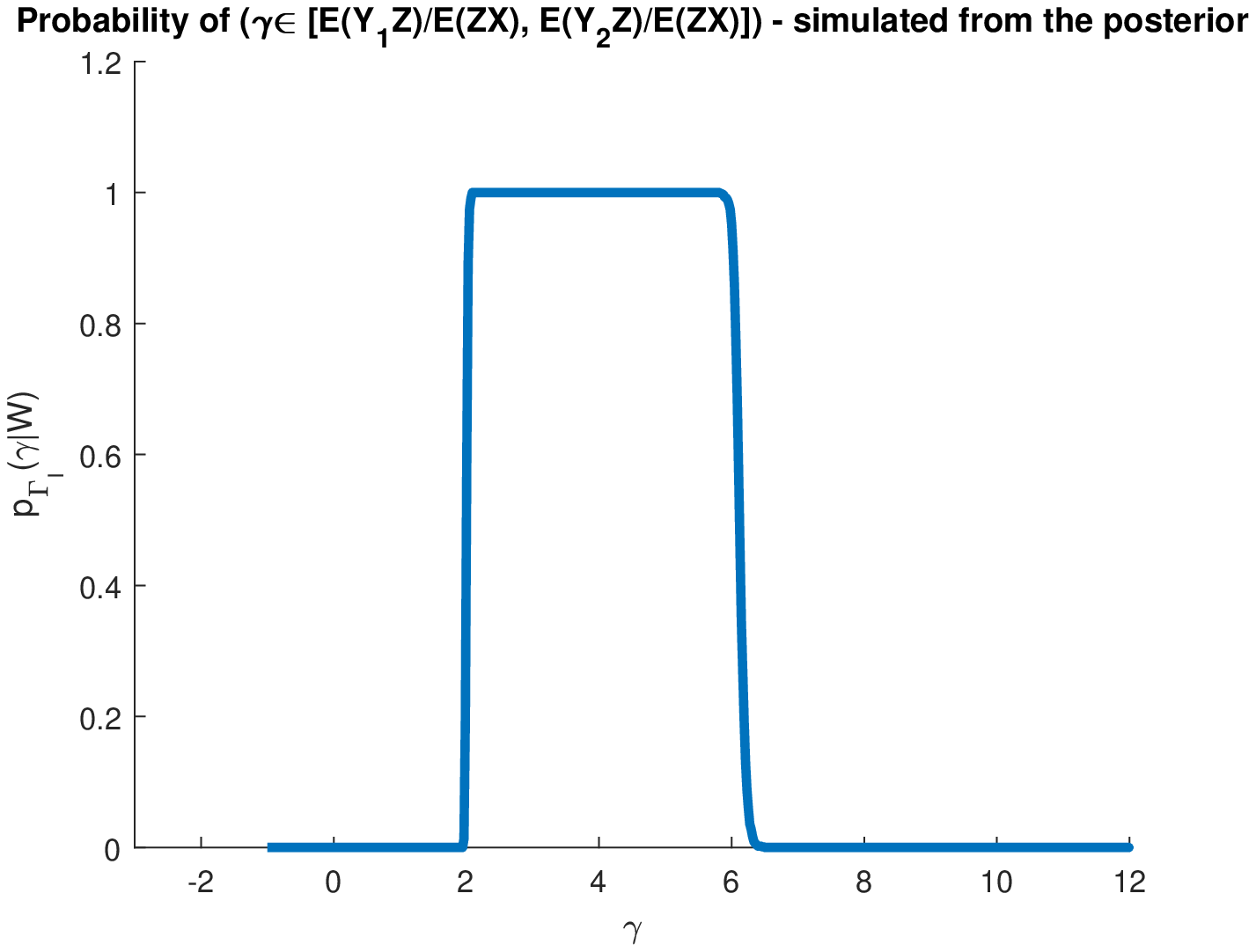}}
  \caption{{\small Interval Regression Model. Prior and posterior coverage functions $p_{\scriptscriptstyle{\Gamma_{I}}}(\cdot)$ and $p_{\scriptscriptstyle{\Gamma_{I}}}(\cdot|\{w_{i}\})$ of $\Gamma_{I}$. The true $\Gamma_I$ is $[2,6]$.}}
  \label{fig_Example_2_Dirichlet_random_set}
\end{figure}

\begin{figure}[!h]
  \centering
  \subfloat[{Prior}]{\label{Example_2_fig_2_a}
      \includegraphics[width=0.4\linewidth]{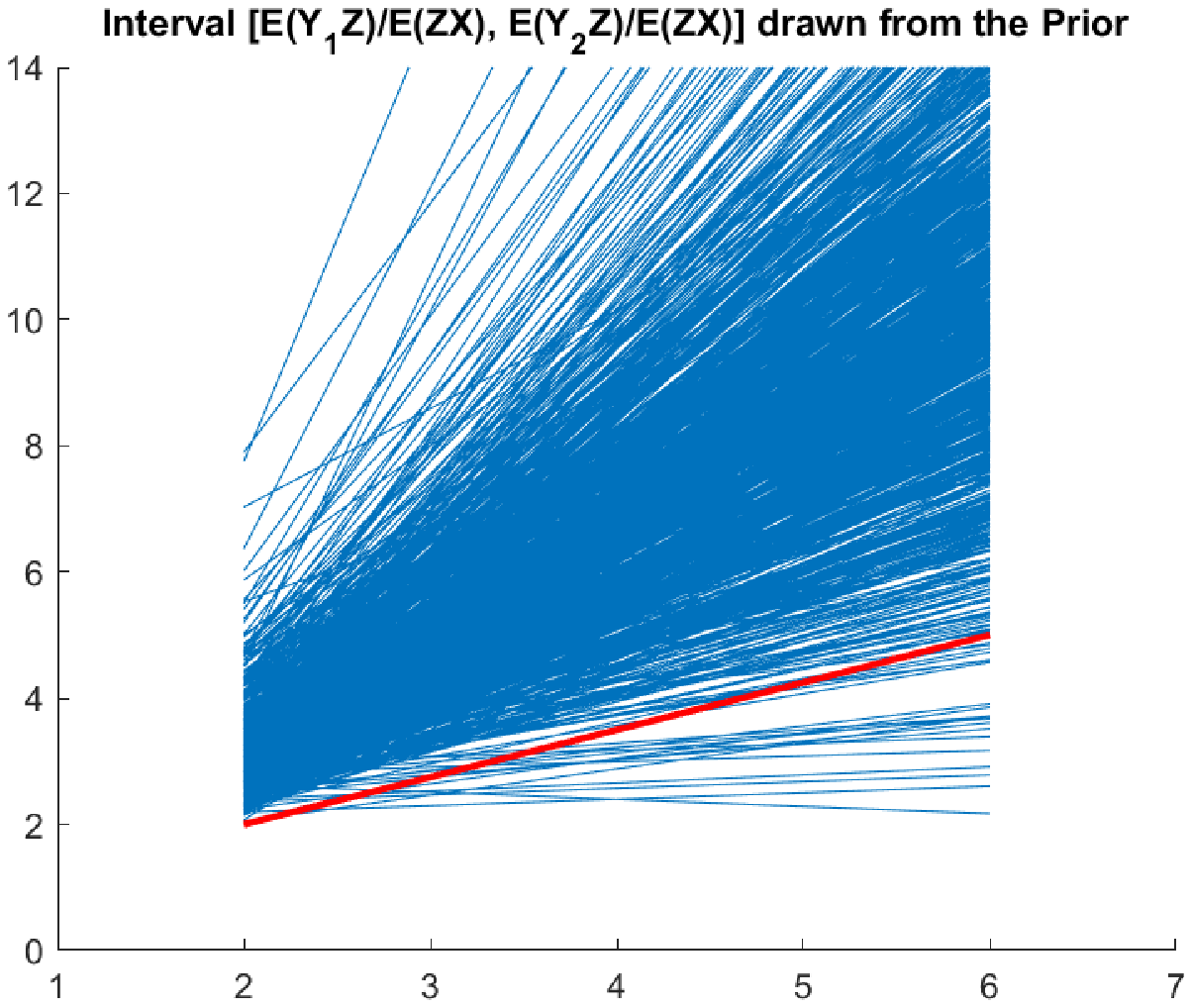}}
  \hspace{0.1\linewidth}
  \subfloat[{Posterior}]{\label{Example_2_fig_2_b}
      \includegraphics[width=0.4\linewidth]{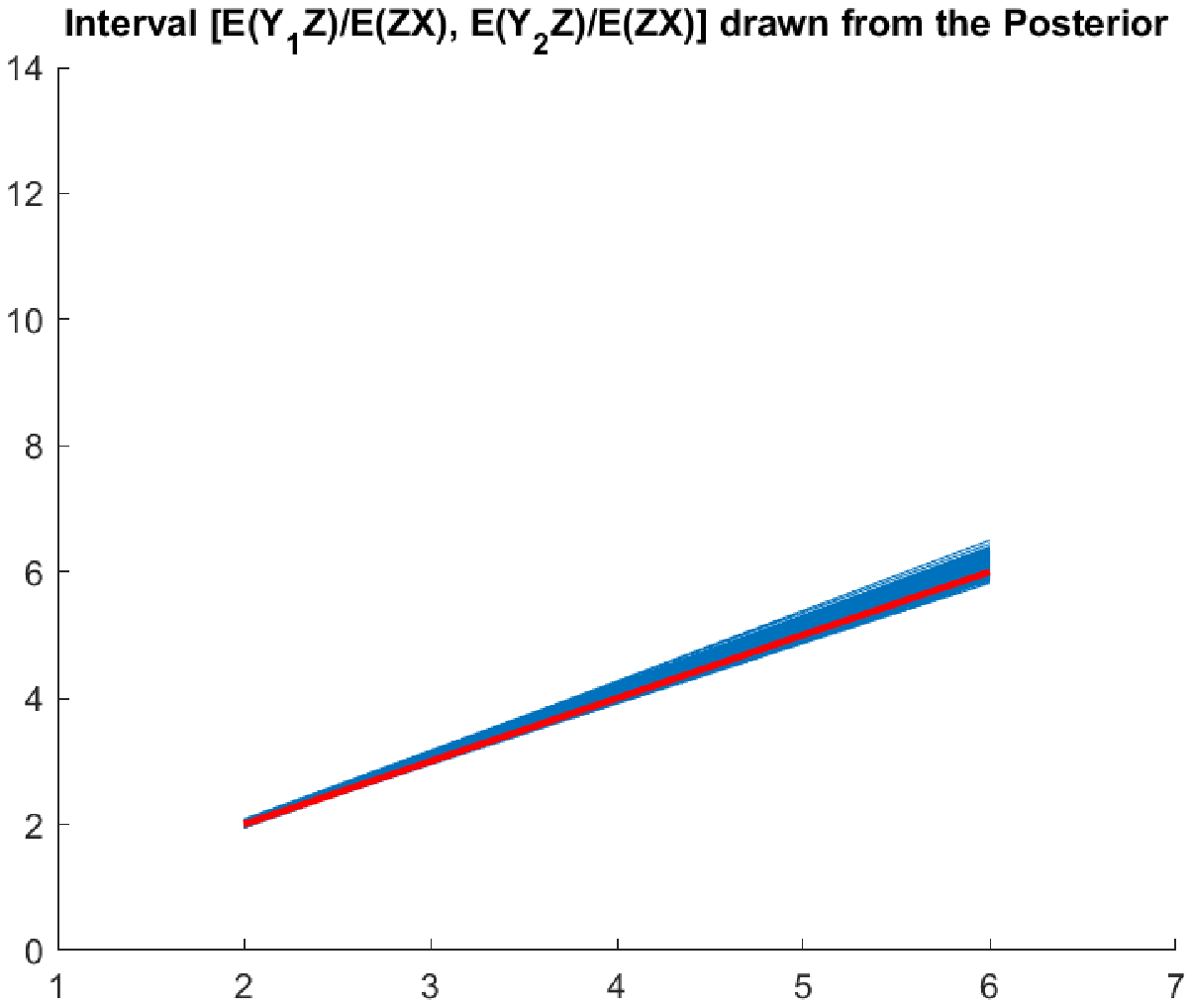}}
  \caption{{\small Interval Regression Model. Representation of the intervals $\Gamma_I(\theta) = [\frac{\mathbf{E}_{F}(Y_{1i}Z_{i})}{\mathbf{E}_{F}(Z_{i}X_{i})}, \frac{\mathbf{E}_{F}(Y_{2i}Z_{i})}{\mathbf{E}_{F}(Z_{i}X_{i})}]$ drawn from the prior (panel \ref{Example_2_fig_2_a}) and posterior (panel \ref{Example_2_fig_2_b}) against the true interval $[2,6]$ (in red).}}
  \label{fig_Example_2_Dirichlet_Figure_2}
\end{figure}
\end{example}

\begin{example}[\textbf{Binary outcome and missing data}]
  Let us consider a random binary variable $Y$ taking values in $\{0,1\}$ and a $n$-random sample of it. This random sample is incomplete in the sense that for some units the outcome $Y$ is not observed due to unknown reasons. Therefore, the observed sample is $x_{i} := (y_{i}d_{i}, d_{i})$, $i=1,\ldots,n$, where $y_{i}$ is the realization of $Y$ for the $i$-th unit and $d_{i}$ is a dummy variable that takes the value $1$ if the corresponding $y_{i}$ is observed and $0$ otherwise. We denote $p_{kj} := \mathbb{P}(y_{i} = k, d_{i}=j)$, $j,k\in\{0,1\}$ and the parameter of interest is $p_{1} := \mathbf{E}(Y) = \mathbb{P}(Y = 1) = p_{10} + p_{11}$. The likelihood of $p := (p_{11}, p_{10}, p_{01}, p_{00})$ is given by
    \begin{equation}
      l(x_{1},\ldots,x_{n}|p) = \frac{n!}{n_{1}!n_{0}!m!}p_{11}^{n_{1}}p_{01}^{n_{0}}(p_{10} + p_{00})^{m},
    \end{equation}
\noindent where $n_{1} := \sum_{i=1}^{n}y_{i}d_{i}$ is the number of observed units with $y_{i} = 1$, $n_{0} := \sum_{i=1}^{n}(1 - y_{i})d_{i}$ is the number of observed units with $y_{i} = 0$ and $m := \sum_{i=1}^{n}(1 - d_{i})$ is the number of units for which we do not observe the corresponding $y_i$ (that is, the units with $d_i=0$). The identified parameter, \textit{i.e.} the minimal sufficient parameter, is $(p_{11},p_{01},\wtl{p}_{\cdot 0})$, where $\wtl{p}_{\cdot 0} := p_{10} + p_{00}$ and the probability of interest $p_{1}$ is not identified. In this example condition (4.1) writes $p_{1}\in[p_{11}, p_{11} + \wtl{p}_{\cdot 0}]$ with $\gamma = p_{1}$, $\theta = (p_{11},p_{11} + \wtl{p}_{\cdot 0})$ and $\Gamma_{I} := [p_{11}, p_{11} + \wtl{p}_{\cdot 0}] \subset [0,1]$. We use a Bayesian nonparametric approach where we specify a Dirichlet prior distribution for $(p_{11},p_{01},\wtl{p}_{\cdot 0})$: $(p_{11},p_{01},\wtl{p}_{\cdot 0})\sim \mathcal{D}ir(\alpha)$ with $\alpha := (\alpha_{1},\alpha_{2},\alpha_{3})$ a vector of positive numbers. The Bayesian model is the following
    \begin{eqnarray}
      x_{1},\ldots,x_{n}|p_{11},p_{01},\wtl{p}_{\cdot 0} & \sim & \mathcal{M}u(n,(p_{11},p_{01},\wtl{p}_{\cdot 0})), \nonumber\\
      (p_{11},p_{01},\wtl{p}_{\cdot 0}) & \sim & \mathcal{D}ir(\alpha), \label{eq:6}
    \end{eqnarray}
\noindent where $\mathcal{M}u$ denotes a joint Multinomial distribution. The prior coverage function is $p_{\scriptscriptstyle{\Gamma_{I}}}(\gamma) = P(p_{11}\leq \gamma \leq p_{11} + \wtl{p}_{\cdot 0})$, $\forall \gamma\in\Gamma$, where $p_{11}\sim \mathcal{B}e(\alpha_{1}, \alpha_{2} + \alpha_{3})$, $(p_{11} + \wtl{p}_{\cdot 0}) \sim \mathcal{B}e(\alpha_{1} + \alpha_{3}, \alpha_{2})$ and $\mathcal{B}e(a_1,a_2)$ denotes a Beta distribution with parameters $a_{1}$ and $a_{2}$. The posterior distribution of the minimal sufficient parameter is $(p_{11},p_{01},\wtl{p}_{\cdot 0})|x_{1},\ldots,x_{n} \sim \mathcal{D}ir(\alpha_{*})$, $\alpha_{*} = (\alpha_{1} + n_{1}, \alpha_{2} + n_{0}, \alpha_{3} + m)$ and the posterior coverage function is defined accordingly. In our simulation exercise we generate an $n$-sample of observations as:
    \begin{eqnarray}
      y_{i} & \sim & \mathcal{B}(p_{y}), \qquad p_{y} = 0.8, \nonumber\\
      d_{i} & \sim & \mathcal{B}(p_{d}), \qquad p_{d} = 0.5 \label{eq:7}
    \end{eqnarray}
\noindent where $\mathcal{B}(q)$ denotes a Bernoulli distribution with probability $q$. Thus, $x_{i} = (y_{i}d_{i}, d_{i})$ and $p_{11} = p_{y}p_{d}$, $p_{10} = p_{y}(1 - p_{d})$, $p_{01} = (1 - p_{y})p_{d}$, $p_{00} = (1 - p_{y})(1 - p_{d})$. The true set $[p_{11},p_{11} + \wtl{p}_{\cdot 0}]$ is $[0.4,0.9]$. The parameter $\alpha$ of the Dirichlet distribution is set equal to $\alpha = (2,3,1)$. We draw $1000$ times $p_{11}$ and $\wtl{p}_{\cdot 0}$ from the prior and posterior distributions. The results are shown in Figures \ref{fig_Example_4_Dirichlet_random_set} and \ref{fig_Example_4_Dirichlet_Figure_2}. Figure \ref{fig_Example_4_Dirichlet_Figure_2} has been obtained by using the first $100$ draws from the prior and posterior distributions and represents the corresponding intervals $\Gamma_{I}$ (on the vertical axes) against the true interval $[0.4,0.9]$. Once the posterior distribution is obtained one could take as an estimator for $[p_{11}, p_{11} + \wtl{p}_{\cdot 0}]$ the interval
    \begin{displaymath}
      [\widehat{p}_{11}, \widehat{p}_{11} + \widehat{\wtl{p}}_{\cdot 0}] := \Big[\frac{\alpha_{1} + n_{1}}{\bar{\alpha} + n}, \frac{\alpha_{1} + \alpha_{2} + n_{1} + m}{\bar{\alpha} + n}\Big],
    \end{displaymath}
\noindent where $\bar{\alpha} = \alpha_{1} + \alpha_{2} + \alpha_{3}$ and where the lower and upper bounds are the posterior means of $p_{11}$ and $p_{11} + \wtl{p}_{\cdot 0}$, respectively.

\begin{figure}[!h]
  \centering
  \subfloat[{Prior coverage function}]{
      \includegraphics[width=0.4\linewidth]{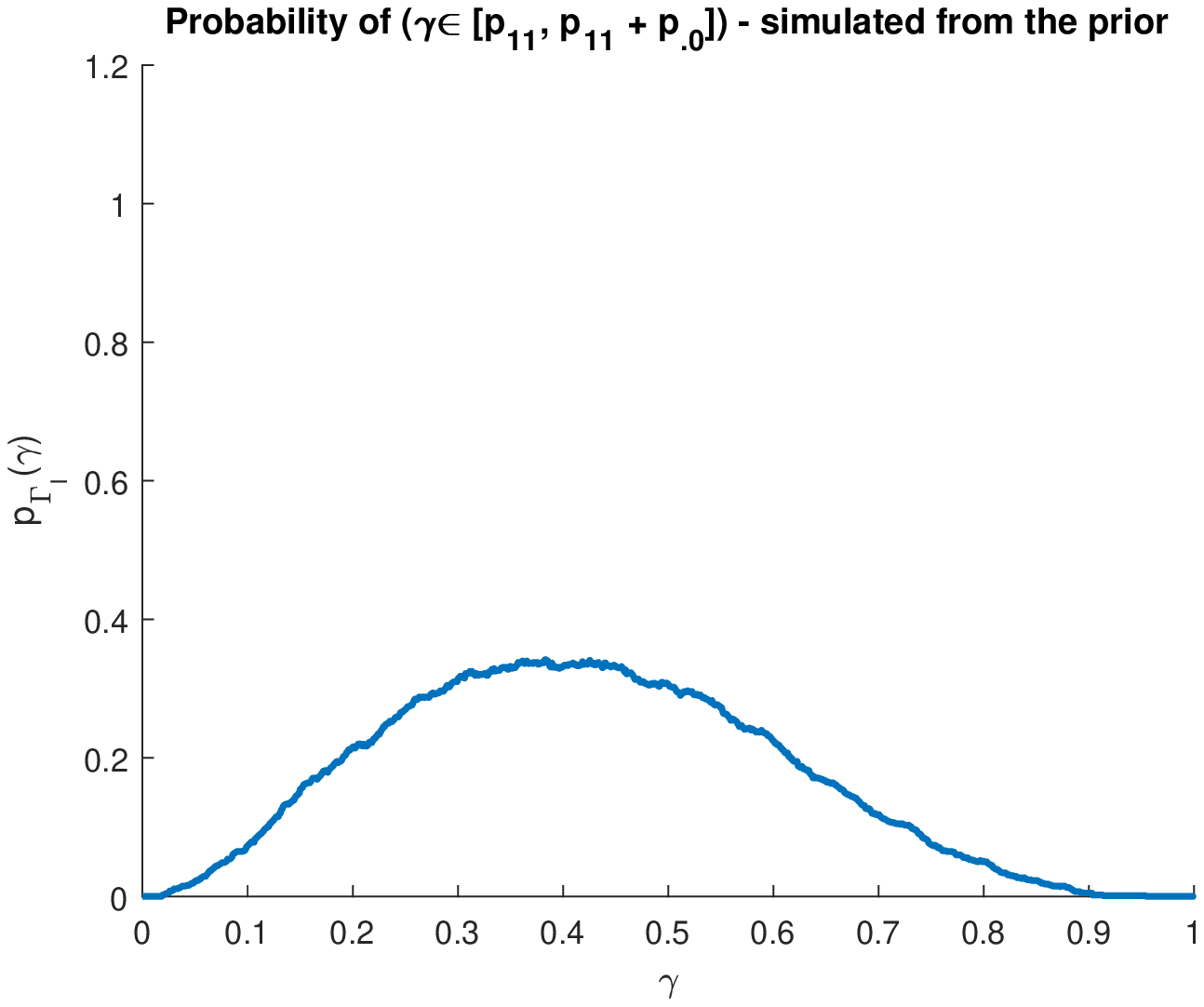}}
  \hspace{0.1\linewidth}
  \subfloat[{Posterior coverage function}]{
      \includegraphics[width=0.4\linewidth]{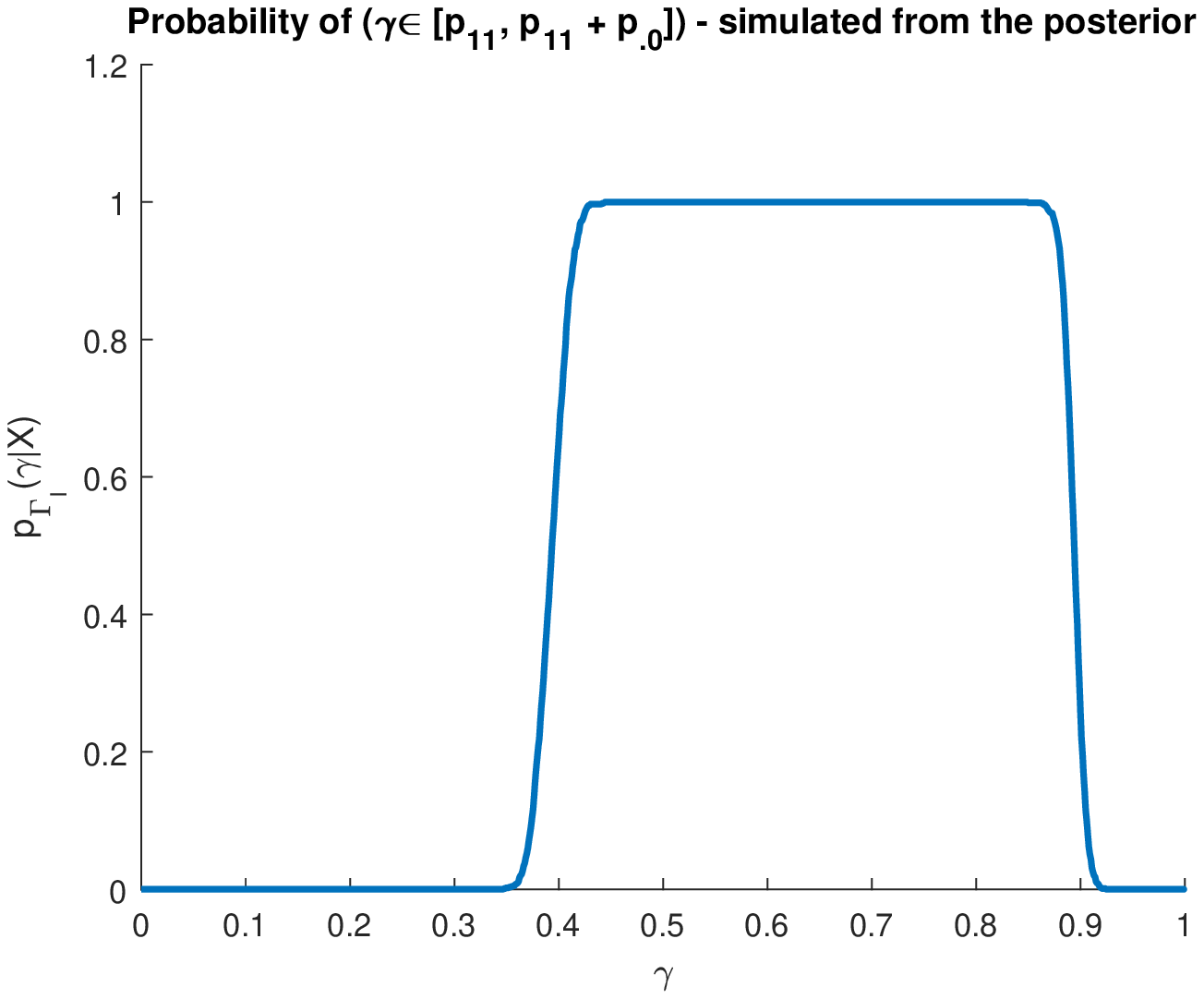}}
  \caption{{\small Binary outcome and missing data. Prior and Posterior coverage functions $p_{\scriptscriptstyle{\Gamma_{I}}}(\cdot)$ and $p_{\scriptscriptstyle{\Gamma_{I}}}(\cdot|\{x_{i}\})$ of $\Gamma_{I}$. The true $\Gamma_I$ is $[0.4,0.9]$.}}
  \label{fig_Example_4_Dirichlet_random_set}
\end{figure}

\begin{figure}[!h]
  \centering
  \subfloat[{Prior}]{\label{Example_4_fig_2_a}
      \includegraphics[width=0.4\linewidth]{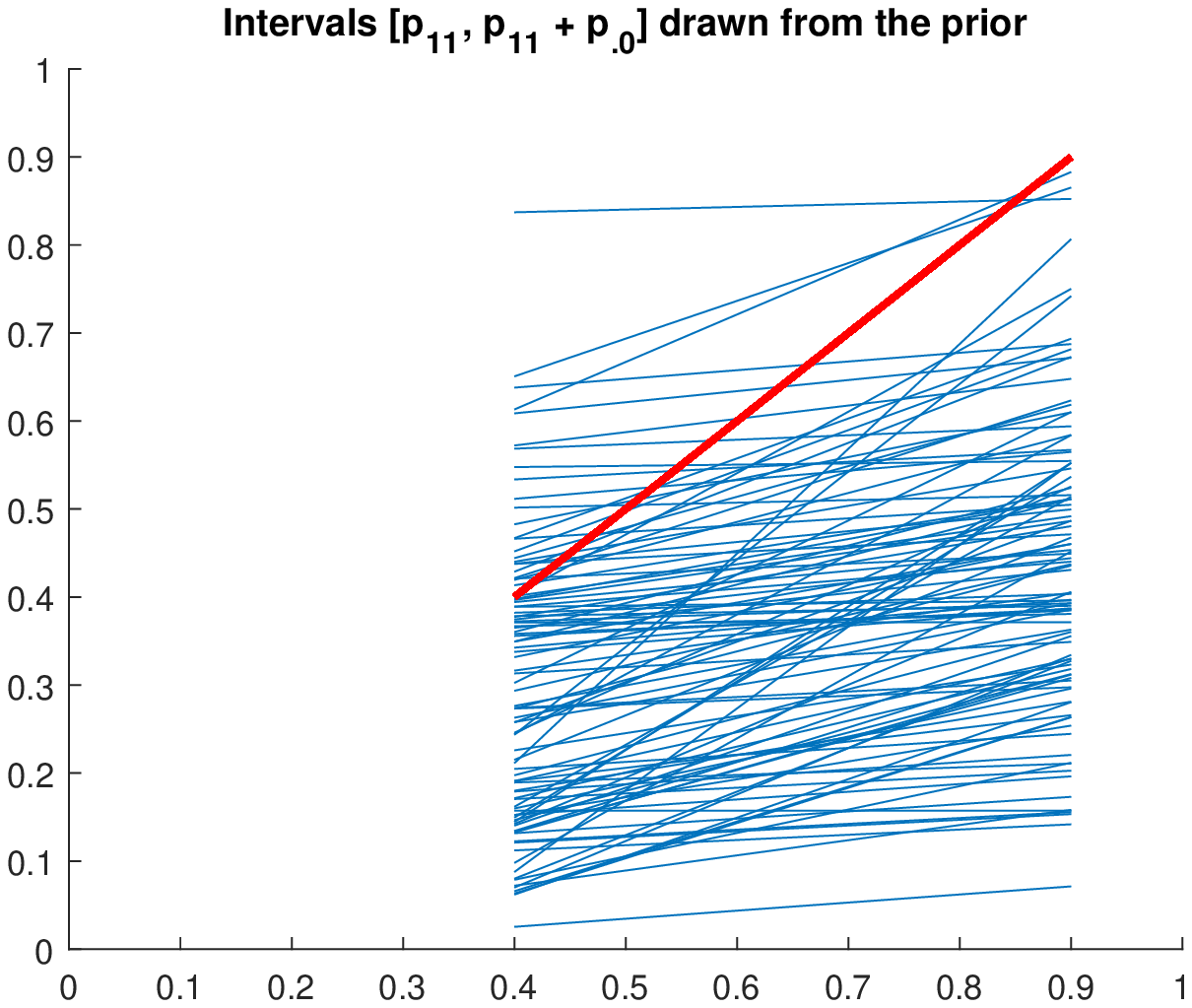}}
  \hspace{0.1\linewidth}
  \subfloat[{Posterior}]{\label{Example_4_fig_2_b}
      \includegraphics[width=0.4\linewidth]{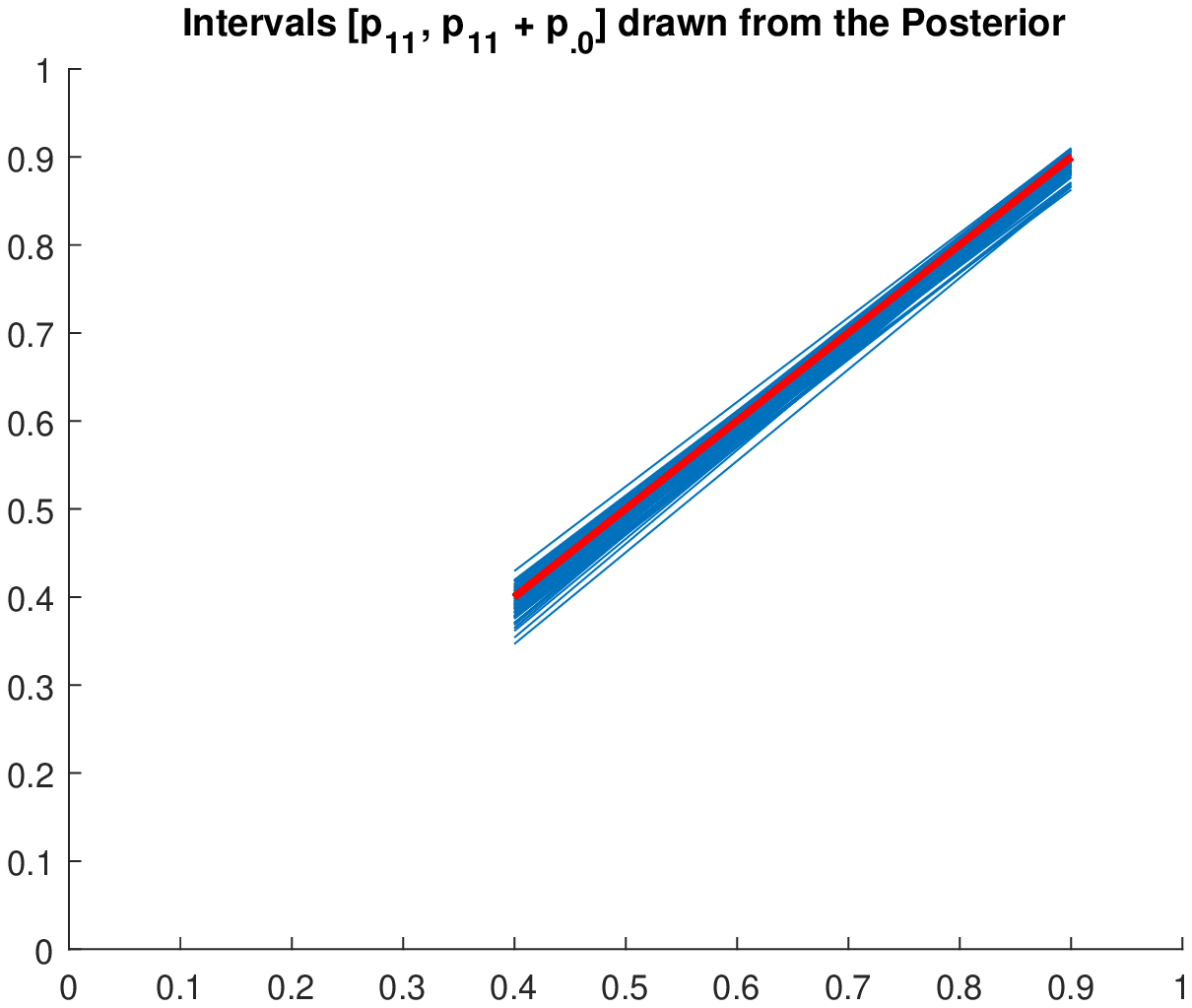}}
  \caption{{\small Binary outcome and missing data. Representation of the intervals $\Gamma_{I} = [p_{11},p_{11} + \wtl{p}_{\cdot 0}]$ drawn from the prior (panel \ref{Example_4_fig_2_a}) and posterior (panel \ref{Example_4_fig_2_b}) against the true interval $[0.4,0.9]$ (in red).}}
  \label{fig_Example_4_Dirichlet_Figure_2}
\end{figure}
\end{example}

\section{Further examples for section 5.3}\label{App:s:Ex:2}
\setcounter{example}{1}
\begin{example}[\textbf{Interval Censored Data} (\textit{continued})]
  Suppose that we are not only interested in the identified region $\Gamma_{I} := [\mathbf{E}_{F_{1}}(Y_{1}), \mathbf{E}_{F_{2}}(Y_{2})]$ but also in the partially identified parameter $\gamma$ itself. The marginal posterior distribution of $\gamma$ is informative about the areas of the identified region $\Gamma_{I}$ where the parameter is more likely. The prior distribution of $\theta := (\mathbf{E}_{F_{1}}(Y_{1}), \mathbf{E}_{F_{2}}(Y_{2}))$ is obtained from a Dirichlet process prior for $F$ denoted by $\mu_{F}$, see section 4 for a description of the method. Under the assumption that $Y_{1}\perp Y_{2}|F$ the Bayesian hierarchical model is as in (4.11) completed with the specification of a prior for $\gamma$ conditional on $F$: $\gamma|F \sim \mu_{\gamma}^{F}$. In our simulation exercise we consider the four specifications (I)-(IV) for $\mu_{\gamma}^{F}$ given in section 5.5 with: $a_0 = \mathbf{E}_{F_{1}}(Y_{1})$, $b_0 = \mathbf{E}_{F_{2}}(Y_{2})$, $c_0 = 1$, and $\wtl{\gamma}_{0}^{i} := \mathbf{E}_{F_{i}}(Y_{i})=\int y F_{i}(dy)$, $i=1,2$. Since $\gamma$ is partially identified in the full model, but identified in the marginal model, its posterior distribution depends on the data only through $F$. This means that the moments $\mathbf{E}_{F_{i}}(Y_{i})$, $i=1,2$, in the prior for $\gamma$ are replaced with the posterior means $\mathbf{E}(Y_{i}|F_{i},y_{i1}, \ldots, y_{in})$, for $i=1,2$ in the posterior distribution. The notation $\mathbf{E}(Y_{i}|F_{i},y_{i1}, \ldots, y_{in})$ is for the mean of $Y_{i}$ drawn from the posterior distribution of $\phi(F)$.\\
\indent We generate an $n$-sample of observations of $(Y_{1},Y_{2})$ as in section 4.2: $Y_{1}\sim\mathcal{N}(0,0.1)$, $Y_{2}\sim\mathcal{N}(5,0.1)$. The parameters are fixed as follows: $n = 1000$, $n_{0}^{1}=10$, $n_{0}^{2}=20$, $F_{0}^{1}= \mathcal{N}(0,1)$, $F_{0}^{2}= \mathcal{N}(10,1)$ and $\tau_{0}^{2} = 1$, $\sigma_{0}^{2}=2$, $p=2$ and $q=2$. The true identified set is $\Gamma_{I} = [0,5]$.\\
\indent We draw $1000$ times from the marginal prior and posterior distributions of $\gamma$. The simulation scheme is the following: for each $1\leq j\leq 1000$, draw $F^{(j)}$ from the prior $\mu(F)$ (resp. the posterior $\mu(F|\{y_{ij}\}_{j=1}^{n},i=1,2)$), compute $\theta_{i}^{(j)} = \mathbf{E}_{F_{i}^{(j)}}(Y_{i})$, $i=1,2$ and draw $\gamma^{(j)}$ from $\mu_{\gamma}^{F^{(j)}}$ (resp. $\mu_{\gamma}^{F^{(j)}}(\gamma|F^{(j)},\{y_{ij}\}_{j=1}^{n},i=1,2)$). Figure \ref{fig_Example_1_Dirichlet_2nd} shows the histograms of the marginal prior (in blue) and posterior (in red) distribution of $\gamma$. Each panel corresponds to one of the four specifications for $\mu_{\gamma}^{F}$. 
\begin{figure}[!h]
  \centering
  \subfloat[{$\gamma|F \sim \mathcal{N}(\gamma_{0}, 1)$ by discarding the draws that are not in $[\mathbf{E}_{F_{1}}(Y_{1}),\mathbf{E}_{F_{2}}(Y_{2})]$}.]{
      \includegraphics[width=0.4\linewidth]{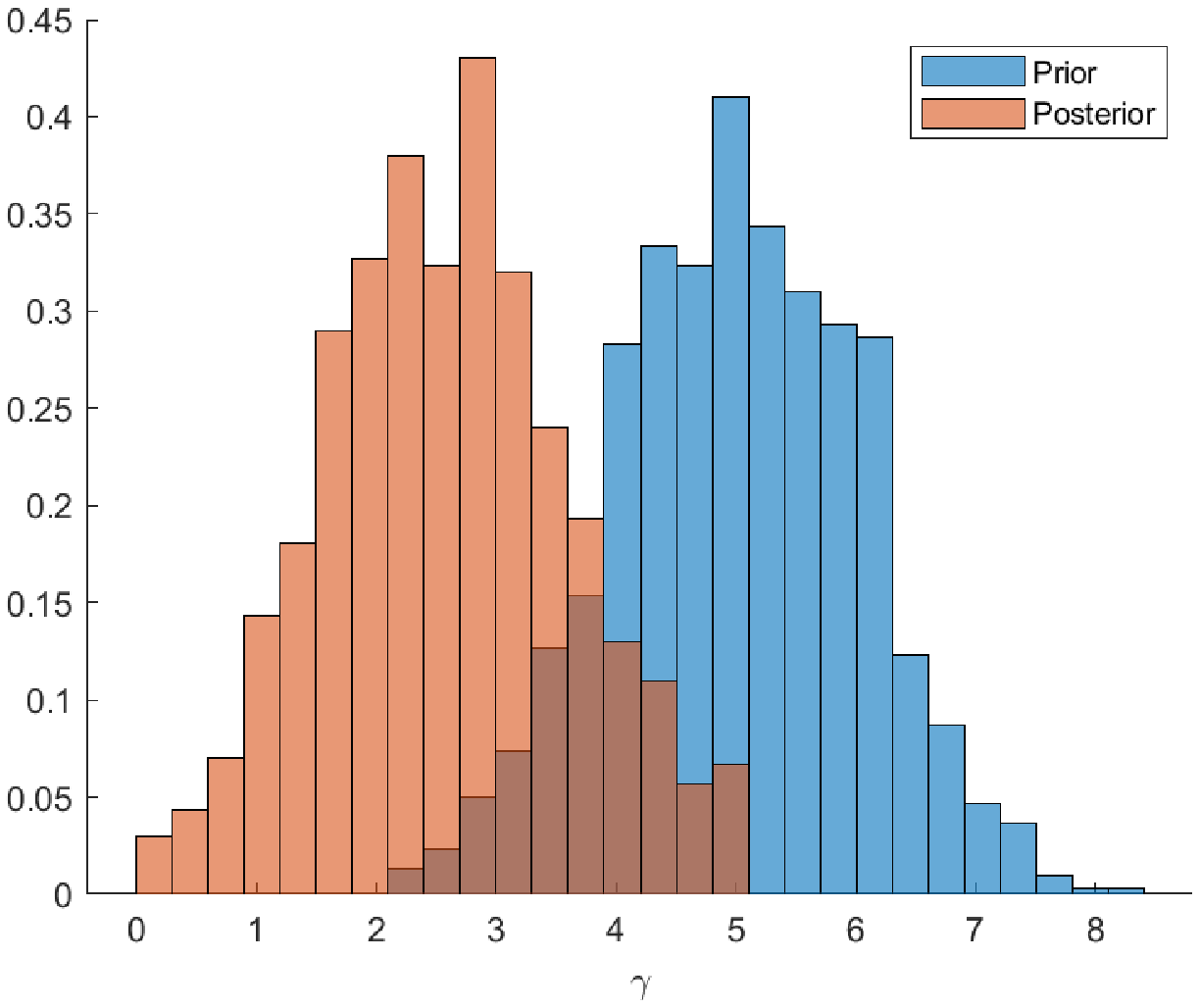}}
  \hspace{0.1\linewidth}
  \subfloat[{$\gamma|F \sim \mathcal{N}(0, 2)$ truncated to $[\mathbf{E}_{F_{1}}(Y_{1}),\mathbf{E}_{F_{2}}(Y_{2})]$}.]{
      \label{density_constraint}
      \includegraphics[width=0.4\linewidth]{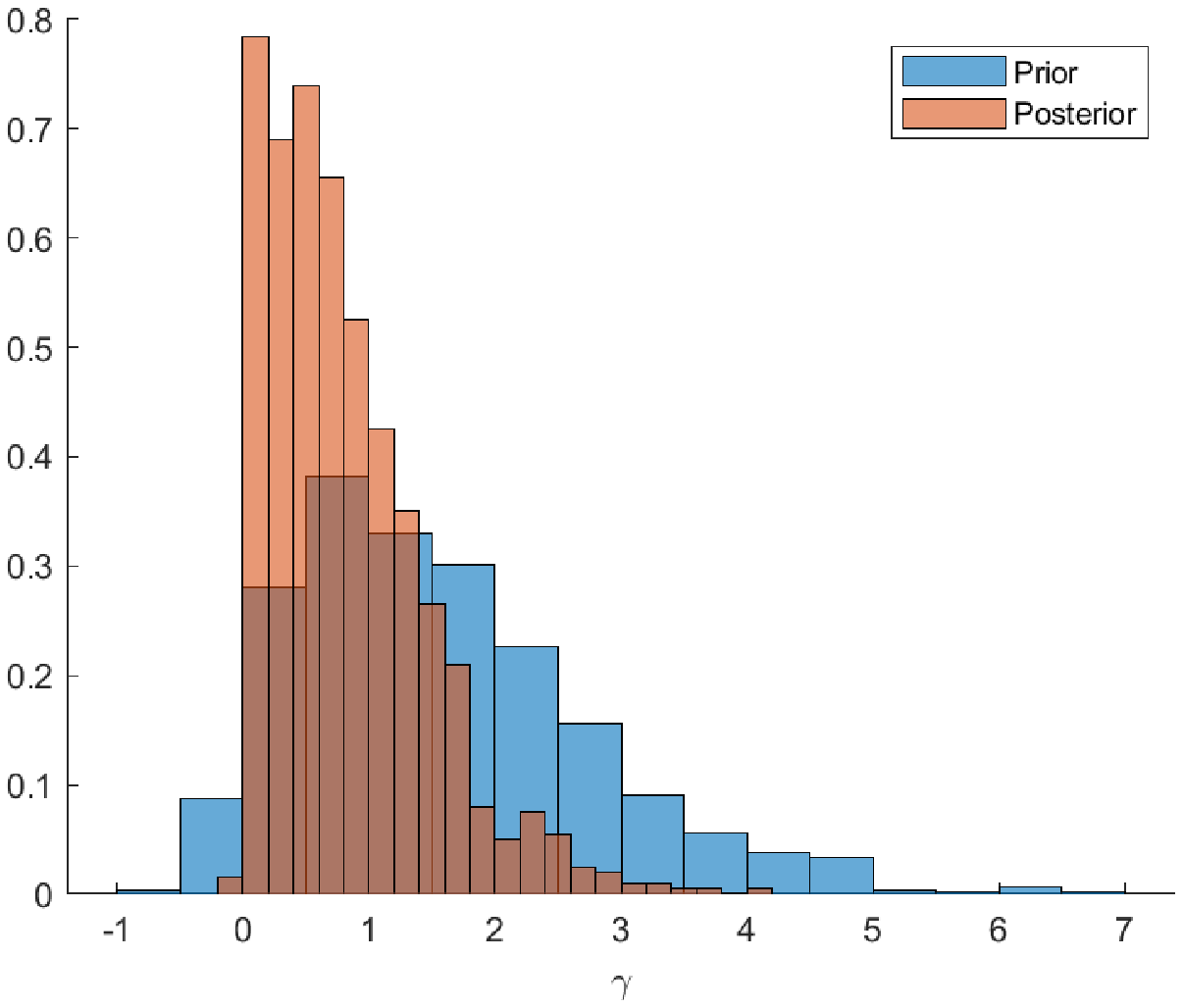}}
  \subfloat[{$\gamma|F \sim \mathcal{U}[\mathbf{E}_{F_{1}}(Y_{1}),\mathbf{E}_{F_{2}}(Y_{2})]$}.]{
      \includegraphics[width=0.4\linewidth]{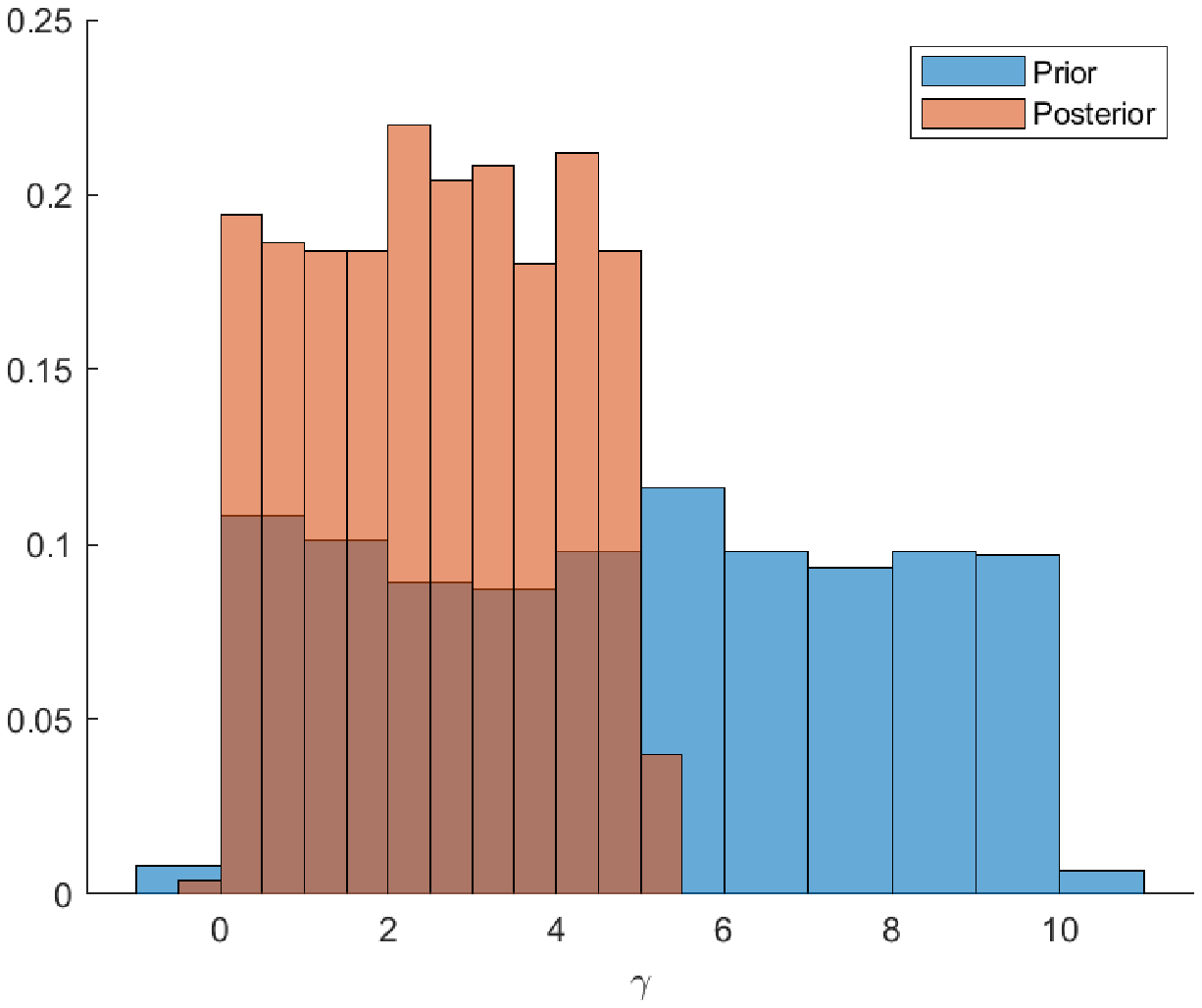}}
  \hspace{0.1\linewidth}
  \subfloat[{$\gamma|F \sim \mathcal{B}eta(\mathbf{E}_{F_{1}}(Y_{1}),\mathbf{E}_{F_{2}}(Y_{2}),2,2)$}.]{
      \label{density_constraint}
      \includegraphics[width=0.4\linewidth]{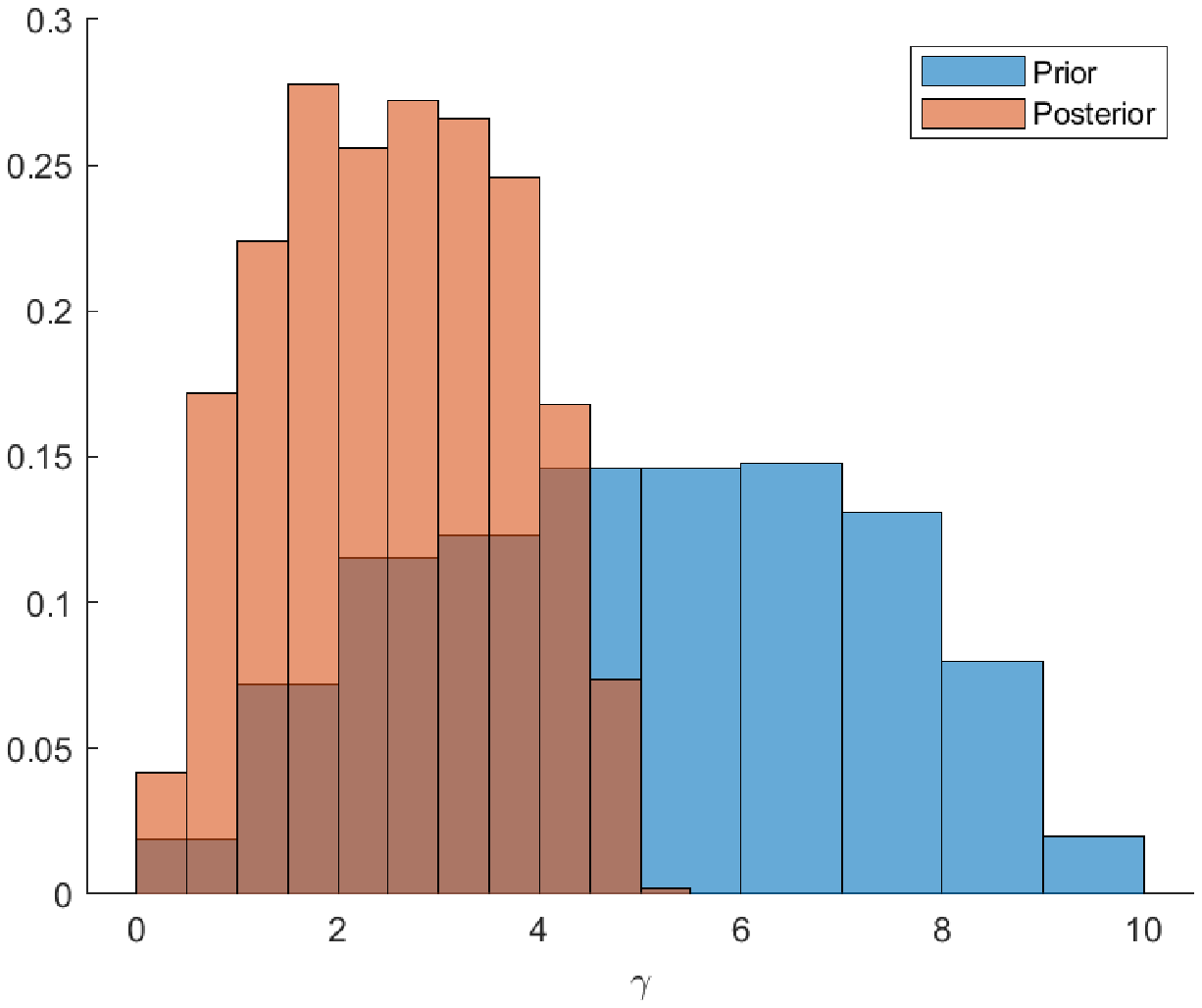}}
  \caption{{\small Interval Censored Data. Histograms of the prior (in blue) and posterior (in red) probability distributions. The true identified set is $\Gamma_{I} = [0,5]$.}}
  \label{fig_Example_1_Dirichlet_2nd}
\end{figure}
\end{example}

\setcounter{example}{4}
\begin{example}[\textbf{Interval Regression Model} (\textit{continued})]
  Let $\gamma\in\Gamma$ be the parameter of interest. The Bayesian hierarchical model \eqref{eq:4} is completed with a conditional prior on $\gamma$, given $F$: $\gamma|F \sim \mu_{\gamma}^{F}$, restricted to satisfy  the constraint \eqref{example_4_2_moment_inequalities}. In our simulation study we take $X$ and $Z$ to be univariate and consider the four specifications (I)-(IV) for $\mu_{\gamma}^{F}$ given in section 5.5 with: $a_0 = \frac{\mathbf{E}_{F}(Y_{1}Z)}{\mathbf{E}_{F}(ZX)}$, $b_0 = \frac{\mathbf{E}_{F}(Y_{2}Z)}{\mathbf{E}_{F}(ZX)}$, $c_0 = \mathbf{E}_{F}(XZ)$, and $\wtl{\gamma}_{0}^{j} := \mathbf{E}_{F}(Y_{j}Z)=\int y_{j} z F(dy_{j},dz)$, $j=1,2$. Since $\gamma$ is not identified in the full model, the prior-to-posterior transformation of $\gamma$ is affected by the data only indirectly through $F$ by replacing the moments $\mathbf{E}_{F}(Y_{i}Z)$ and $\mathbf{E}_{F}(ZX)$ with the posterior moments $\mathbf{E}(Y_{i}Z|F,\{w_{i}\}_{i=1}^{n})$ and $\mathbf{E}(ZX|F,\{w_{i}\}_{i=1}^{n})$, for $i=1,2$, respectively. The notations $\mathbf{E}(Y_{i}Z|F,\{w_{i}\}_{i=1}^{n})$ and $\mathbf{E}(ZX|F,\{w_{i}\}_{i=1}^{n})$ refer to the mean of $Y_{i}Z$, $i=1,2$, and of $ZX$, respectively, with respect to a distribution drawn from the posterior of $F$.\\
\indent This simulation exercise allows to see to what extent the support of the marginal prior and posterior distributions overlaps with the true identified set. We generate an $n$-sample of realizations of $(Y_{1},Y_{2},X,Z)$ according to the data generating process in \eqref{eq:5}. Hence, the true identified set is $[2,6]$. The parameters of the Dirichlet process are fixed as $n_{0} = 20$ and $F_0$ as in \eqref{ex:3:Dir:hyperparameter}. Moreover, we set: $n = 1000$, $\tau_{0}^{2}=1$, $\sigma_{0}^{2}=2$, $p=1$ and $q = 0.5$. \\
\indent We draw $1000$ times from the marginal prior and posterior distributions of $\gamma$. The simulation scheme is the following: for each $1\leq j\leq 1000$, draw $F^{(j)}$ from the prior $\mu(F)$ (resp. the posterior $\mu(F|\{w_{i}\}_{i=1}^{n})$), compute $\mathbf{E}_{F^{(j)}}(Y_{i}Z)$, $i=1,2$ and $\mathbf{E}_{F^{(j)}}(ZX)$, and draw $\gamma^{(j)}$ from $\mu_{\gamma}^{F^{(j)}}$ (resp. $\mu_{\gamma}^{F^{(j)}}(\gamma|F^{(j)},\{w_{i}\}_{i=1}^{n})$). Figure \ref{fig_Example_4_2_Dirichlet_2nd} shows the histograms of the marginal prior (in blue) and posterior (in red) probability distributions corresponding to the four prior specifications for $\mu_{\gamma}^{F}$ given above.

\begin{figure}[!h]
  \centering
  \subfloat[{$\gamma \sim \mathcal{N}(\gamma_{0}, 1)$ by discarding the draws that are not in $\left[\frac{\mathbf{E}_{F}(Y_{1}Z)}{\mathbf{E}_{F}(ZX)},\frac{\mathbf{E}_{F}(Y_{2}Z)}{\mathbf{E}_{F}(ZX)}\right]$}.]{\label{fig_Example_4_2_Dirichlet_Normal}
      \includegraphics[width=0.4\linewidth]{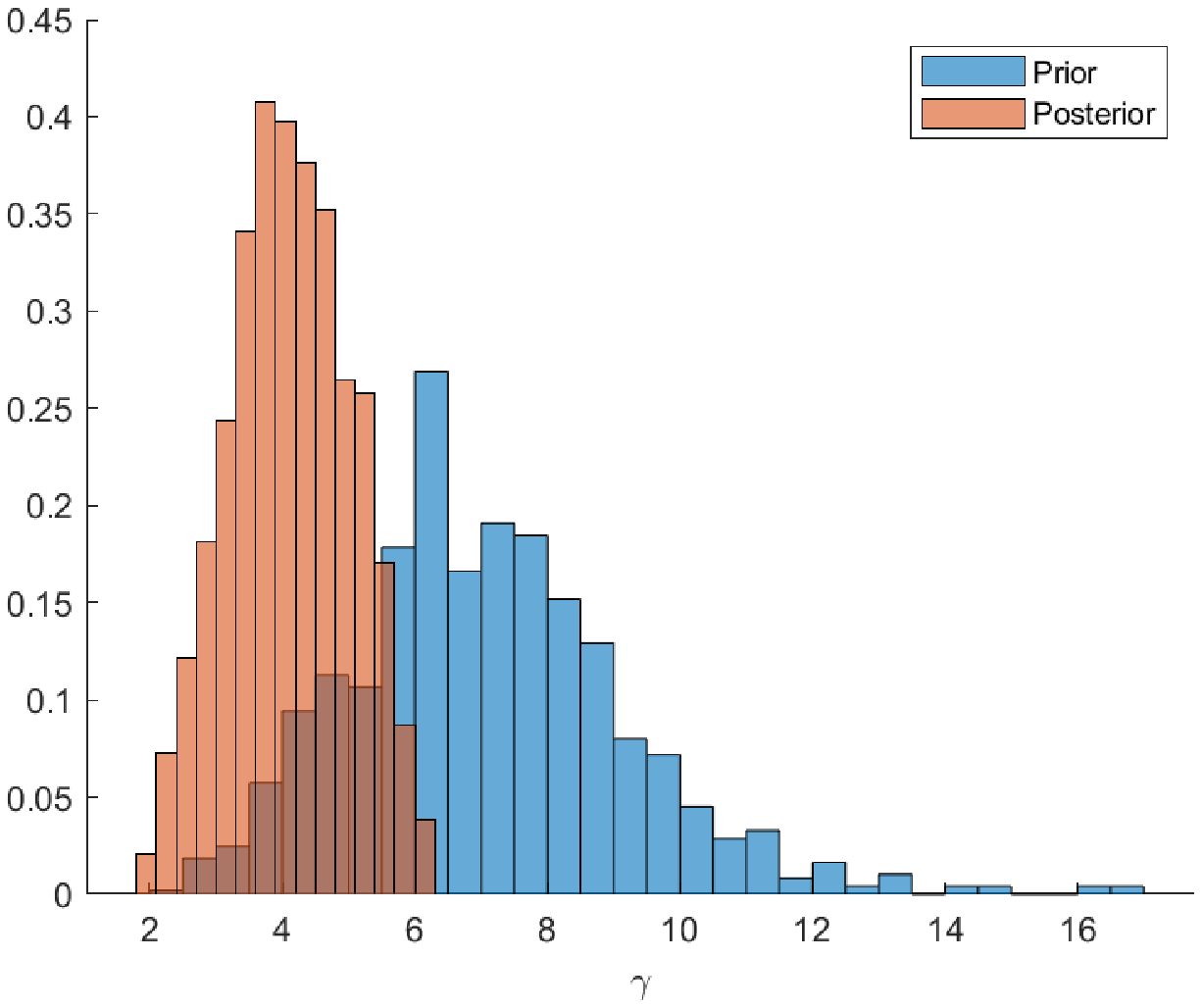}}
  \hspace{0.1\linewidth}
  \subfloat[{$\gamma \sim \mathcal{N}(0, 2)$ truncated to $\left[\frac{\mathbf{E}_{F}(Y_{1}Z)}{\mathbf{E}_{F}(ZX)},\frac{\mathbf{E}_{F}(Y_{2}Z)}{\mathbf{E}_{F}(ZX)}\right]$}.]{\label{fig_Example_4_2_Dirichlet_Truncated_Normal}
      \includegraphics[width=0.4\linewidth]{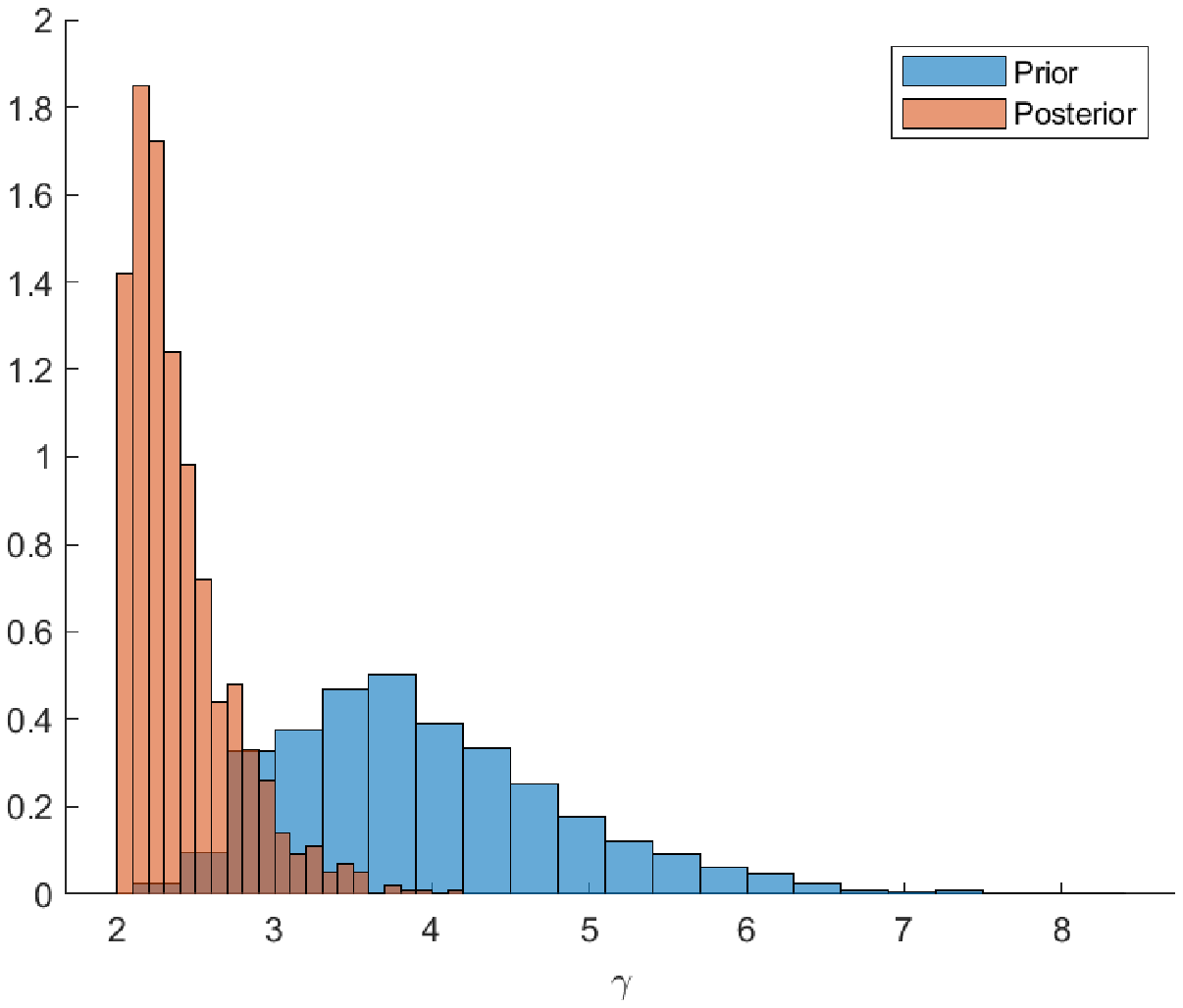}}

  \subfloat[{$\gamma \sim \mathcal{U}\left[\frac{\mathbf{E}_{F}(Y_{1}Z)}{\mathbf{E}_{F}(ZX)},\frac{\mathbf{E}_{F}(Y_{2}Z)}{\mathbf{E}_{F}(ZX)}\right]$}.]{\label{fig_Example_4_2_Dirichlet_Uniform}
      \includegraphics[width=0.4\linewidth]{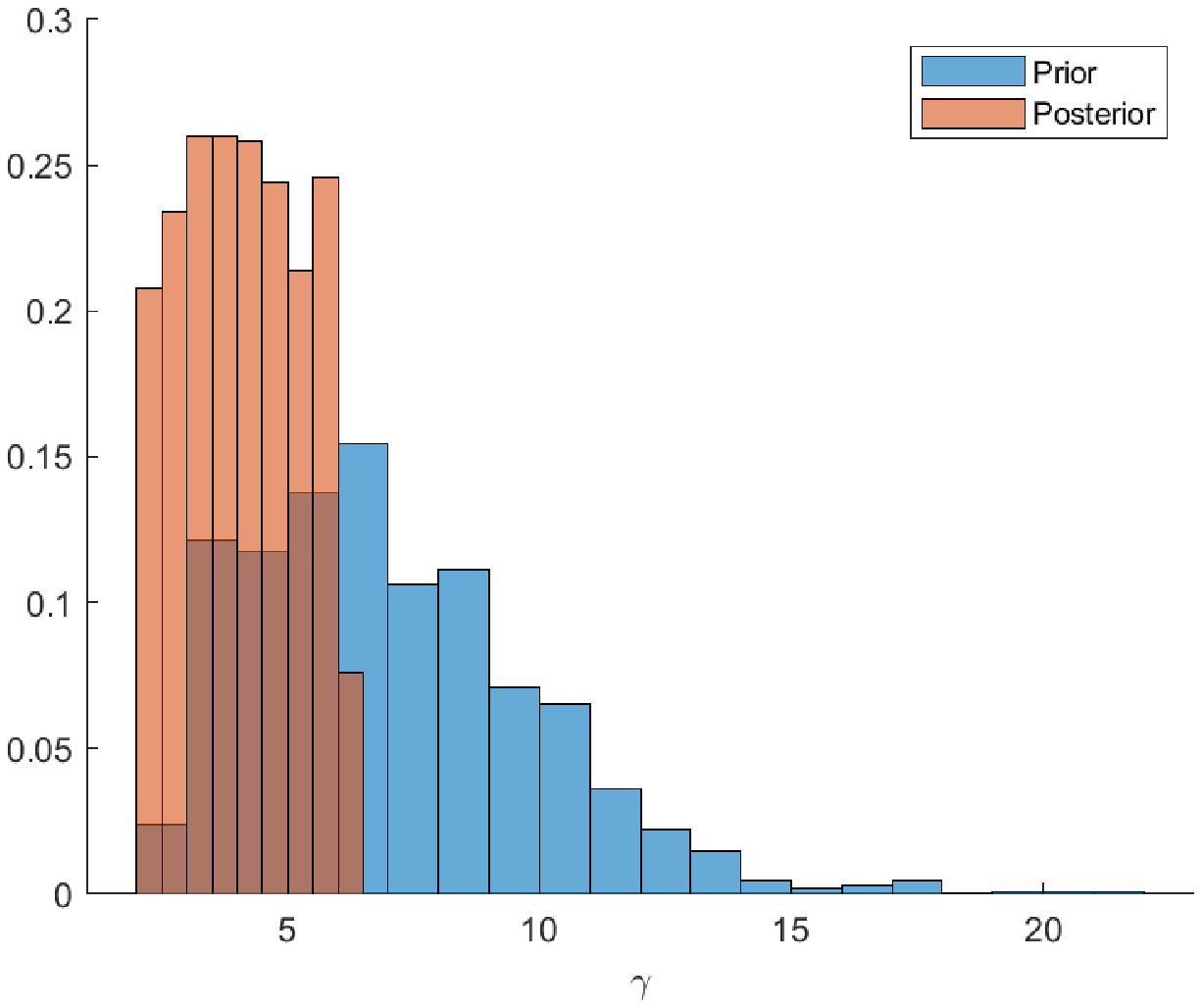}}
  \hspace{0.1\linewidth}
  \subfloat[{$\gamma \sim \mathcal{B}eta\left(\frac{\mathbf{E}_{F}(Y_{1}Z)}{\mathbf{E}_{F}(ZX)},\frac{\mathbf{E}_{F}(Y_{2}Z)}{\mathbf{E}_{F}(ZX)},1,0.5\right)$}.]{\label{fig_Example_4_2_Dirichlet_Beta}
      \label{density_constraint}
      \includegraphics[width=0.4\linewidth]{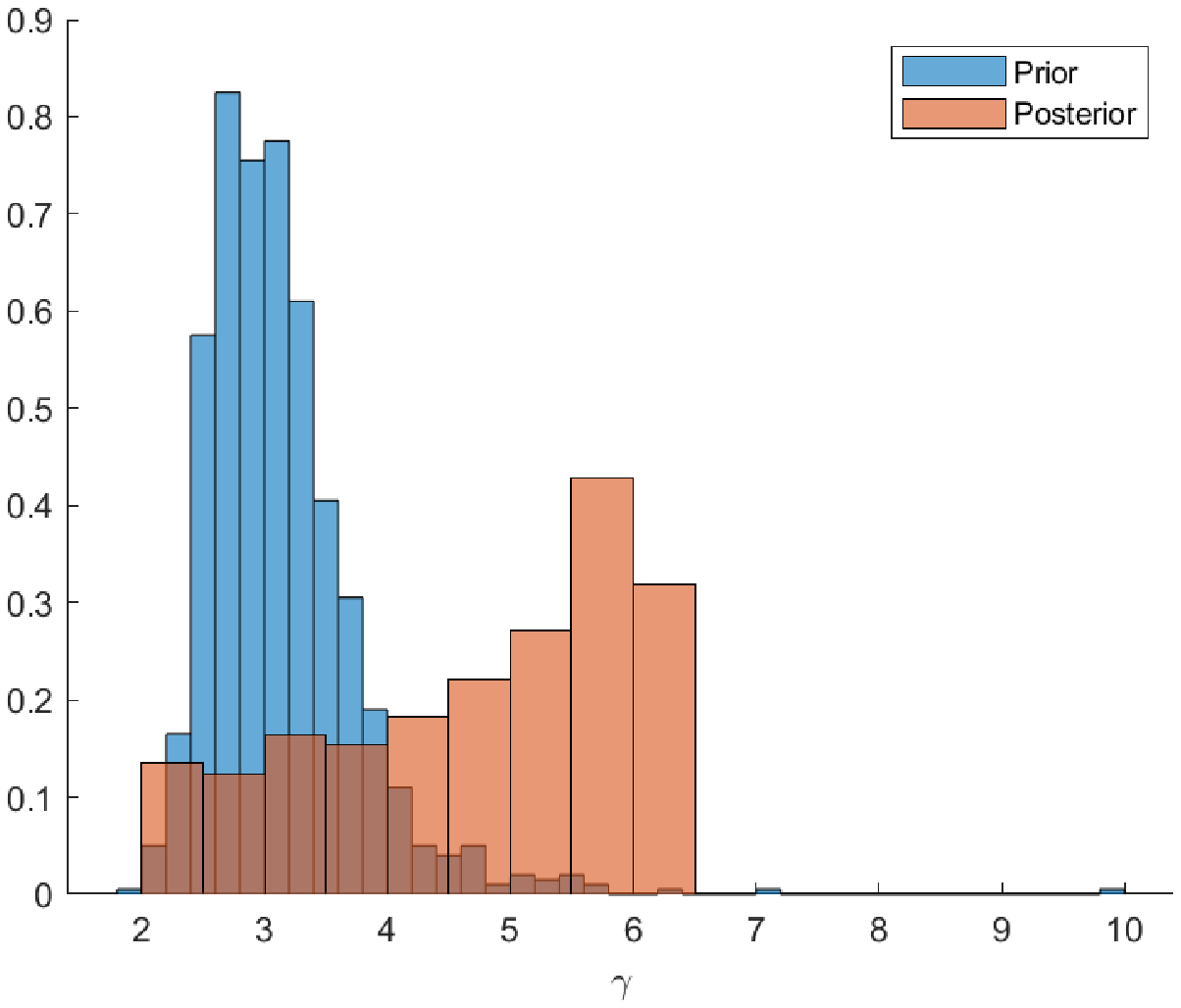}}
  \caption{{\small Interval Regression Model. Histograms of the prior (in blue) and posterior (in red) probability distributions. The true identified set is $\Gamma_I = [2,6]$.}}
  \label{fig_Example_4_2_Dirichlet_2nd}
\end{figure}
\end{example}

\begin{example}[\textbf{Binary outcome and missing data (\textit{continued})}]
  The parameter of interest is the probability $p_{1} := p_{11} + p_{10}$ which is not identified but is known to belong to the interval $\Gamma_{I} := [p_{11}, p_{11} + \wtl{p}_{\cdot 0}]$. We provide $p_{1}$ with a prior distribution conditional on $(p_{11},\wtl{p}_{\cdot 0})$: $p_{1}|p_{11}, \wtl{p}_{\cdot 0} \sim \mu(p_{1}|p_{11}, \wtl{p}_{\cdot 0})$, and the Bayesian hierarchical model is as in \eqref{eq:6} enriched with this prior. In our simulation we consider the four specifications (I)-(IV) given in section 5.5 with $\mu_{\gamma}^F$ replaced by $\mu(p_{1}|p_{11}, \wtl{p}_{\cdot 0})$, $F$ replaced by $(p_{11}, \wtl{p}_{\cdot 0})$, $\gamma_{0} = p_{11} + \frac{\wtl{p}_{\cdot 0}}{2}$, $\tau_0 = 1$, $a_0 = p_{11}$, $b_0 = p_{11} + \wtl{p}_{\cdot 0}$, $\sigma_0^2 = 2$. The data generating process is the same as in \eqref{eq:7} so that the true identified set is $[0.4,0.9]$ and the parameter $\alpha$ of the Dirichlet distribution is set equal to $\alpha = (2,3,1)$. The simulation scheme is the following: for each $1\leq j\leq 1000$, draw $(p_{11}^{(j)},p_{01}^{(j)},\wtl{p}_{\cdot 0}^{(j)})$ from $\mathcal{D}ir(\alpha)$ (resp. $\mathcal{D}ir(\alpha_{*})$) and draw $p_{1}^{(j)}$ from $\mu(p_{1}|p_{11}, \wtl{p}_{\cdot 0})$ (resp. from $\mu(p_{1}|p_{11}, \wtl{p}_{\cdot 0},\{x_{i}\}_{i=1}^{n})$). In Figure \ref{fig_Example_4_4_Dirichlet_2nd} each panel shows the histograms corresponding to one of the four specifications above for $\mu(p_{1}|p_{11}, \wtl{p}_{\cdot 0})$.

\begin{figure}[!h]
  \centering
  \subfloat[{$\gamma \sim \mathcal{N}(\gamma_{*}, 1)$ by discarding the draws that are not in $[p_{11},p_{11} + \wtl{p}_{\cdot 0}]$.}]{\label{fig_Example_4_2_Dirichlet_Normal}
      \includegraphics[width=0.4\linewidth]{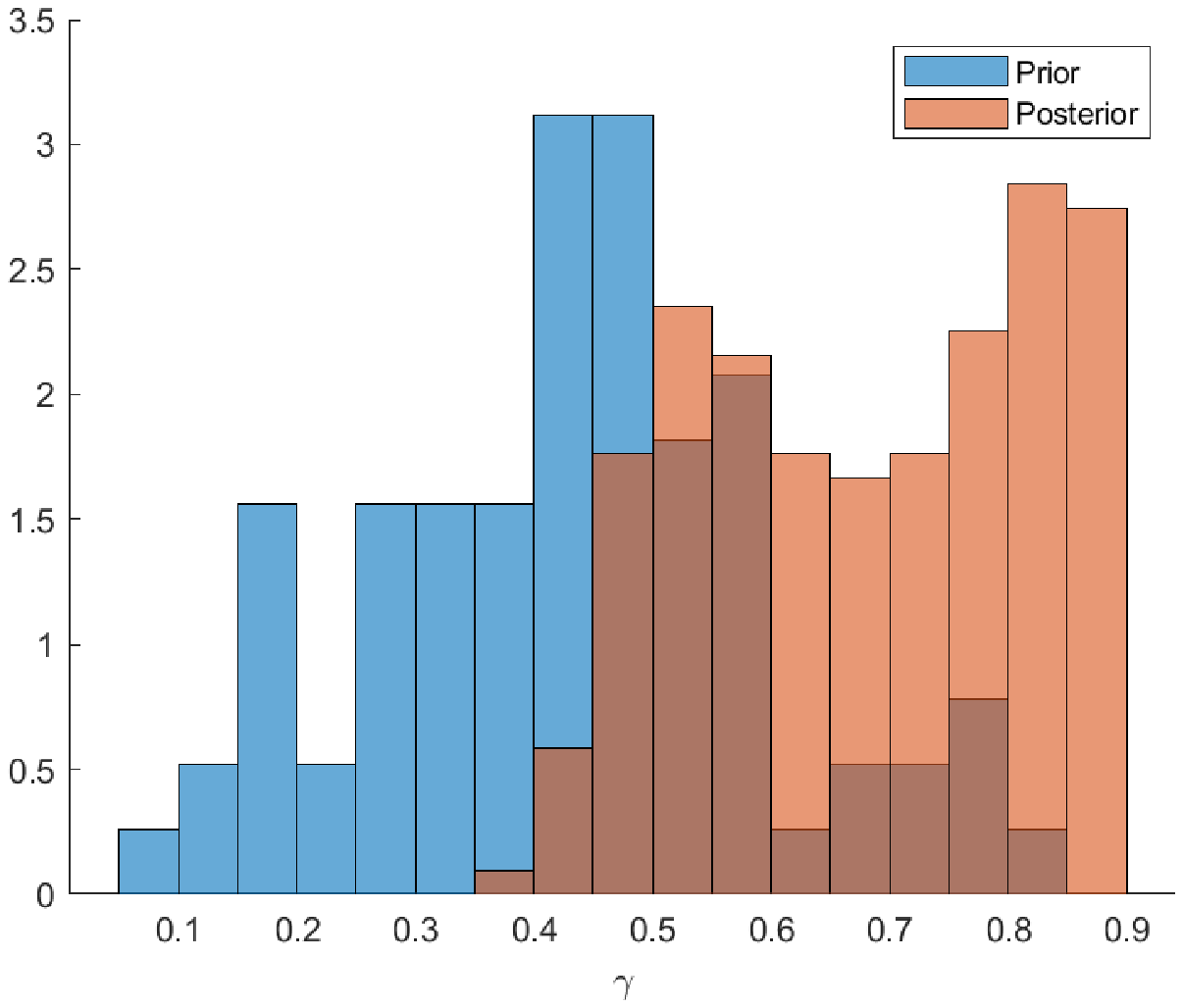}}
  \hspace{0.1\linewidth}
  \subfloat[{$\gamma \sim \mathcal{N}(0, 2)$ truncated to $[p_{11},p_{11} + \wtl{p}_{\cdot 0}]$.}]{\label{fig_Example_4_2_Dirichlet_Truncated_Normal}
      \includegraphics[width=0.4\linewidth]{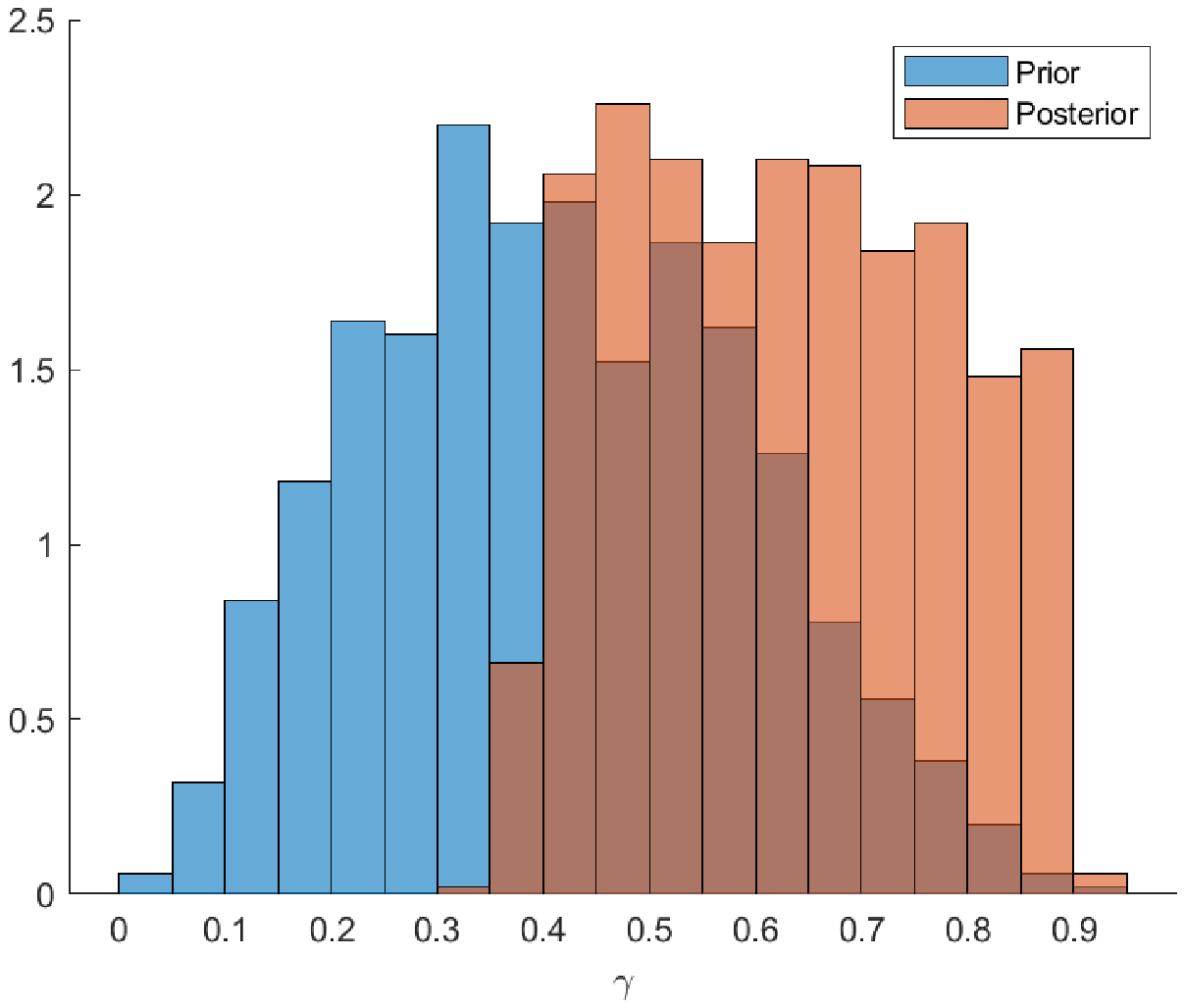}}

  \subfloat[{$\gamma \sim \mathcal{U}[p_{11},p_{11} + \wtl{p}_{\cdot 0}]$.}]{\label{fig_Example_4_2_Dirichlet_Uniform}
      \includegraphics[width=0.4\linewidth]{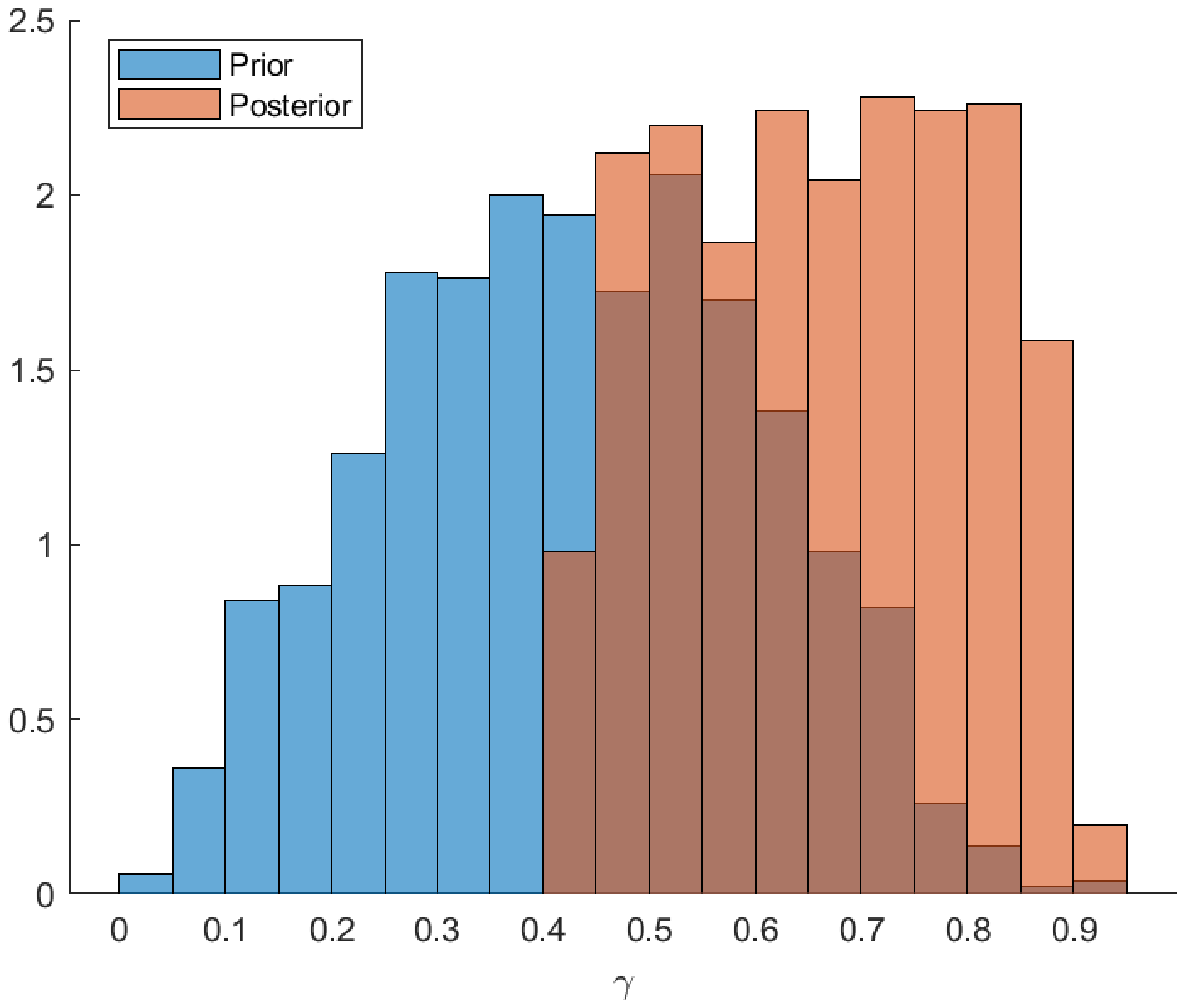}}
  \hspace{0.1\linewidth}
  \subfloat[{$\gamma \sim \mathcal{B}eta(p_{11},p_{11} + \wtl{p}_{\cdot 0},1,0.5)$.}]{\label{fig_Example_4_2_Dirichlet_Beta}
      \label{density_constraint}
      \includegraphics[width=0.4\linewidth]{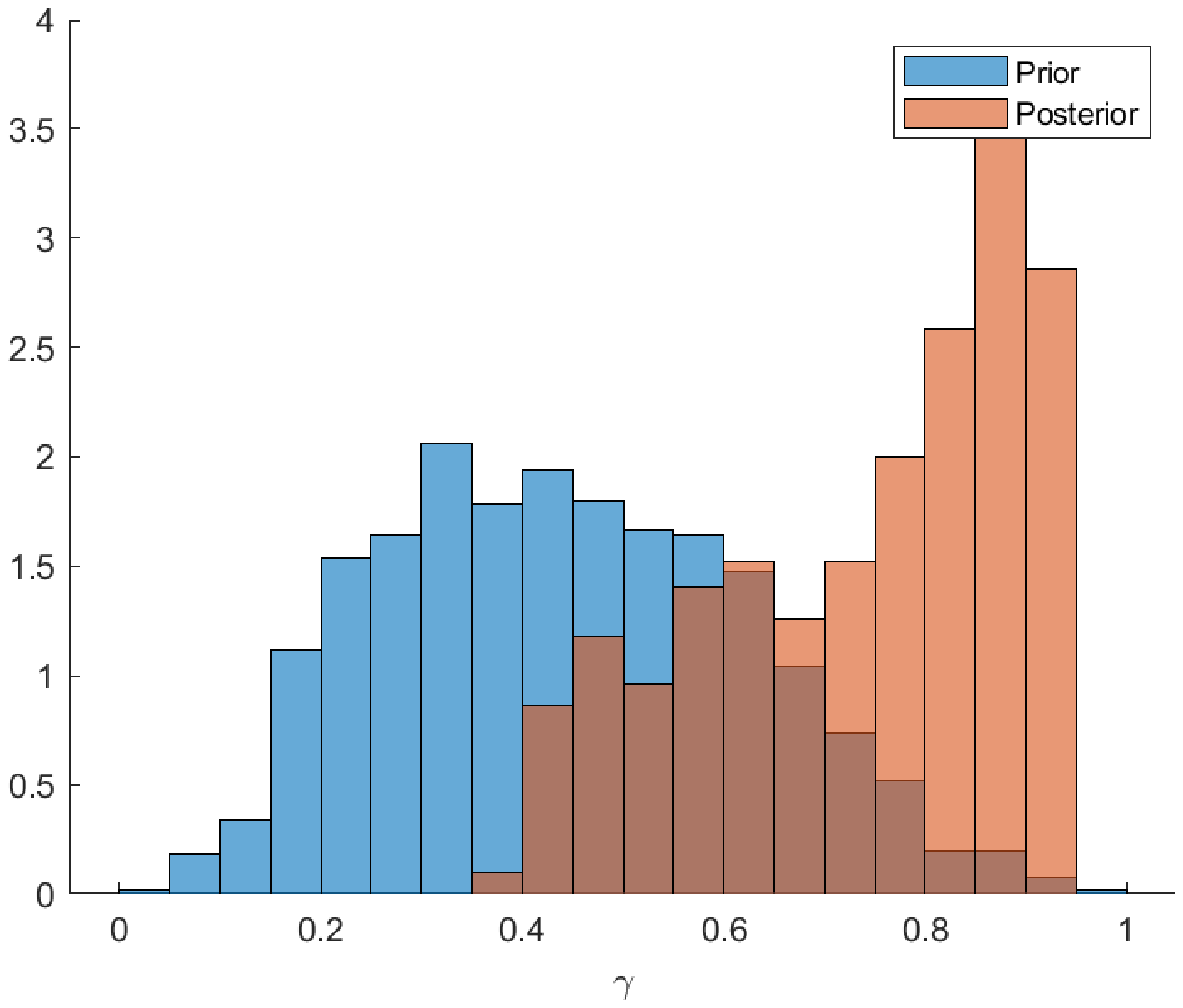}}
  \caption{{\small Binary outcome and missing data. Histograms of the prior (in blue) and posterior (in red) probability distributions. The true identified set is $\Gamma_I = [0.4,0.9]$.}}
  \label{fig_Example_4_4_Dirichlet_2nd}
\end{figure}
\end{example}
\section{Representation of the Dirichlet Process}\label{a_Dirichlet_process}
In this appendix we recall the main results concerning the representation of the Dirichlet process used for simulations and we refer to \cite{FlorensRolin1994} and \cite{Florens2002} for more details.

\textsc{1. Stick-breaking representation of the Dirichlet process (\cite{Sethuraman1994})}. Consider the Dirichlet process $F\sim \mathcal{D}ir(n_{0},F_{0})$ and suppose for simplicity that $F_{0}$ is a diffuse probability, \textit{i.e.} $F_{0}(x)=0$, $\forall x$. The probability distribution $F$ is generated by a $\mathcal{D}ir(n_{0},F_{0})$ process if and only if $F$ can be written as
    \begin{equation}\label{Sethuraman_Rolin_decomposition}
      F = \sum_{j=1}^{\infty}\alpha_{j}\delta_{\xi_{j}},
    \end{equation}
\noindent where $\delta_{\xi_{j}}$ is a Dirac mass at the value $\xi_{j}$ and the sequences $\{\alpha_{j}\}$ and $\{\xi_{j}\}$ are generated as
    \begin{itemize}
      \item $\xi_{j} \sim \:i.i.d. \: F_{0}$;
      \item $\alpha_{j} = v_{j}\prod_{k = 1}^{j - 1}v_{k}$, with $v_{k} \sim \: i.i.d.\: \mathcal{B}e(1,n_{0})$ for $k = 1, \ldots,j$;
      \item $\{\xi_{j}\}$ and $\{v_{j}\}$ are independent.
    \end{itemize}
\noindent Suppose now that we have observed an $n$-sample $x_{1},\ldots,x_{n}$ of realizations of $X$, where $X|F \: \sim \: F$, and denote by $F_{n}$ the empirical cumulative distribution function. The representation of the posterior Dirichlet process $F\sim\mathcal{D}ir(n_{*},F_{*})$, with $n_{*} = n_{0} + n$ and $F_{*} = n_{0}F_{0} + nF_{n}$, is
    \begin{equation}
      F = (1 - \gamma)\sum_{j=1}^{\infty}\alpha_{j}\delta_{\xi_{j}} + \gamma\sum_{j=1}^{n}\beta_{j}\delta_{x_{j}}
    \end{equation}
\noindent where the sequences $\{\alpha_{j}\}$ and $\{\xi_{j}\}$ are generated as above, $\gamma \sim\mathcal{B}e(n,n_{0})$ and $(\beta_{1},\ldots,\beta_{n})$ follows a Dirichlet distribution with parameter $(1,\ldots,1)$ over the simplex $S_{n-1}$.\\

\textsc{2. Draw from the Dirichlet process.} To draw a trajectory from a Dirichlet process we use the stick-breaking representation described above. Remark that the infinite sum in \eqref{Sethuraman_Rolin_decomposition} cannot be implemented in practice since it requires an infinity of draws. Instead, one can truncate the sum at some $K<\infty$ and normalize:
    \begin{displaymath}
      F_{K}= \frac{1}{\sum_{k=1}^{K}\alpha_{k}}\sum_{k=1}^{K}\alpha_{k}\delta_{\xi_{k}}.
    \end{displaymath}
\indent Once $\{\xi_{k}\}$ and $\{\alpha_{k}\}$ have been drawn for the Dirichlet process, one can compute any functional of the Dirichlet, for instance $\mathbf{E}_{F}(Y)$:
    \begin{displaymath}
      \mathbf{E}_{F}(Y) = \frac{1}{\sum_{k=1}^{K}\alpha_{k}}\sum_{k=1}^{K}\alpha_{k}\xi_{k}.
    \end{displaymath}
\indent The error made by approximating $F$ by $F_{K}$ is measured by $\varepsilon = \sup_{B}|F(B) - F_{K}(B)|$, where $B$ is a Borel set of $\mathbb{R}^{m}$ (if $F$ is a distribution on $\mathbb{R}^{m}$), and is equal to $\sum_{j=K+1}^{\infty}\alpha_{j}$. The probability distribution of $\varepsilon$ is known and has a density function equal to
%
      $\frac{n_{0}^{K}}{\Gamma(K)}\Big(\ln\frac{1}{\varepsilon}\Big)^{K-1}\varepsilon^{n_{0} - 1}$
%
\noindent over $[0,1]$.
%
\section{Proofs}\label{App:proofs}
\subsection{Proof of Proposition 2.1}
\begin{proof}
Let $\{\mathcal{B}_j,\,j\in J\}$, be the family of all sufficient sub-$\sigma$-fields of subsets of $\mathcal{A}$ in $\mathcal{E}_m$. 
By \cite[Proposition 0.2.2]{FlorensMouchartRolin1990} the intersection $\mathcal{A}_* := \bigcap_{j\in J}\mathcal{B}_j$ is a sub-$\sigma$-field of $\mathcal{A}$ and is the largest $\sigma$-field contained in each $\mathcal{B}_j$, $j\in J$. We now have to show that $\mathcal{A}_*$ is minimal sufficient. For this, we simply verify conditions \textit{(i)} and \textit{(ii)} in Definition 2.2 in the manuscript. To verify \textit{(i)}, since $\mathcal{A}_*$ is the largest $\sigma$-field contained in each $\mathcal{B}_j$, $j\in J$, then there exists a $j_*\in J$ such that $\mathcal{B}_{j_*} = \mathcal{A}_*$. Condition \textit{(ii)}  follows from the fact that $\{\mathcal{B}_j,\,j\in J\}$ contains all sufficient sub-$\sigma$-fields of $\mathcal{A}$ in $\mathcal{E}_m$ and the fact that $\forall j\in J$, $\mathcal{A}_*\subseteq \mathcal{B}_j$.
\end{proof}

\subsection{Proof of Theorem 2.1}
\begin{proof}
Part (2) of the theorem follows from \cite[Theorem 4.6.18]{FlorensMouchartRolin1990}. Here, we only provide the proof of part (1). This proof follows from \cite[Theorems 0.2.16 ]{FlorensMouchartRolin1990} and the Blackwell theorem, see \cite{DellacherieMeyer1975} and is made of two steps. First, let us recall the definition of the atoms of a $\sigma$-field. If $\mathcal{B}$ is a $\sigma$-field on $\Theta$ we define the equivalence relation:
$$\theta_1\sim\theta_2 \qquad \Leftrightarrow \qquad \mathbbm{1}\{\theta_1\in B\} = \mathbbm{1}\{\theta_2\in B\},\quad \forall B\in\mathcal{B}.$$
The atoms of $\mathcal{B}$ (non necessarily elements of $\mathcal{B}$) are the equivalence classes for this relation. In the first step of the proof one shows that the atoms of $\mathcal{A}_*$ are exactly the equivalence classes defined by the sampling probabilities. This follows by \cite[Theorem 4.6.16]{FlorensMouchartRolin1990} and measurable identification of the function $a(\cdot)$.\\
\indent In the second step of the proof we notice that a real measurable function is constant on the atoms of the $\sigma$-field but the reciprocal requires the technical conditions given in the theorem, see \cite[Theorem III 2.6]{DellacherieMeyer1975}. Under these conditions a function is $\mathcal{A}_*$-measurable if and only if it is constant on the atoms and therefore constant on the equivalence class and then identified.
\end{proof}

\subsection{Proof of Proposition 2.2}
\begin{proof}
By definition of exact estimability, $\mathcal{B}\subset \overline{\mathcal{X}}$. Then, by using \cite[Proposition 4.5.2 (ii)]{FlorensMouchartRolin1990} with $\mathcal{M}_1 = \mathcal{B}$, $\mathcal{M}_2 = \mathcal{X}$ and $\mathcal{M}_3 = \{\emptyset, X\}$ we know that $\mathcal{B}\subset \overline{\mathcal{X}}$ implies $\mathcal{BS} = \mathcal{B}$. Definition 2.7 allows to conclude.
\end{proof}

\subsection{Proof of Theorem 2.3}
\begin{proof}
  It follows from \cite[Definition 4.7.1]{FlorensMouchartRolin1990} and the discussion below it.
\end{proof}
\subsection{Proof of Theorem 2.4}
\begin{proof}
  Remember that $\mathcal{AX}=\overline{\mathcal{A}}_{*}\cap\mathcal{A}$. First, assume $\mathcal{B}$ identified, then $\mathcal{B}\subset \mathcal{AX}$ and, as $\mathcal{AX}\subset \overline{\mathcal{X}}$, we have $\mathcal{B}\subset \overline{\mathcal{X}}$ and $\mathcal{B}$ is exactly estimable.\\
  Reciprocally, let $\mathcal{B}$ be exactly estimable. As $\mathcal{AX}$ is the minimal sufficient $\sigma$-field we have
    \begin{displaymath}
      \mathcal{X}\perp \mathcal{B}|\mathcal{AX}.
    \end{displaymath}
  \noindent This conditional independence implies
  %
      $\overline{\mathcal{X}}\cap \overline{\mathcal{B}}\subset\overline{\mathcal{AX}}$
  %
  \noindent by Corollary 2.2.9 in \cite{FlorensMouchartRolin1990}. In particular, $\mathcal{B}\subset \overline{\mathcal{AX}}$ and $\mathcal{B}$ is identified.
\end{proof}

\subsection{Proof of Proposition 4.1}
\begin{proof}
  Consider the random function $\{A(\cdot,\gamma),\: \gamma\in\Gamma\}$. By definition of random function we have
    \begin{displaymath}
      A^{-1}(A_{0},\gamma) = \{\theta;\: A(\cdot,\gamma)\in A_{0}\}\in \mathcal{A}, \quad \forall A_{0} \in \mathcal{B}(\Phi) \textrm{ and } \forall \gamma\in\Gamma.
    \end{displaymath}
  \noindent Theorem 2.6 in \cite{Molchanov2005} allows to conclude.
\end{proof}

\subsection{Proof of Theorem 5.1}
\begin{proof}
  By the martingale convergence theorem $\mathbf{E}(c|\mathcal{X}_{n}) \rightarrow \mathbf{E}(c|\mathcal{X}_{\infty})$, $\Pi$-a.s. Moreover, as $\mathcal{A}\perp\mathcal{X}_{\infty}|\mathcal{AX}_{\infty}$ we have, see   \cite{FlorensMouchartRolin1990}:
    \begin{displaymath}
      \mathbf{E}(c|\mathcal{X}_{\infty}) = \mathbf{E}(\mathbf{E}(c|\mathcal{AX}_{\infty})|\mathcal{X}_{\infty}) = \mathbf{E}(c|\mathcal{AX}_{\infty}), \quad \Pi-a.s.
    \end{displaymath}
  \noindent due to the exact estimability of $\mathcal{AX}_{\infty}$.
\end{proof}
\begin{spacing}{1}
\setlength{\bibsep}{0.2cm}

\bibliography{AnnaBib}
\end{spacing}